\documentclass[journal,10pt]{IEEEtran}

\usepackage{graphicx}
\usepackage{caption}
\usepackage{subcaption}
\usepackage{mathtools}
\usepackage{bm,amsmath, amssymb}
\usepackage{amsmath}
\usepackage{amsfonts}
\usepackage {amssymb,amsmath}
\usepackage{cite}
\usepackage{epstopdf}
\usepackage{color}
\usepackage{microtype}
\DeclareGraphicsExtensions{.pdf,.png,.jpg}

\newcommand {\Define} {\stackrel {\Delta} {=}  }
\newcommand{\mya}{\mathrel{\overset{\makebox[0pt]{{\tiny(a)}}}{=}}}
\newcommand{\myb}{\mathrel{\overset{\makebox[0pt]{{\tiny(b)}}}{=}}}

\newtheorem{theorem}{Theorem}

\newtheorem{lemma}{Lemma}

\usepackage{cite}

\begin{document}
	\title{\vspace{-8mm} Zak-OTFS to Integrate Sensing the I/O Relation and Data Communication}
\author{\IEEEauthorblockN{Muhammad Ubadah\IEEEauthorrefmark{1}, Saif Khan Mohammed\IEEEauthorrefmark{1},
Ronny Hadani\IEEEauthorrefmark{2},
Shachar Kons\IEEEauthorrefmark{3},
Ananthanarayanan Chockalingam\IEEEauthorrefmark{4} and 
Robert Calderbank\IEEEauthorrefmark{5}~\IEEEmembership{Fellow,~IEEE}}
\\
\IEEEauthorblockA{\IEEEauthorrefmark{1}Department of Electrical Engineering, Indian Institute of Technology Delhi, India}\\
\IEEEauthorblockA{\IEEEauthorrefmark{2}Department of Mathematics, University of Texas at Austin, USA}\\
\IEEEauthorblockA{\IEEEauthorrefmark{3}Cohere Technologies Inc., CA, USA}\\
\IEEEauthorblockA{\IEEEauthorrefmark{4} Department of Electrical Communication Engineering, Indian Institute of Science Bangalore, India}\\
\IEEEauthorblockA{\IEEEauthorrefmark{5}Department of Electrical and Computer Engineering, Duke University, USA}\\
\thanks{This work has been submitted to the IEEE for possible publication. Copyright may be transferred without notice, after which this version may no longer be accessible.}
\thanks{S. K. Mohammed is also associated with the Bharti School of Telecom. Tech. and Management (BSTTM), IIT Delhi. The work of S. K. Mohammed was supported in part by the Prof. Kishan and Pramila Gupta Chair at I.I.T. Delhi, and also in part by the Jai Gupta Chair at I.I.T. Delhi.}
}
 
	\maketitle
	
	\begin{abstract}   
 The Zak-OTFS input/output (I/O) relation is predictable and non-fading when the delay and Doppler periods are greater than the effective channel delay and Doppler spreads, a condition which we refer to as the crystallization condition. The filter taps can simply be read off from the response to a single Zak-OTFS point (impulse) pulsone waveform, and the I/O relation can be reconstructed for a sampled system that operates under finite duration and bandwidth constraints. Predictability opens up the possibility of a model-free mode of operation. The time-domain realization of a Zak-OTFS point pulsone is a pulse train modulated by a tone, hence the name, pulsone. The Peak-to-Average Power Ratio (PAPR) of a pulsone is about $15$ dB, and we describe a general method for constructing a spread pulsone for which the time-domain realization has a PAPR of about 6dB. We construct the spread pulsone by applying a type of discrete spreading filter to a Zak-OTFS point pulsone. The self-ambiguity function of the point pulsone is supported on the period lattice ${\Lambda}_{p}$, and by applying a discrete chirp filter, we obtain a spread pulsone with a self-ambiguity function that is supported on a rotated lattice ${\Lambda^*}$. We show that if the channel satisfies the crystallization conditions with respect to ${\Lambda^*}$ then the effective DD domain filter taps can simply be read off from the cross-ambiguity between the channel response to the spread pulsone and the transmitted spread pulsone. If, in addition, the channel satisfies the crystallization conditions with respect to the period lattice ${\Lambda}_{p}$, then in an OTFS frame consisting of a spread pilot pulsone and point data pulsones, after cancelling the received signal corresponding to the spread pulsone, we can recover the channel response to any data pulsone.
 We integrate sensing the I/O relation and data communication
 within a single OTFS subframe using geometric properties of a lattice ${\Lambda}_{p}$ used for data transmission and a rotated lattice ${\Lambda^*}$ used for sensing. The spread pilot pulsone looks like noise to the point data pulsones, and it is this incoherence that makes it possible to integrate communications and sensing without time-sharing delay-Doppler resources. We demonstrate that it is possible to increase effective throughput by integrating spread pilot pulsones and data communication.
	\end{abstract} 
	
 \vspace{-2mm}
	\normalsize
	\section{Introduction}
	\label{secintro}
In the 1990s, we witnessed the transition from narrowband Time-Division Multiple Access (TDMA) systems like IS-136 to Code-Division Multiple Access (CDMA). Spreading produced a waveform with excellent Peak-to-Average Power Ratio (PAPR), but the method of spreading was not connected to the dynamics of the channel, making equalization very challenging. Huge demand for high-speed data then led to the transition from CDMA to Orthogonal Frequency-Division Multiplexing (OFDM), and though we lost the benefits of PAPR, we gained the benefits of measuring and adapting to instantaneous channel gains. In 6G propagation environments, as we encounter Doppler spreads measured in KHz, especially for non-terrestrial networks, it is becoming more and more difficult to estimate channels, and the standard channel-model-dependent approach to wireless communication is starting to break down. This paper describes a physical layer that is designed to integrate sensing the I/O relation and communicating data (ISAC) in challenging propagation environments, typical of 6G networks.

4G and 5G wireless communication networks use OFDM. Here, a cyclic prefix is used to create shared eigenfunctions of the group of time shifts, and we have shared eigenfunctions because the
time shift group is commutative. Note that pulses in the time domain serve as geometric modes that are moved around by the Linear Time-Invariant (LTI) channel. In this paper, we focus on doubly-spread channels, where there are no shared eigenfunctions because delay shifts and Doppler shifts do not commute. Nevertheless, pulses in the delay-Doppler domain enable sensing, since they are geometric modes that are moved around by the Linear Time-Variant (LTV) channel. These pulses are the Orthogonal Time Frequency Space (OTFS) carrier waveforms,
their time domain (TD) realizations are called pulsones,
and Section \ref{sec_zakotfs} describes the OTFS system model (see Chapter $2$ in \cite{otfsbook} for an exhaustive treatment of Zak-OTFS modulation). 

A pulse in the delay-Doppler DD domain is a quasi-periodic localized function, defined by a delay period $\tau_p$ and a Doppler period $\nu_p$. We have shown that the Zak-OTFS Input / Output (I/O) relation is predictable and non-fading when the delay period $\tau_p$ is greater than the effective channel delay spread, and the Doppler period $\nu_p$ is greater than the effective Doppler spread \cite{zakotfs1, zakotfs2}. We refer to this condition as the crystallization condition with respect to the period lattice and speak of operating in the crystalline regime. We have explained how non-predictability and fading result from aliasing in the DD domain, and why the crystallization condition prevents aliasing (see Section II in \cite{zakotfs2}). In the crystalline regime we have shown that the effective DD domain channel filter taps can simply be read off from the response to a single Zak-OTFS point pulsone \cite{zakotfs1, zakotfs2}.\footnote{\footnotesize{In \cite{zakotfs1, zakotfs2}, point pulsones enabled sensing and data transmission in separate Zak-OTFS subframes. However, in this paper we consider Zak-OTFS subframes which consists of both sensing and data pulsones.}} 
Some Zak-OTFS implementations proposed in literature are based on the discrete Zak transform (see Chapter $8$ in \cite{otfsbook} and also \cite{Lampel22}) and those based on TF windowing \cite{Hanly23}.

Why OTFS rather than OFDM? The first reason is that 6G propagation environments are changing the balance between time-frequency methods characteristic of OFDM and delay-Doppler methods. In OFDM, once the I/O relation is known, equalization is relatively simple, at least when there is no inter-carrier interference (ICI) \cite{Nee2000}, \cite{Wang2006}. However, acquisition of the I/O relation is non-trivial and model-dependent. In contrast, equalization is more involved in OTFS, due to intersymbol interference, but acquisition of the I/O relation is simple and model-free (it can be read off from the response to a single point pulsone \cite{zakotfs1, zakotfs2}). Acquisition becomes more critical in 6G, as Doppler spreads measured in KHz make it more and more challenging to estimate channels. 

The second reason is that it is possible to integrate sensing the I/O relation with data communication without compromising effective capacity. This paper describes how to spread an OTFS point pulsone so that sensing is simple, and so that pilot waveform and data can coexist in the same OTFS subframe.

Section \ref{zakotfspointpilot} introduces sensing the I/O relation through a discussion of practical limitations on pilot design with a point pulsone. A pulsone is a pulse train modulated by a tone, and this is a waveform with high PAPR. High PAPR necessitates high-power linear amplifiers, and unfortunately these are power-inefficient. Also, when we use the same lattice for communication and sensing we end up dedicating delay Doppler resources to either sensing or communication. For example, when we combine point data symbols with a point (impulse) pilot in a single OTFS subframe, we need to avoid interference between data symbols and pilot. The standard approach is to introduce a guard band, but this is an overhead that reduces spectral efficiency, and the greater the channel spread, the greater is the overhead. This tradeoff motivates our use of different lattices for communications and sensing in Section \ref{zakotfsspread}.

We also need to solve the issue of high PAPR which plagues multicarrier modulation. For example, OFDM signals are generated by adding many subcarrier components, and they can have high peak values in the time domain. As a consequence, OFDM systems suffer from high PAPR compared with single carrier systems.  Many techniques have been proposed for reducing PAPR in OFDM (see Chapter 7 in \cite{YR2013} for a review) but none avoid degrading BER performance. Over the past several years several variants of OTFS have been reported in the literature \cite{Thaj2022}. A multicarrier approximation to Zak-OTFS, which we refer to as MC-OTFS, has been the focus of most research attention so far \cite{Hadani2017}, \cite{Hadani2018}, \cite{Bestreadings2022}. MC-OTFS suffers from high PAPR, techniques for PAPR reduction are presented in \cite{Surabhi2019}, \cite{Wei2022}, \cite{Hossain2020}, and again none avoid degrading BER performance.

Zak-OTFS modulation starts with a quasi-periodic Dirac-delta DD domain pulse, and we apply a DD domain pulse shaping filter. The time-domain (TD) representation of the filtered signal is a TD pulsone with time duration $T$ and bandwidth $B$ inversely proportional to the Doppler and delay spread respectively of the filter (see Fig.~\ref{fig4_paper1}). We transmit information using non-overlapping DD domain pulses spaced $1/T$ apart along the delay axis and $1/B$ apart along the Doppler axis. Since each pulse repeats quasi-periodically, there are $M = \tau_p/(1/B) = B \tau_p$ pulse locations along the delay axis and $N = \nu_p/(1/T ) = T \nu_p$ pulse locations along the Doppler axis. The number of distinct non-overlapping information carriers is the time-bandwidth product $BT = MN$, and the pulse locations are the points in the information lattice. 

A discrete quasi-periodic DD domain signal is periodic along both delay and Doppler axes with period $MN$. In Section \ref{secfiltering}, we define a discrete DD domain filter to be a discrete periodic DD domain function with period $MN$ along both delay and Doppler axes ($M^2N^2$ degrees of freedom). Section \ref{secfiltering} develops the fundamentals of filtering in the discrete DD domain. We describe how we obtain a spread pilot by applying a discrete spreading filter to a DD domain signal localized at a point in the information lattice. This is a general method, but we restrict our attention to a proof-of-concept demonstration using a discrete chirp filter which distributes energy equally to all lattice points. Different lattice points correspond to pulsones with different delays and Doppler shifts; hence we observe a TD signal with almost constant amplitude. The PAPR of the spread pulsone is about $5$ dB, significantly less than the PAPR of the point pulsone, which is about $15$ dB.  

Section \ref{zakotfsspread} describes how we recover the effective channel filter by de-spreading the received signal. We estimate the taps of the discrete effective channel filter from the discrete cross-ambiguity function between the received DD signal and the transmitted spread pilot signal. The self-ambiguity function of a point pulsone is a rectangular lattice $\Lambda_{z,p}$, and we show that the self-ambiguity function of the spread pulsone is obtained by rotating $\Lambda_{p}$ to obtain a lattice $\Lambda^*$.  If the channel satisfies the crystallization conditions defined by $\Lambda_{p}$ then we show that we are able to accurately estimate the discrete effective channel taps using a point pulsone. If the channel satisfies the crystallization conditions defined by  $\Lambda^*$ then we show that we are able to accurately estimate the discrete effective channel taps using the spread pulsone. The method we propose for spreading the pilot is connected to the dynamics of the sensing environment.

We emphasize that we consider transmission of pilot and data within a single OTFS subframe with no division of DD domain resources between sensing and communication. Data is carried by point pulsones, the data signal interferes with channel estimation using a spread pulsone, and this interference adversely affects the estimation of certain discrete effective channel filter taps. Section \ref{integdatasense} explores how the significance of this interference depends on the ratio of pilot power to data power. After equalization, the spread pulsone interferes with data carried by the point pulsones, and again the degree of interference depends on the ratio of pilot power to data power. Section \ref{integdatasense} identifies how to choose this ratio
to integrate sensing the I/O relation and data communication within a single OTFS subframe and avoid time-sharing of delay-Doppler resources.

Passive radar is an important instance of integrated sensing and communication. Passive radar exploits readily available, non-cooperative sources of radio energy to measure reflections from the environment and targets of interest. When a suitable illuminator is available, covert surveillance becomes possible, without the need for deployment and operation of a dedicated transmitter. Terrestrial digital television transmissions (DVB–T) provide an especially attractive opportunity for radar \cite{Palmer2013}. We describe how to design the illuminator in a passive radar to be a geometric mode of the radar scene (a spread pulsone). We describe how this choice simplifies acquisition of the radar scene, since the scene can be simply read off from the received signal, unlike the DVB-T system. The radar application also benefits from the noise-like characteristics of the spread pulsone which increase the fraction of energy on target. We now highlight our main contributions:

\textbf{Pilot Design}: We have described a method of constructing a spread pulsone, where the self-ambiguity function of the spread pulsone is supported on a lattice $\Lambda^*$ obtained by rotating the period lattice $\Lambda_{p}$ associated with the data-bearing pulsones. We have shown that if the channel satisfies the crystallization conditions with respect to $\Lambda^*$ then the effective DD domain filter taps can simply be read off from the channel response. If the channel also satisfies the crystallization conditions with respect to the $\Lambda_{p}$ then given the I/O response at one point in the OTFS subframe, it is possible to predict the I/O response at all other points in the subframe. We have also shown that the spread pulsone is a noise-like waveform with excellent PAPR.

\textbf{Sensing the I/O Relation and Data Communication}: We have integrated the two functions within a single OTFS subframe using geometric properties of a lattice $\Lambda_{p}$ used for data transmission and a rotated lattice $\Lambda^*$ used for sensing. The data pulsones looks like noise to the sensing pulsones, and we have demonstrated that this incoherence makes it possible to integrate sensing and communication without the loss in spectral efficiency associated with time-sharing delay-Doppler resources. We have demonstrated that effective throughput is increased
by avoiding time-sharing of delay-Doppler resources. An important application is the design of a passive radar system.

\vspace{-1mm}
\section{System model} 	\label{secsystem}
\label{sec_zakotfs}
A time-domain (TD) pulse is an ideal waveform for delay-only channels (where paths induce zero Doppler shift) since
it is possible to separate signals received along different paths according to their path length/distance. Similarly, a frequency-domain (FD) pulse is an ideal waveform for Doppler-only channels
since it is possible to separate signals received along different paths according to the Doppler shift induced on the transmitted signal. However, neither a TD pulse nor a FD pulse is suited for doubly-spread channels where paths induce both delay and Doppler shift.

In this section, we describe how a pulse in the DD domain is matched to doubly-spread channels.
A pulse in the DD domain is a quasi-periodic localized function, defined by a delay period $\tau_p$ and Doppler period $\nu_p = 1/\tau_p$. In the \emph{period lattice} $\Lambda_p = \{ \left(n \tau_p, m \nu_p \right) \, | \, n,m \in {\mathbb Z} \}$, there is only one pulse within the \emph{fundamental region} ${\mathcal D}_0 = \{ (\tau, \nu) \, | \, 0 \leq \tau < \tau_p, 0 \leq \nu < \nu_p \}$, and there are infinitely many replicas along the delay and Doppler axes given by
\begin{eqnarray}
\label{eqn_qperiodic}
x_{\mbox{\scriptsize{dd}}}(\tau + n \tau_p, \nu + m \nu_p) & = & e^{j 2 \pi n \nu \tau_p} \, x_{\mbox{\scriptsize{dd}}}(\tau , \nu )
\end{eqnarray}for all $n,m \in {\mathbb Z}$. Only quasi-periodic DD domain functions can have a TD representation. When viewed in the time domain, this function is realized as a pulse train modulated by a tone (see Fig.~\ref{fig4_paper1}), hence the name \emph{pulsone}. The DD domain pulse is the carrier waveform for Zak-OTFS modulation \cite{zakotfs1, zakotfs2, zakotfsbook}.

\subsection{Zak-OTFS modulation}

\begin{figure}[h]
\centering
\includegraphics[width=5.5cm, height=7.2cm]{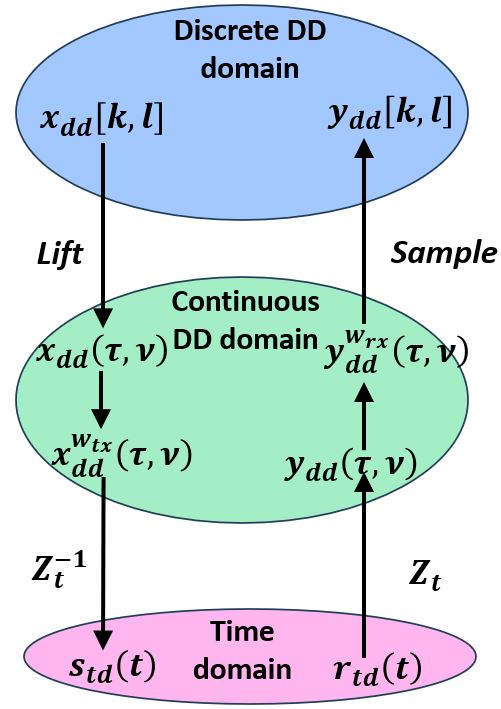}
\caption{
Zak-OTFS transceiver signal processing.}
\label{fig0}
\end{figure}

We consider Zak-OTFS modulation with modulation parameters $(\tau_p, \nu_p), \nu_p = 1/\tau_p$.
The transmitted TD Zak-OTFS frame is limited to a time duration $T = N \tau_p$
and bandwidth $B = M \nu_p$. Fig.~\ref{fig0} illustrates Zak-OTFS transceiver processing.

Let $x[k,l]$, $k=0,1,\cdots, M -1$, $l=0,1,\cdots, N-1$, denote the $B \, T = M \, N$ information symbols, each having unit average energy, i.e., ${\mathbb E} [ \vert x[k,l] \vert^2 ] = 1$. These $MN$ information symbols are encoded into a discrete DD domain
information signal $x_{\mbox{\scriptsize{dd}}}[\cdot,\cdot]$ given by
\begin{eqnarray}
\label{eqn_01}
x_{\mbox{\scriptsize{dd}}}[k + nM,l +mN] & \Define & e^{j 2 \pi \frac{n l}{N}} \, x[k,l] \, 
\end{eqnarray}for all $k=0,1,\cdots, M -1$, $l=0,1,\cdots, N-1$, $n,m \in {\mathbb Z}$. From this it follows that
\begin{eqnarray}
    \label{eqn_qp}
    x_{\mbox{\scriptsize{dd}}}[k + nM,l + mN] & = &  e^{j 2 \pi \frac{ n l}{N}} \, x_{\mbox{\scriptsize{dd}}}[k ,l ] 
\end{eqnarray}for all $k,l,n,m \in {\mathbb Z}$. In other words, the information signal $x_{\mbox{\scriptsize{dd}}}[k,l]$
is quasi-periodic, with period $M$ along the delay axis and $N$ along the Doppler axis. The encoding of the $MN$ information symbols to $x_{\mbox{\scriptsize{dd}}}[k,l]$ is carried out as in (\ref{eqn_01}), so that $x_{\mbox{\scriptsize{dd}}}[k,l]$ is quasi-periodic, since only quasi-periodic DD functions have TD realizations. 

The information signal $x_{\mbox{\scriptsize{dd}}}[k ,l ]$ is then converted to a continuous quasi-periodic DD signal by lifting it to the \emph{information lattice} $\Lambda_{\mbox{\scriptsize{dd}}} = \{ (k \tau_p/M, l \nu_p/N) \, | \, k,l \in {\mathbb Z} \}$, i.e.\footnote{\footnotesize{In this paper we use $\delta$ with a square bracket i.e., $\delta[\cdot]$ to denote a Dirac-delta function in a discrete domain and we use $\delta(\cdot)$ to denote a Dirac-delta function in a continuous domain.}}
\begin{eqnarray}
\label{eqn_discreteinformationsig}
x_{\mbox{\scriptsize{dd}}}(\tau, \nu) & \hspace{-2mm} \Define &  \hspace{-2mm} \sum\limits_{k,l \in {\mathbb Z}}  x_{\mbox{\scriptsize{dd}}}[k ,l ]  \, \delta(\tau - k \tau_p/M) \, \delta(\nu - l \nu_p/N).
\end{eqnarray}This signal is quasi-periodic in the continuous DD domain with periods $\tau_p$ and $\nu_p$ respectively along the delay and Doppler axis, i.e.
for all $n,m \in {\mathbb Z}$ (see (\ref{prelim_2}) in Appendix \ref{prelim_zak})
\begin{eqnarray}
    \label{eqn_qp_continuous}
    x_{\mbox{\scriptsize{dd}}}(\tau + n \tau_p, \nu + m \nu_p) & = & e^{j 2 \pi n \nu \tau_p} \, x_{\mbox{\scriptsize{dd}}}(\tau , \nu ).
\end{eqnarray}Pulse shape filtering in the DD domain is implemented by twisted convolution.\footnote{\footnotesize{Twisted convolution is denoted by $*_{\sigma}$. Twisted convolution between two DD functions $a(\tau, \nu)$ and $b(\tau, \nu)$ is given by $c(\tau, \nu) = a(\tau, \nu) *_{\sigma} b(\tau, \nu) = \iint a(\tau', \nu') \, b(\tau - \tau', \nu - \nu') \, e^{j 2 \pi \nu' (\tau - \tau')} \, d\tau' \, d\nu'$. Twisted convolution is associative, i.e., $b(\tau, \nu) *_{\sigma} ( c(\tau, \nu) *_{\sigma} d(\tau, \nu)) = (b(\tau, \nu) *_{\sigma} c(\tau,\nu) ) *_{\sigma} d(\tau, \nu)$. However it is not commutative, i.e., $a(\tau, \nu) *_{\sigma} b(\tau, \nu) \ne b(\tau, \nu) *_{\sigma} a(\tau, \nu)$.}} Twisted convolution of $x_{\mbox{\scriptsize{dd}}}(\tau, \nu)$ with a pulse shaping filter $w_{tx}(\tau, \nu)$ gives
\begin{eqnarray}
\label{eqnxddwtx56}
    x_{\mbox{\scriptsize{dd}}}^{w_{tx}}(\tau , \nu ) & \Define & w_{tx}(\tau, \nu) \, *_{\sigma} \, x_{\mbox{\scriptsize{dd}}}(\tau , \nu ).
\end{eqnarray}Since twisted convolution preserves quasi-periodicity, $x_{\mbox{\scriptsize{dd}}}^{w_{tx}}(\tau , \nu )$ is also quasi-periodic. Also, appropriate pulse shaping guarantees that the transmit TD Zak-OTFS modulated signal has time duration $T$ and bandwidth $B$. 

The inverse Zak-transform of $x_{\mbox{\scriptsize{dd}}}^{w_{tx}}(\tau , \nu )$ gives the transmit TD Zak-OTFS modulated signal\footnote{\footnotesize{A brief introduction to the Zak transform and its inverse, and to quasi-periodicity is provided in Appendix \ref{prelim_zak}.}}
\begin{eqnarray}
\label{eqn_stdt}
    s_{\mbox{\scriptsize{td}}}(t) & = & {\mathcal Z_t}^{-1}{\Big (}  x_{\mbox{\scriptsize{dd}}}^{w_{tx}}(\tau , \nu ) {\Big )}.
\end{eqnarray}
Equivalently, the transmit TD signal $s_{\mbox{\scriptsize{td}}}(t)$ is given by (see Theorem \ref{thm0} in Appendix \ref{app_pulsones})
\begin{eqnarray}
    s_{\mbox{\scriptsize{td}}}(t) & = & {\mathcal Z_t}^{-1}{\Big (}  x_{\mbox{\scriptsize{dd}}}^{w_{tx}}(\tau , \nu ) {\Big )} \nonumber \\
    & = & \sum\limits_{k=0}^{M-1} \sum\limits_{l=0}^{N-1} x[k,l] \,  s_{\mbox{\scriptsize{td}},k,l}(t)
\end{eqnarray}where $s_{\mbox{\scriptsize{td}},k,l}(t)$ is the carrier waveform for the $(k,l)$-th information symbol $x[k,l]$.
A detailed derivation and discussion of the TD, FD and DD domain realization of these Zak-OTFS carrier waveforms is provided in Appendix \ref{app_pulsones}.
Fig.~\ref{fig4_paper1} provides an illustration of the Zak-OTFS carrier waveform.
In the DD domain it is a quasi-periodic pulse. In both the TD/FD domains, the carrier waveform is a pulse train modulated by a tone and is therefore called a TD/FD pulsone.
\begin{figure*}[!h]
\centering
\includegraphics[width=14cm, height=8.0cm]{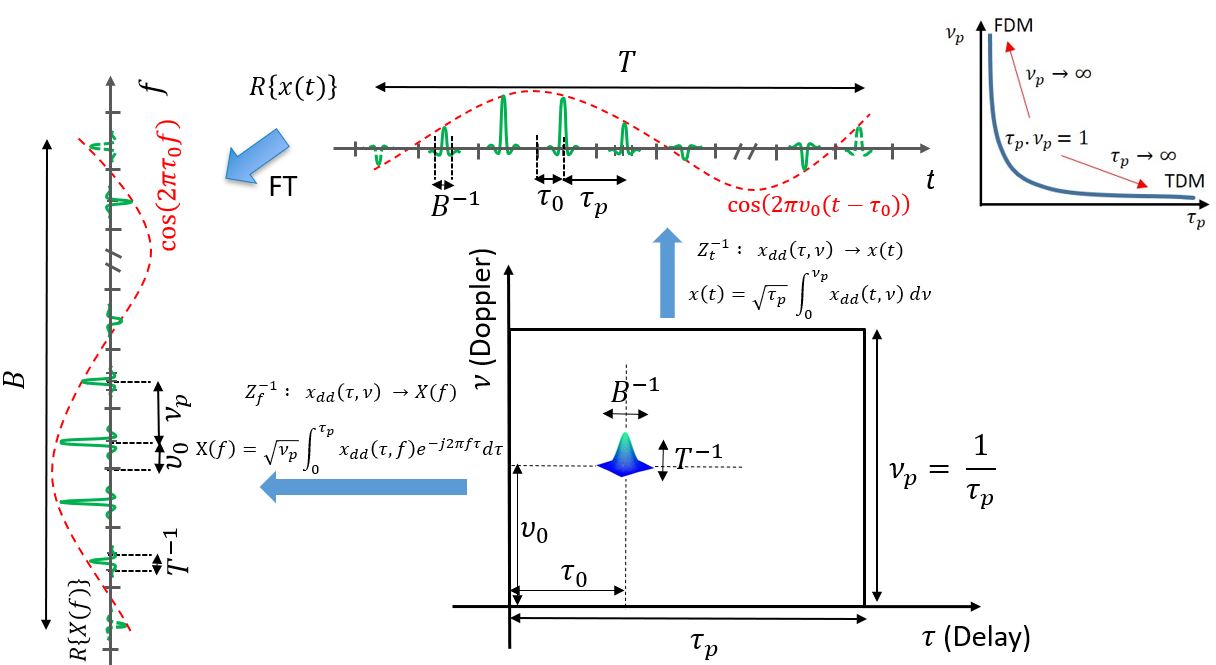}
\caption{The $(k,l)$-th quasi-periodic DD domain pulse located at $(\tau_0 , \nu_0) = {\Big(} \frac{k \tau_p}{M} \,,\, \frac{l \nu_p}{N} {\Big )}$ and its TD/FD realizations referred to as TD/FD pulsone. The TD pulsone comprises of a finite duration pulse train modulated by a TD tone. The FD pulsone comprises of a finite bandwidth pulse train modulated by a FD tone. The location of the pulses in the TD/FD pulse train and the frequency of the modulated TD/FD tone is determined by the location of the DD domain pulse $(\tau_0,\nu_0)$. The time duration and bandwidth of a pulsone are inversely proportional to the characteristic width of the DD domain pulse along the Doppler axis and the delay axis, respectively. As $\tau_p \rightarrow \infty$, the TD pulsone approaches a single TD pulse which is suited for delay-only channels. Similarly, as $\nu_p \rightarrow \infty$, the FD pulsone approaches a single FD pulse which is suited for carrying information in Doppler-only channels.  Zak-OTFS is therefore a family of modulations parameterized by $\tau_p$ that interpolates between TD pulse modulation (TDM) and FD pulse modulation (FDM). Source: Fig.~$4$ in \cite{zakotfs1}.}
\label{fig4_paper1}
\end{figure*}

\subsection{Zak-OTFS receiver}
\label{seczakotfsreceiver}
The received TD signal is given by
\begin{eqnarray}
\label{eqn_rtdt}
    r_{\mbox{\scriptsize{td}}}(t) & \hspace{-3.5mm} = & \hspace{-3.5mm} \iint \hspace{-1mm} h_{\mbox{\scriptsize{phy}}}(\tau, \nu) \, s_{\mbox{\scriptsize{td}}}(t - \tau)  \, e^{j 2 \pi \nu (t - \tau)} \, d\tau \, d\nu \, + \, n_{\mbox{\scriptsize{td}}}(t)
\end{eqnarray}where $h_{\mbox{\scriptsize{phy}}}(\tau, \nu)$ is the channel DD spreading function and $n_{\mbox{\scriptsize{td}}}(t)$ is the AWGN at the receiver.
The DD representation $y_{\mbox{\scriptsize{dd}}}(\tau, \nu)$ of  $r_{\mbox{\scriptsize{td}}}(t)$ is given by $y_{\mbox{\scriptsize{dd}}}(\tau, \nu) = {\mathcal Z_t} {\Big (} r_{\mbox{\scriptsize{td}}}(t) {\Big )}$.
Matched filtering with the receive DD pulse $w_{rx}(\tau, \nu)$ results in the signal
\begin{eqnarray}
    y_{\mbox{\scriptsize{dd}}}^{w_{rx}}(\tau, \nu) = w_{rx}(\tau, \nu) \, *_{\sigma} \, y_{\mbox{\scriptsize{dd}}}(\tau, \nu).
\end{eqnarray}Note that $y_{\mbox{\scriptsize{dd}}}^{w_{rx}}(\tau, \nu)$ is quasi-periodic because twisted convolution preserves quasi-periodicity. Sampling $y_{\mbox{\scriptsize{dd}}}^{w_{rx}}(\tau, \nu)$ on the information lattice $\Lambda_{\mbox{\scriptsize{dd}}}$ gives
\begin{eqnarray}
\label{eqn_yddkl}
    y_{\mbox{\scriptsize{dd}}}[k,l] & = & y_{\mbox{\scriptsize{dd}}}^{w_{rx}}\left( \tau = \frac{k \tau_p}{M}, \nu = \frac{l \nu_p}{N}  \right)
\end{eqnarray}for all $k,l \in {\mathbb Z}$.

\subsection{Zak-OTFS I/O relation}
Combining all equations above, from (\ref{eqn_discreteinformationsig}) to (\ref{eqn_yddkl}), gives the I/O relation
\begin{eqnarray}
\label{eqn_io_relation}
y_{\mbox{\scriptsize{dd}}}[k,l] & = & h_{\mbox{\scriptsize{eff}}}[k,l] \, *_{\sigma} \, x_{\mbox{\scriptsize{dd}}}[k,l] \, + \, n_{\mbox{\scriptsize{dd}}}[k,l],
\end{eqnarray}
where the effective channel $h_{\mbox{\scriptsize{eff}}}[k,l]$ is given by
\begin{eqnarray}
    h_{\mbox{\scriptsize{eff}}}[k,l] & = & h_{\mbox{\scriptsize{eff}}}{\Big (}\tau = k \frac{\tau_p}{M}, \nu = l \frac{\nu_p}{N} {\Big )}, \,\,\, \mbox{\small{where}} \nonumber \\
    h_{\mbox{\scriptsize{eff}}}(\tau, \nu) & = & w_{rx}(\tau, \nu) \, *_{\sigma} \, h_{\mbox{\scriptsize{phy}}}(\tau, \nu) \, *_{\sigma} \, w_{tx}(\tau, \nu).
\end{eqnarray}In (\ref{eqn_io_relation}), $n_{\mbox{\scriptsize{dd}}}[k,l]$ is the discrete DD domain noise signal. Also, in (\ref{eqn_io_relation}), $*_{\sigma}$ denotes twisted convolution between two discrete DD domain functions $h_{\mbox{\scriptsize{eff}}}[k,l]$ and $x_{\mbox{\scriptsize{dd}}}[k,l]$
which is explicitly given by
\begin{eqnarray}
    \label{twistedconveqn}
    h_{\mbox{\scriptsize{eff}}}[k,l] \, *_{\sigma} \, x_{\mbox{\scriptsize{dd}}}[k,l] & &  \nonumber \\
    & & \hspace{-31mm} = \sum\limits_{k', l' \in {\mathbb Z}} h_{\mbox{\scriptsize{eff}}}[k',l'] \, x_{\mbox{\scriptsize{dd}}}[k - k',l - l'] \, e^{j 2 \pi \frac{l'(k - k')}{MN}}.
\end{eqnarray}

Quasi-periodicity of $y_{\mbox{\scriptsize{dd}}}[k,l]$ implies that we may detect the transmitted information symbols $x[k,l], k=0,1,\cdots, M-1$, $l=0,1,\cdots, N-1$ using the symbols $y_{\mbox{\scriptsize{dd}}}[k,l], k=0,1,\cdots, M-1$, $l=0,1,\cdots, N-1$. This yields the matrix-vector form of the I/O relation where the $MN \times 1$ vector of these $MN$ received symbols is given by the product of an $MN \times MN$ effective channel matrix with the $MN \times 1$ vector of information symbols. It is possible to detect transmitted Zak-OTFS information symbols using techniques developed for MIMO equalization since the I/O relations share the same form.

In this paper, for DD domain pulse shaping we only consider sinc and root raised cosine (RRC) pulses which are respectively given by
\begin{eqnarray}
\label{eqn_sincwtx}
w_{tx}(\tau, \nu) = \sqrt{B T } \, sinc (\tau B) \, sinc(\nu T),
\end{eqnarray}and
\begin{eqnarray}
\label{eqn_rrcwtx}
w_{tx}(\tau, \nu) & = & \sqrt{BT} \, rrc_{_{\beta_{\tau}}}( B \tau ) \,  rrc_{_{\beta_{\nu}}}( T \nu ), \nonumber \\
\end{eqnarray}where for $0 \leq \beta \leq 1$
\begin{eqnarray}
\label{rrceqn1}
rrc_{_{\beta}}(x) & \hspace{-3mm} = & \hspace{-3mm} \frac{\sin(\pi x (1 - \beta)) + 4 \beta x \cos(\pi x (1 + \beta))}{\pi x \left( 1 -( 4 \beta x )^2 \right)}.
\end{eqnarray}Note that $\beta = 0$ coincides with a sinc filter. Choosing $\beta > 0$ trades energy leakage outside the main lobe and  expansion in time/bandwidth. The receiver pulse shaping filter for both sinc and RRC waveforms is given by
\begin{eqnarray}
\label{eqn_wrx}
    w_{rx}(\tau, \nu) & = & e^{j 2 \pi \nu \tau} \, w_{tx}^*(-\tau, -\nu).
\end{eqnarray}

\section{Zak-OTFS with point pilot pulsone}
\label{zakotfspointpilot}
We have shown (see \cite{zakotfs1} and \cite{zakotfs2}) that the Zak-OTFS I/O relation is predictable and non-fading when the delay period $\tau_p$ is greater than the effective channel delay spread (the spread of $h_{\mbox{\scriptsize{eff}}}(\tau, \nu)$ along the delay axis) and the Doppler period is greater than the effective channel Doppler spread (the spread of $h_{\mbox{\scriptsize{eff}}}(\tau, \nu)$ along the Doppler axis). This is the \emph{crystallization condition}. In the \emph{crystalline regime} (when the crystallization condition holds), we have shown that it is possible to read off the taps of the effective DD domain channel filter $h_{\mbox{\scriptsize{eff}}}[k,l]$ from the response to a single Zak-OTFS pulsone.
We often refer to this pilot waveform as a \emph{point pilot} since it's DD domain realization is a localized quasi-periodic pulse. While simple and effective, this is not the maximum likelihood (ML) estimate. In this Section, we connect to the active sensing community by describing the ML method for estimating the taps of $h_{\mbox{\scriptsize{eff}}}[k,l]$ in terms of ambiguity functions.

We begin by reviewing the model-dependent and model-free approaches
to acquiring the I/O relation in the crystalline regime (see \cite{zakotfs2}
for more details). 
In the model-dependent
approach we impose a model on the DD spreading function $h_{\mbox{\scriptsize{phy}}}(\tau, \nu)$, typically by prescribing a finite number of paths and constraining their delay and Doppler shifts. Given this model, the receiver estimates $h_{\mbox{\scriptsize{phy}}}(\tau, \nu)$. The accuracy of this estimate is limited by the time and bandwidth constraints on the pilot signal and by any mismatch between the channel model and the physical channel.

In the model-free approach, the receiver estimates the taps of the effective
DD domain channel filter $h_{\mbox{\scriptsize{eff}}}[k,l]$ from the response to a pilot waveform. This approach is similar in spirit to estimating the taps of an effective discrete-time LTI channel instead of estimating the continuous-time impulse response of the underlying physical channel. 
\begin{table}
\caption{Power Delay Profile of Doubly-spread Veh-A Channel.}
\centering
\begin{tabular}{ | c || c |  c |  c | c |  c |  c |} 
  \hline
   Path no. $i$ & $1$ & $2$ & $3$ & $4$ & $5$ & $6$  \\
   \hline
  Rel. Delay $\tau_i$ ($\mu s$) & $0$ & $0.31$ & $0.71$ & $1.09$ & $1.73$ & $2.51$  \\ 
  \hline
  Rel. Power $\frac{{\mathbb E}[\vert h_i \vert^2]}{{\mathbb E}[\vert h_1 \vert^2]}$ (dB)      & $0$   &  $-1$  &  $-9$ &  $-10$ & $-15$  &  $-20$ \\
  \hline
\end{tabular}
\label{tab1_paper2}
\end{table}

Here and throughout this paper, we consider the Veh-A channel model \cite{EVAITU} which consists of six channel paths. The delay-Doppler spreading function is given by
\begin{eqnarray}
\label{paper2_eqn13}
h_{_{\mbox{\scriptsize{phy}}}}(\tau,\nu) & = & \sum\limits_{i=1}^6 h_i \, \delta(\tau - \tau_i) \, \delta(\nu - \nu_i).
\end{eqnarray}
where $h_i, \tau_i,$ and $\nu_i$ respectively
denote the gain, delay, and Doppler shift of the $i$-th channel path.
Table-\ref{tab1_paper2} lists the power-delay profile for the six channel paths. The channel delay spread is $\tau_{max} \Define \max_i \tau_i \, - \, \min_i \tau_i = 2.5 \, \mu s$. For a maximum Doppler shift of $\nu_{max} = 815$ Hz, the Doppler spread is $2 \times 0.815 = 1.63$ KHz.  
The Doppler shift of the $i$-th path is modeled as $\nu_i = \nu_{max} \cos(\theta_i)$, where the variables $\theta_i, i=1,2,\cdots, 6$ are independent and distributed uniformly in the interval $[0 \,,\, 2\pi)$.

We focus on the Veh-A channel first because it is representative of real propagation
environments, and second because it is very difficult to make the model-dependent
mode of operation work.
As an example, for a channel bandwidth $B = 0.96$ MHz, the delay domain resolution is $1/B \approx 1.04 \, \mu s$, and the first three paths introduce delay shifts in the interval $[ 0 \,,\, 0.71] \, \mu s$ which is less than the delay domain resolution. These paths are therefore not separable, and so cannot be estimated accurately. One can choose a higher bandwidth so that the resolution is smaller than the path difference between any two paths thereby making the paths separable. However, this would only work for a particular path delay profile. In real scenarios, the path delays can change, however bandwidth cannot be increased indefinitely. 

\begin{figure}[h]
\includegraphics[width=9cm, height=6.5cm]{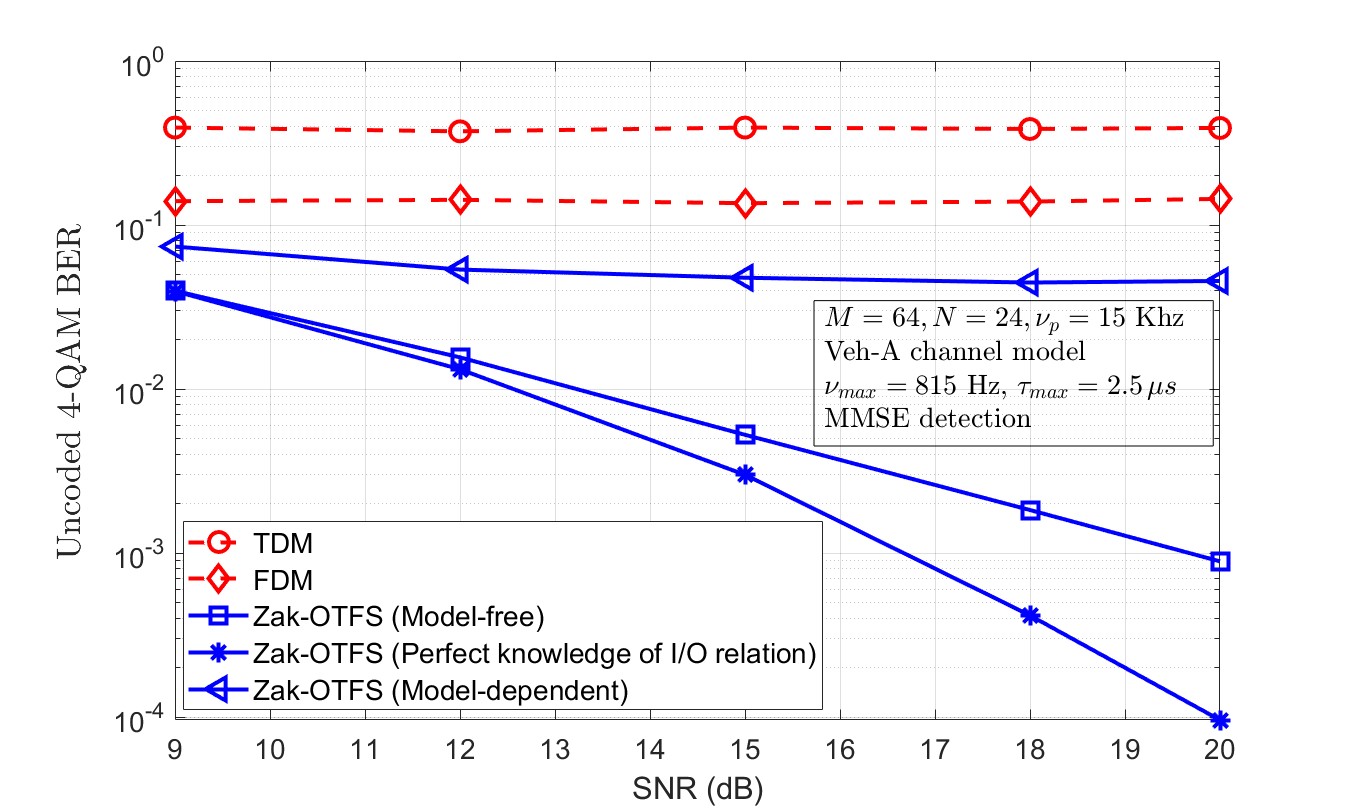}
\caption{BER performance of Zak-OTFS in model-dependent and model-free mode of operation. In the crystalline regime ($\nu_p = 15$ KHz), BER performance of the model-free mode of Zak-OTFS is only slightly worse than performance with perfect knowledge of the I/O relation. Source:  Fig.$10$ in \cite{zakotfs2}.}
\label{fig4}
\end{figure}

By contrast, model-free operation is always feasible irrespective of whether the paths are separable or not. Fig.~\ref{fig4} appears as Fig.$10$ in \cite{zakotfs2}. 
It compares BER performance of model-dependent and model-free modes of operation. We consider the Veh-A channel with $\nu_p = 15$ KHz, $B = 0.96$ MHz, $T = 1.6$ ms. In \cite{zakotfs2} we dedicate separate Zak-OTFS subframes 
to channel sensing and data transmission in order to focus on the difference between model-dependent and model-free modes of operation. We {emphasize}
that this is NOT integrated sensing and communication.  
In the model-free mode, we estimate the taps of $h_{\mbox{\scriptsize{eff}}}[k,l]$ directly from the received DD domain symbols received in response to a point pulsone situated in the middle of ${\mathcal D}_0$. In the model-dependent mode, the receiver first estimates the parameters $(\tau_i, \nu_i, h_i)$ of the underlying physical channel, which is then used to estimate the taps of $h_{\mbox{\scriptsize{eff}}}[k,l]$. We employ MMSE equalization of the matrix-vector form of the Zak-OTFS I/O relation to detect information symbols at the receiver. We observe that BER performance of the model-dependent mode exhibits a high error floor. This is due to inaccurate estimates of the parameters of the underlying physical channel, which is a consequence of insufficient \emph{subframe} bandwidth and duration. On the other hand, BER performance of the model-free mode of Zak-OTFS is considerably better, only slightly worse than performance with perfect knowledge of the Zak-OTFS I/O relation.

Here and throughout this paper we focus on the model-free mode of operation in the crystalline regime.
We now describe how to integrate sensing and communication within a single Zak-OTFS subframe.
The discrete DD domain point pulsone indexed by $(k_p, l_p)$ is given by

{\vspace{-4mm}
\small
\begin{eqnarray}
\label{eqn_imp_pilot}
    x_{\mbox{\scriptsize{p,dd}}}[k,l] & \hspace{-3mm} = & \hspace{-4.5mm} \sum\limits_{n,m \in {\mathbb Z}} \hspace{-2.5mm}  \, e^{j 2 \pi \frac{n l}{N}} \, \delta[k - k_p - nM] \, \delta[l - l_p - mN].
\end{eqnarray}\normalsize}Quasi-periodicity results in impulses at $(k_p + nM , l_p + mN)$
for $n,m \in {\mathbb Z}$. In the absence of data, the received response is given by
\begin{eqnarray}
\label{eqnpaper3_21}
    y_{\mbox{\scriptsize{p,dd}}}[k,l] & = & h_{\mbox{\scriptsize{eff}}}[k,l] \, *_{\sigma} \, x_{\mbox{\scriptsize{p,dd}}}[k,l] \, + \, n_{\mbox{\scriptsize{dd}}}[k,l] \nonumber \\
    &  &  \hspace{-22mm} =  \hspace{-3mm} \sum\limits_{n,m \in {\mathbb Z} } \hspace{-2mm} h_{\mbox{\scriptsize{eff}}}[k - k_p - nM,l - l_p - mN] \, e^{j 2 \pi \frac{nl}{N}} \, e^{j 2 \pi \frac{k_p(l - l_p -mN)}{MN}} \nonumber \\
    & &  \hspace{-22mm}  + n_{\mbox{\scriptsize{dd}}}[k,l]
\end{eqnarray}where the last equation follows from (\ref{twistedconveqn}).

The support set ${\mathcal S}$ of $h_{\mbox{\scriptsize{eff}}}[k,l]$ consists of all pairs $(k,l)$ for which $h_{\mbox{\scriptsize{eff}}}[k,l] \ne 0$. The
crystallization condition is satisfied when the delay spread of ${\mathcal S}$ is less than the delay period $M$ and the Doppler spread is less than the Doppler period $N$.
In the crystalline regime, we can read off the taps of $h_{\mbox{\scriptsize{eff}}}[k,l]$ from  
$(n,m) = (0,0)$ term in (\ref{eqnpaper3_21}).
The method we presented in \cite{zakotfs2} was simply
to read off the point pulsone response within the support set of $h_{\mbox{\scriptsize{eff}}}[k - k_p, l - l_p]$
and multiply by $e^{-j 2 \pi \frac{k_p (l - l_p)}{MN}}$. This method gives a much better estimate than the model-dependent approach,
but it is not the ML estimation method. 
The ML estimator is given by the samples
of the cross-ambiguity between the received point pulsone $y_{\mbox{\scriptsize{p,dd}}}[k,l]$ and the point pulsone $x_{\mbox{\scriptsize{p,dd}}}[k,l]$, when observed only within the support set ${\mathcal S}$ of ${h}_{\mbox{\scriptsize{eff}}}[k,l]$.
Given $(k,l) \in {\mathcal S}$ we have

{\vspace{-4mm}
\small
\begin{eqnarray}
\label{eqnmlest}
{\widehat h}_{\mbox{\scriptsize{eff}}}[k,l] & \hspace{-2mm} = & \hspace{-2mm} A_{y_p,x_p}[k,l] \nonumber \\
& & \hspace{-15mm} = \sum\limits_{k'=0}^{M-1} \sum\limits_{l'=0}^{N-1}  y_{\mbox{\scriptsize{p,dd}}}[k',l'] \, x_{\mbox{\scriptsize{p,dd}}}^*[k'-k,l'-l] \, e^{-j 2 \pi \frac{l(k' - k)}{MN}}.
\end{eqnarray}\normalsize}The cross-ambiguity function $A_{a,b}[k,l]$ between any two quasi-periodic discrete DD signals 
$a[k,l]$ and $b[k,l]$ is defined in Appendix \ref{app_prop_ambig}.  
It follows from Theorem \ref{thm_crossambig2} in Appendix \ref{app_prop_ambig} that
the noise-free cross-ambiguity $A_{y_p,x_p}[k,l]$ is given by
\begin{eqnarray}
\label{ayxpeqn2}
A_{y_p,x_p}[k,l] & = & h_{\mbox{\scriptsize{eff}}}[k,l] \, *_{\sigma} \, A_{x_p,x_p}[k,l]
\end{eqnarray}where $A_{x_p,x_p}[k,l]$ is the self-ambiguity function of the point pulsone
$x_{\mbox{\scriptsize{p,dd}}}[k,l]$. From Appendix \ref{app_ppambig} it follows that the self-ambiguity function of a point pulsone is given by
\begin{eqnarray}
\label{eqn673510}
    A_{x_p, x_p}[k,l] & \hspace{-2.5mm} = & \hspace{-3.5mm} \sum\limits_{n,m \in {\mathbb Z}} \hspace{-2mm} e^{j 2 \pi \frac{(nM l_p - mN k_p)}{MN}} \, \delta[k - nM] \, \delta[l - mN] \nonumber \\
\end{eqnarray}from which it is clear that it is supported on the period lattice $\Lambda_{p}$, where
\begin{eqnarray}
    \Lambda_{p}  & \Define & M {\mathbb Z} \, \times \, N {\mathbb Z} \, = \, \{(nM, mN) \, | \, n,m \in {\mathbb Z} \}.
\end{eqnarray}Using (\ref{eqn673510}) in (\ref{ayxpeqn2}) gives

{\vspace{-4mm}
\small
\begin{eqnarray}
    A_{y_p,x_p}[k,l] & & \nonumber \\
    & & \hspace{-20mm} = h_{\mbox{\scriptsize{eff}}}[k,l] *_{\sigma} \hspace{-2mm} \sum\limits_{n,m \in {\mathbb Z}} \hspace{-2.5mm} e^{j 2 \pi \frac{(nM l_p - mN k_p)}{MN}} \, \delta[k - nM] \, \delta[l - mN] \nonumber \\
    & & \hspace{-20mm} = \sum\limits_{n,m \in {\mathbb Z}} \hspace{-2.5mm} e^{j 2 \pi \frac{(nM l_p - mN k_p)}{MN}} \, e^{j 2 \pi \frac{n l}{N} } \, h_{\mbox{\scriptsize{eff}}}[k - nM,l - mN],
\end{eqnarray}\normalsize}where the $(n,m) = (0,0)$ term is simply $h_{\mbox{\scriptsize{eff}}}[k,l]$. In the crystalline regime, the supports of terms in the RHS do not overlap, and the $(0,0)$ term is simply equal to $A_{y_p,x_p}[k,l]$ for $(k,l) \in {\mathcal S}$.

\begin{figure}[h]
\centering
\includegraphics[width=7.5cm, height=5.2cm]{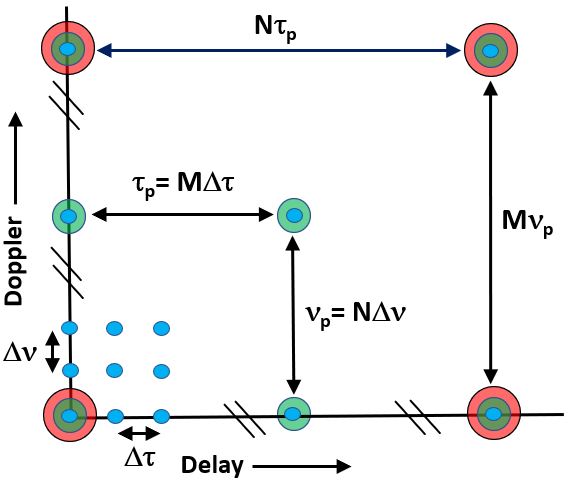}
\caption{Points on the information Lattice $\Lambda_{\mbox{\scriptsize{dd}}}$, period lattice $\Lambda_{\mbox{\scriptsize{p}}}$ and the dual of information lattice $\Lambda_{\mbox{\scriptsize{dd}}}^{\perp}$ shown by blue, green and red dots respectively. Adjacent points on the information lattice are separated by $\Delta \tau = \tau_p/M$ and $\Delta \nu = \nu_p/N$ along the delay and Doppler axis respectively.
Adjacent points on the dual of the information lattice are separated
by $MN \Delta \tau = N \tau_p$ and $MN \Delta \nu = M \nu_p$ along the delay and Doppler axis respectively.
} 
\label{fig44}
\end{figure}

Fig.~\ref{fig44} shows points on the information lattice ${\Lambda}_{dd}$,
the period lattice $\Lambda_p$ and the dual of the information lattice ${\Lambda}_{dd}^\perp$. If we scale down the delay and Doppler coordinates of these lattices by $\tau_p/M$ and $\nu_p/N$ respectively, then we may identify
${\Lambda}_{dd}$ with ${\mathbb Z}^2$ and ${\Lambda}_{dd}^\perp$ with $MN {\mathbb Z} \times MN {\mathbb Z}$.
Now consider any shift $(nMN, mMN)$
of a quasi-periodic signal $x_{\mbox{\scriptsize{dd}}}[k ,l ]$
in the discrete DD domain.
It follows from (\ref{eqn_qp}) that 
\begin{eqnarray}
\label{eqn925316e}
    x_{\mbox{\scriptsize{dd}}}[k + nMN,l + mMN] & = & e^{j 2 \pi \frac{n MN l}{N}} \, x_{\mbox{\scriptsize{dd}}}[k ,l ] \nonumber \\
    & = &  x_{\mbox{\scriptsize{dd}}}[k ,l ].
\end{eqnarray}Signals that are quasi-periodic with respect to the information lattice are periodic with
respect to the dual of the information lattice. Section \ref{secfiltering} introduces filters
in the discrete DD domain where this interplay will be important.

We conclude this Section by discussing
two practical issues that limit the use of point sensing pulsones.

\subsection{High Peak-to-Average-Power Ratio (PAPR) }
\begin{figure}[h]
\centering
\includegraphics[width=9.5cm, height=5.2cm]{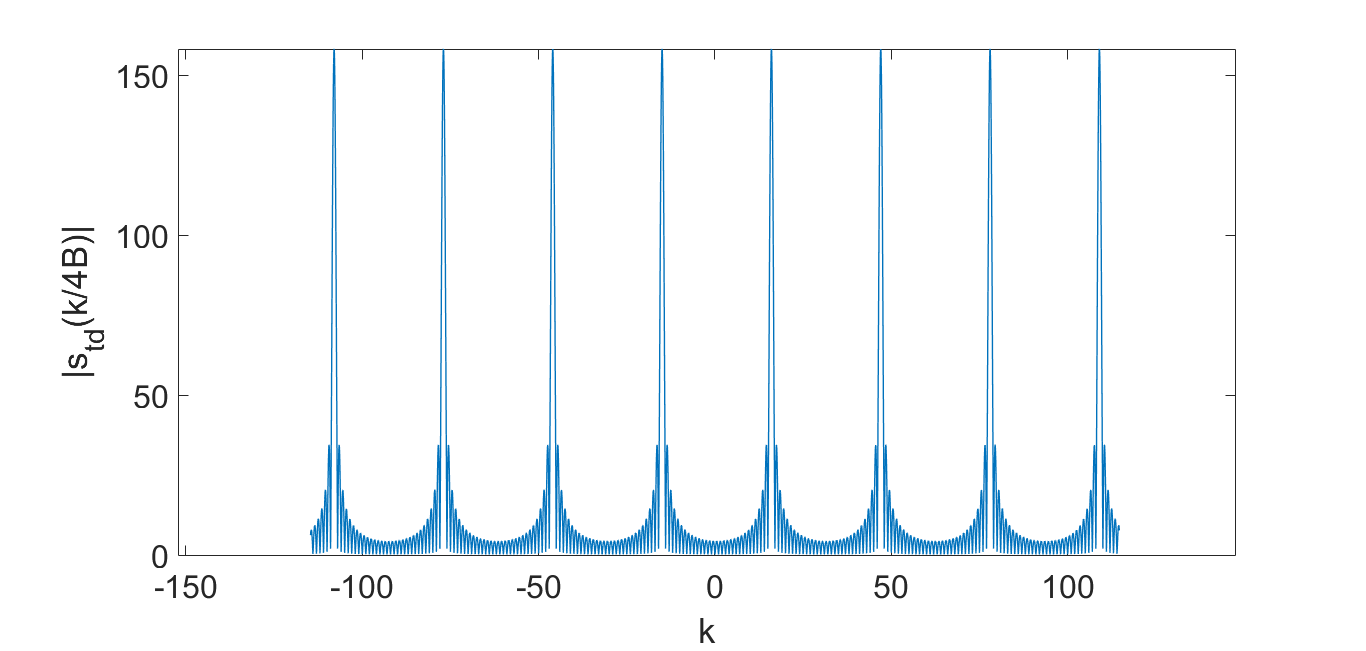}
\caption{
TD realization of an point pulsone with Doppler period $\nu_p = 30$ KHz, $M = 31, N=37$. Point pulsone is located at $(k_p, l_p) = ( (M+1)/2, (N+1)/2)$.
Only part of the entire TD pulsone is shown here, with samples taken every $\frac{1}{4 B}$ seconds where $B = M \nu_p = 930$ KHz. RRC pulse shaping with $\beta_{\tau} = \beta_{\nu} = 0.6$.} 
\label{fig3}
\end{figure}

Fig.~\ref{fig3} displays the magnitude of the TD realization
of a point sensing pulsone. We observe a train of 
narrow pulses, exhibiting sharp peaks at the pulse locations, leading to high PAPR. This requires the use of highly linear power amplifiers which are typically power-inefficient. 

\subsection{Interference between Data and Sensing Pulsones}
Data and sensing pulsones combine at the transmitter to produce the noise-free DD domain signal given by
\begin{eqnarray}
x_{\mbox{\scriptsize{dd}}}[k, l] & = & \sqrt{E_d} \underbrace{x_{\mbox{\scriptsize{d,dd}}}[k, l]}_{\mbox{\scriptsize{data pulsones}}} \, +  \, \sqrt{E_p}  \underbrace{x_{\mbox{\scriptsize{p,dd}}}[k, l]}_{\mbox{\scriptsize{pilot pulsone}}}.
\end{eqnarray}
We will first show that the received data signal results from the action of a discrete DD domain filter
\begin{eqnarray}
\sqrt{E_d} \sum\limits_{k'=0}^{M-1} \sum\limits_{l' = 0}^{N-1}  x[k' , l'] {\Bigg (} h_{\mbox{\scriptsize{eff}}}[k, l] *_{\sigma} \delta[k - k'] \delta[l - l'] {\Bigg )} 
\end{eqnarray}on the discrete quasi-periodic pulse at the origin, denoted by $x_{\mbox{\scriptsize{0,dd}}}[k, l]$. We will then show that the received pilot pulsone signal results from the action of a second discrete DD domain filter
\begin{eqnarray}
   \sqrt{E_p} h_{\mbox{\scriptsize{eff}}}[k, l] *_{\sigma} ( \delta[k - k_p] \delta[l - l_p] )
\end{eqnarray}on the same quasi-periodic pulse
$x_{\mbox{\scriptsize{0,dd}}}[k, l]$. This description makes it clear that data and sensing pulsones will interfere at the receiver.
We emphasize that it is only possible to express this interference in terms of operators because signal processing takes place in the DD domain where twisted convolution is associative.

An equivalent expression for the DD domain information signal in (\ref{eqn_01}) is that the data symbol $x[k',l']$ is carried by a discrete
quasi-periodic DD domain pulse at $(k',l')$ so that

{\vspace{-4mm}
\small
\begin{eqnarray}
\label{eqn90217t4}
    x_{\mbox{\scriptsize{d,dd}}}[k, l] & \hspace{-3mm} = & \hspace{-3mm} \sum\limits_{k'=0}^{M-1} \sum\limits_{l' = 0}^{N-1} {\Bigg (} x[k' , l'] \, \nonumber \\
    & & \hspace{5mm} \underbrace{\sum\limits_{n,m \in {\mathbb Z}} \hspace{-1mm} e^{j 2 \pi \frac{n l'}{N}} \, \delta[k - k' - nM] \, \delta[l - l' - mN]}_{\mbox{\scriptsize{Quasi-periodic DD pulse at}} \, (k',l')} \, {\Bigg )} \nonumber \\
    & & \hspace{-10mm} \mya \sum\limits_{k'=0}^{M-1} \sum\limits_{l' = 0}^{N-1}  x[k' , l'] {\Bigg (} (\delta[k - k'] \delta[l - l'] ) \, *_{\sigma} \, x_{\mbox{\scriptsize{0,dd}}}[k, l] {\Bigg )}.
\end{eqnarray}\normalsize}
Step (a) follows from the fact that a quasi-periodic DD pulse at $(k',l')$ is simply $x_{\mbox{\scriptsize{0,dd}}}[k, l]$ shifted by $(k',l')$, i.e.

{\vspace{-4mm}
\small
\begin{eqnarray}
    \sum\limits_{n,m \in {\mathbb Z}} \hspace{-1mm} e^{j 2 \pi \frac{n l'}{N}} \, \delta[k - k' - nM] \, \delta[l - l' - mN]  &  &  \nonumber \\
    & & \hspace{-60mm} = (\delta[k - k'] \delta[l - l'] ) \, *_{\sigma} \, \underbrace{\sum\limits_{n,m \in {\mathbb Z}} \hspace{-1mm} \delta[k - nM] \, \delta[l - mN]}_{= x_{\mbox{\tiny{0,dd}}}[k, l]}.
\end{eqnarray}\normalsize}
Since the pilot signal $x_{\mbox{\scriptsize{p,dd}}}[k, l]$ is a discrete
quasi-periodic DD domain pulse at $(k_p, l_p)$ it follows that
\begin{eqnarray}
\label{xpddprobeeqn}
x_{\mbox{\scriptsize{p,dd}}}[k, l] & \hspace{-3mm} = & \hspace{-3mm} (\delta[k - k_p] \delta[l - l_p]) *_{\sigma} x_{\mbox{\tiny{0,dd}}}[k, l] \nonumber \\
&  &  \hspace{-16mm} = \sum\limits_{n,m \in {\mathbb Z}} \hspace{-1mm} e^{j 2 \pi \frac{n l_p}{N}} \, \delta[k - k_p - nM] \, \delta[l - l_p - mN].
\end{eqnarray}Since twisted convolution is
associative, it follows from (\ref{eqn90217t4}) that

{\vspace{-4mm}
\small
\begin{eqnarray}
    x_{\mbox{\scriptsize{d,dd}}}[k, l] & \hspace{-3mm} = & \hspace{-3mm} {\Bigg (} \sum\limits_{k'=0}^{M-1} \sum\limits_{l' = 0}^{N-1}  \hspace{-2mm} x[k' , l'] \delta[k - k'] \delta[l - l'] {\Bigg )} *_{\sigma} x_{\mbox{\tiny{0,dd}}}[k, l], \nonumber \\
    \end{eqnarray}\normalsize} so that the received noise-free signal is given by

\begin{figure}[h]
\centering
\includegraphics[width=9.5cm, height=5.2cm]{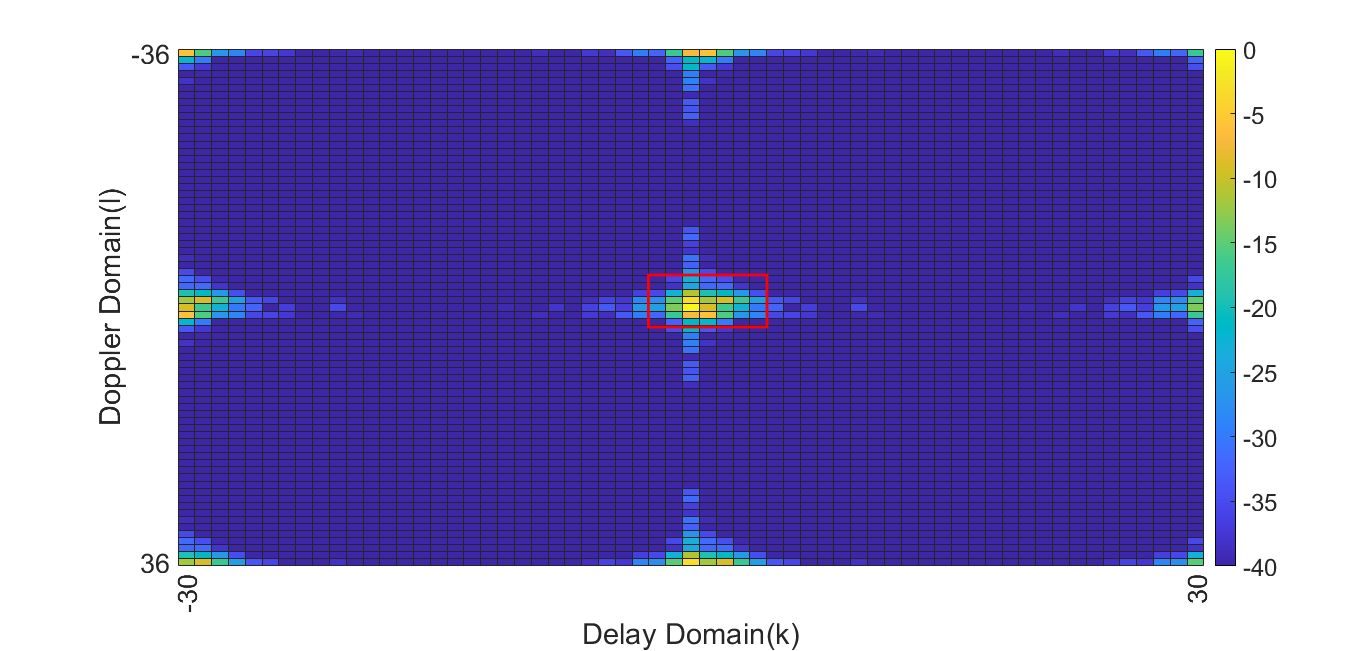}
\caption{Magnitude of the cross-ambiguity between the transmitted point pilot signal and the received point pilot signal, i.e., $\vert h_{\mbox{\scriptsize{eff}}}[k, l] *_{\sigma}  \, A_{x_p, x_p}[k,l] \vert$
for EVA channel with $\nu_{max} = 815$ Hz. $M = 31, N = 37, \nu_p = 30$ KHz. RRC pulse shaping $\beta_{\tau} = \beta_{\nu} = 0.6$. $(k_p , l_p) = ((M+1)/2, (N+1)/2)$. The support set of $h_{\mbox{\scriptsize{eff}}}[k, l]$ is approximately the region within the red rectangle. 
} 
\label{figcrossambig_815}
\end{figure}

{\vspace{-4mm}
\small
\begin{eqnarray}
\label{eqn87254}
y_{\mbox{\scriptsize{dd}}}[k, l] &  & \nonumber \\
& & \hspace{-15mm} = \sqrt{E_d} \, h_{\mbox{\scriptsize{eff}}}[k, l] *_{\sigma} x_{\mbox{\scriptsize{d,dd}}}[k, l] \, +  \, \sqrt{E_p} h_{\mbox{\scriptsize{eff}}}[k, l] *_{\sigma} x_{\mbox{\scriptsize{p,dd}}}[k, l] \nonumber \\
& & \hspace{-15mm} =  \sqrt{E_d} \, {\Bigg [} {\Bigg (} \sum\limits_{k'=0}^{M-1} \sum\limits_{l' = 0}^{N-1}  \hspace{-2mm} x[k' , l']  \, h_{\mbox{\scriptsize{eff}}}[k, l] *_{\sigma} \, (\delta[k - k'] \delta[l - l']) {\Bigg )}  \nonumber \\
& &  *_{\sigma}  \, x_{\mbox{\tiny{0,dd}}}[k, l]  {\Bigg ]} \nonumber \\
& & \hspace{-14mm} \, + \, \, \sqrt{E_p} \, {\Bigg (} h_{\mbox{\scriptsize{eff}}}[k, l] *_{\sigma} (\delta[k - k_p] \, \delta[l - l_p]) {\Bigg )}  \, *_{\sigma} \, x_{\mbox{\tiny{0,dd}}}[k, l].
\end{eqnarray}\normalsize}
Suppose that we dedicate separate Zak-OTFS subframes to channel sensing and data transmission. The ML estimate for the taps of $h_{\mbox{\scriptsize{eff}}}[k, l]$ is simply the cross-ambiguity function of the received pilot with the transmitted pilot (see (\ref{eqnmlest})), as shown in Fig.~\ref{figcrossambig_815}.

\begin{figure}[h]
\centering
\includegraphics[width=9.5cm, height=5.2cm]{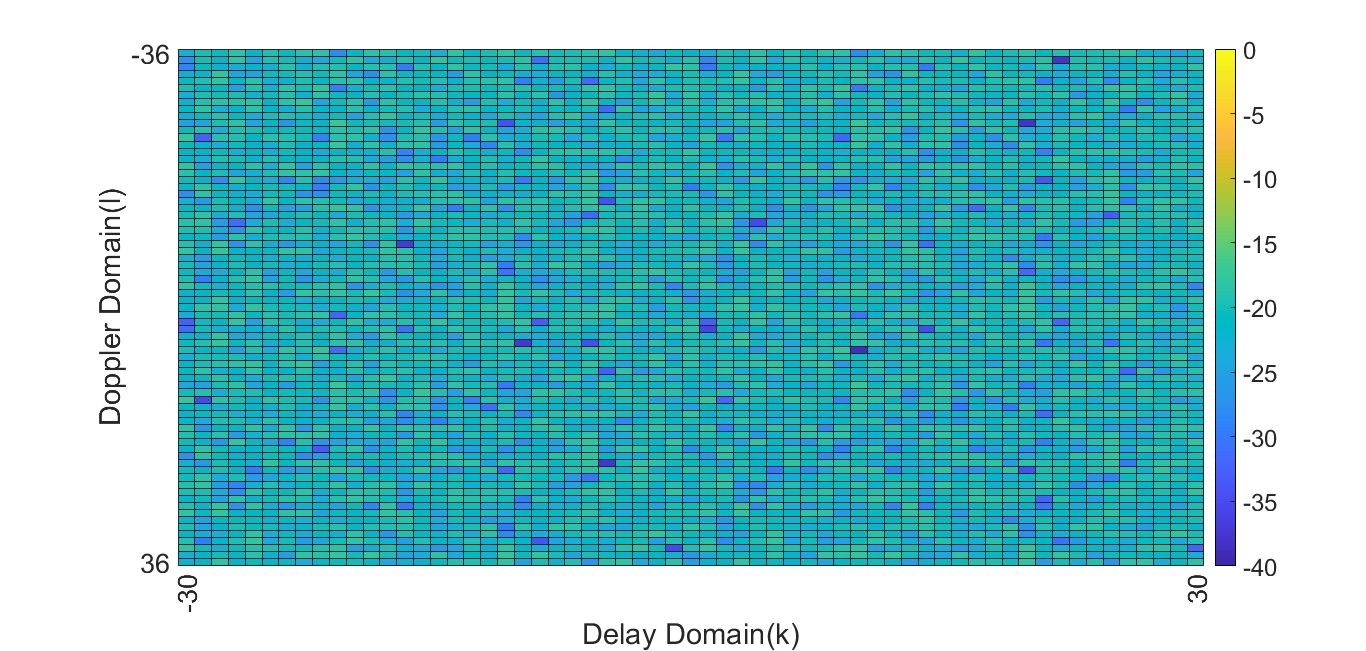}
\caption{Magnitude of the cross-ambiguity between the transmitted point pilot signal and the received signal which includes contributions from pilot and data ($\vert A_{y, x_p}[k,l] \vert$).
\emph{EVA} channel with $\nu_{max} = 815$ Hz. $M = 31, N = 37, \nu_p = 30$ KHz. RRC pulse shaping $\beta_{\tau} = \beta_{\nu} = 0.6$. $(k_p , l_p) = ((M+1)/2, (N+1)/2)$.
\emph{Interference prevents channel sensing.}
} 
\label{figcrossambigintrf}
\end{figure}

Now suppose that we integrate channel sensing
and data transmission within the same Zak-OTFS subframe for which the received discrete noise-free DD signal is given by (\ref{eqn87254}). The ML estimates for the taps of
$h_{\mbox{\scriptsize{eff}}}[k, l]$ are now given by the cross-ambiguity function

{\vspace{-4mm}
\small
\begin{eqnarray}
\label{eqn96244_1}
    A_{y,x_p}[k,l] &  & \nonumber \\
    & & \hspace{-18mm} = \sqrt{E_d} {\Bigg [} {\Bigg (} \sum\limits_{k'=0}^{M-1} \sum\limits_{l' = 0}^{N-1}  \hspace{-2mm} x[k' , l']  \, h_{\mbox{\scriptsize{eff}}}[k, l] *_{\sigma} \, (\delta[k - k'] \delta[l - l']) {\Bigg )} \nonumber \\
    & &  \hspace{-6mm} *_{\sigma} \, A_{x_0, x_p}[k,l] {\Bigg ]}  \, + \, \sqrt{E_p} \, h_{\mbox{\scriptsize{eff}}}[k, l] *_{\sigma}  \, A_{x_p, x_p}[k,l].
\end{eqnarray}\normalsize}The second term
in (\ref{eqn96244_1}) is the ML estimate for the taps of $h_{\mbox{\scriptsize{eff}}}[k, l]$ in the absence of data. Fig.~\ref{figcrossambigintrf} illustrates how the first term, which represents
interference from the data signal, obscures the second term. Within the first term, $A_{x_0, x_p}[k,l]$ is the cross-ambiguity function
between the quasi-periodic pulse at the origin and
the quasi-periodic pilot signal.

\begin{figure}[h]
\centering
\includegraphics[width=9.5cm, height=5.2cm]{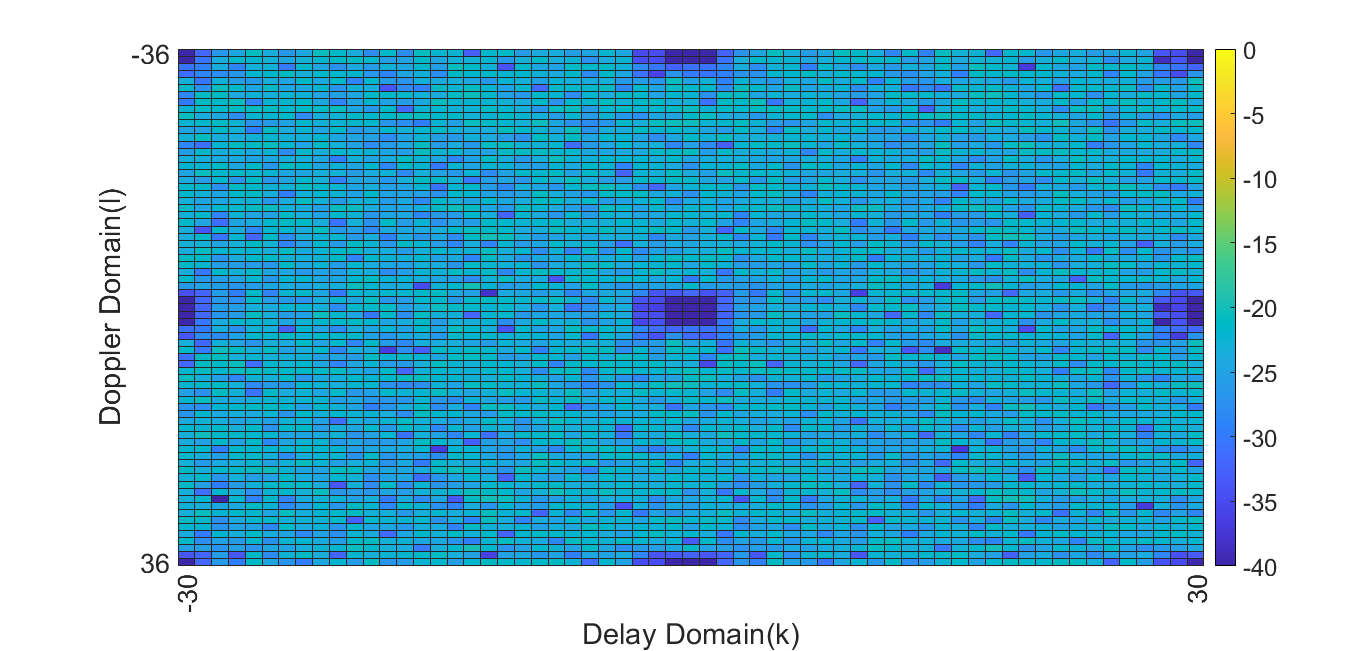}
\caption{Data/information symbols are not transmitted in a guard region ${\mathcal G}$ (a $7 \times 7$ rectangle around the location $(k_p , l_p) = ((M+1)/2, (N+1)/2)$ of the pilot signal) which is the location of the point sensing signal. Magnitude of the cross-ambiguity between the transmitted pilot signal and the data component of the received signal, i.e., $\vert A_{y, x_p}[k,l] - \sqrt{E_p} h_{\mbox{\scriptsize{eff}}}[k, l] *_{\sigma}  \, A_{x_p, x_p}[k,l] \vert$ (see (\ref{eqn96244_1})).
Veh-A channel with $\nu_{max} = 815$ Hz. $M = 31, N = 37, \nu_p = 30$ KHz. RRC pulse shaping $\beta_{\tau} = \beta_{\nu} = 0.6$. Interference preventing channel sensing has been suppressed.
} 
\label{figcrossambigguard}
\end{figure}
The cross-ambiguity functions $A_{x_0, x_p}$
and $A_{x_p, x_p}$ are both supported
on the period lattice. This leads to interference between data and pilot signals
at the receiver which prevents channel sensing
by completely obscuring the second term.
Fig.~\ref{figcrossambigguard}
illustrates how we make the second term visible by introducing a \emph{guard band} ${\mathcal G}$ around the pilot signal, in this case a $7 \times 7$
rectangle. Guard bands cannot be avoided when the cross-ambiguity functions $A_{x_0, x_p}[k,l]$ and $A_{x_p, x_p}[k,l]$ are both supported on the same lattice.  

In Section \ref{zakotfsspread} we describe how to filter the point pilot signal so that $A_{x_p, x_p}[k,l]$ is supported on a lattice
different from the period lattice. We will show that it is possible to choose the filter so that the cross-ambiguity
function $A_{x_0, x_p}[k,l]$ appears as noise to the self-ambiguity function $A_{x_p, x_p}[k,l]$ of the filtered pilot signal. We seek a filter that results in cross-ambiguity functions that are \emph{unbiased} or \emph{incoherent}.
This eliminates the need for a guard band at the cost of reducing the SNR of channel sensing. In addition the energy of the filtered pilot signal will be spread uniformly in the discrete DD domain,
leading to a TD realization with excellent PAPR.

Finally, we define the data signal power to noise power ratio, the pilot signal power to noise power ratio, and the pilot signal power to data signal power ratio.

From (\ref{eqn90217t4}), it follows that the average energy of the data signal is $E_d$, i.e.
\begin{eqnarray}
   {\mathbb E}\left[   \sum\limits_{k=0}^{M-1} \sum\limits_{l=0}^{N-1}   \left\vert \sqrt{E_d} \, x_{\mbox{\tiny{d,dd}}}[k, l]   \right\vert^2 \right] & = & E_d.
\end{eqnarray}
Since the point pilot signal $x_{\mbox{\scriptsize{p,dd}}}[k, l]$ is a discrete quasi-periodic DD pulse localized at $(k_p, l_p)$,
from (\ref{xpddprobeeqn}) it follows that the energy of the point pilot signal is $E_p$, i.e.
\begin{eqnarray}
    \sum\limits_{k=0}^{M-1} \sum\limits_{l=0}^{N-1} \left\vert \sqrt{E_p} \, x_{\mbox{\scriptsize{p,dd}}}[k, l] \right\vert^2 & = & E_p.
\end{eqnarray}AWGN $n_{\mbox{\scriptsize{td}}}(t)$ at the receiver has PSD $N_0$ Watt/Hz. Therefore, from (\ref{eqn12845529}) in Appendix \ref{prfnddklenergy} it follows that
the average total energy of the discrete DD domain noise signal $n_{\mbox{\scriptsize{dd}}}[k,l]$ (see (\ref{eqn_io_relation})) is
\begin{eqnarray}
    {\mathbb E}\left[ \sum\limits_{k=0}^{M-1} \sum\limits_{l=0}^{N-1} \left\vert n_{\mbox{\scriptsize{dd}}}[k,l] \right\vert^2 \right]
    & = & MN N_0,
\end{eqnarray}The effective channel gain is $\sum\limits_{(k,l) \in {\mathcal S}} \left\vert h_{\mbox{\scriptsize{eff}}}[k,l] \right\vert^2 $ and therefore the received energy of data/information pulsones and that of the pilot pulsone are $E_d \sum\limits_{(k,l) \in {\mathcal S}} \left\vert h_{\mbox{\scriptsize{eff}}}[k,l] \right\vert^2 $ and $E_p \sum\limits_{(k,l) \in {\mathcal S}} \left\vert h_{\mbox{\scriptsize{eff}}}[k,l] \right\vert^2 $ respectively. Hence the ratio of the power of the received data pulsones to the noise power (data SNR) is given by
\begin{eqnarray}
\label{eqnrhoddef}
    \rho_d & \Define & \frac{E_d \sum\limits_{(k,l) \in {\mathcal S}} \hspace{-1mm} \left\vert h_{\mbox{\scriptsize{eff}}}[k,l] \right\vert^2 }{MN N_0},
\end{eqnarray}and the ratio of the power of the received pilot pulsone to that of noise (i.e., pilot SNR) is given by
\begin{eqnarray}
\label{rhopdef}
     \rho_p & \Define & \frac{E_p \sum\limits_{(k,l) \in {\mathcal S} } \hspace{-1mm} \left\vert h_{\mbox{\scriptsize{eff}}}[k,l] \right\vert^2} {MN N_0}.
\end{eqnarray}The ratio of the power of the received data pulsones to the power of the received pilot pulsones is therefore given by $\frac{\rho_d}{\rho_p}$ which is subsequently referred to as the pilot power to data power ratio (PDR).

\section{Filtering in the discrete delay Doppler domain}
\label{secfiltering}
In this Section, we describe how to design a filter in the discrete DD domain to spread a point pilot signal over the $MN$ pulsones located on the information grid, so that each pulsone contributes a fraction $1/(MN)$ of the spread pilot energy. The discrete DD domain filter $w_s[k,l]$ acts on the point pilot $x_{\mbox{\scriptsize{p,dd}}}[k,l]$ by twisted convolution to produce a spread pilot signal $x_{\mbox{\scriptsize{s,dd}}}[k,l]$ given by
\begin{eqnarray}
\label{eqn_20}
     x_{\mbox{\scriptsize{s,dd}}}[k,l] & = & w_s[k,l] \, *_{\sigma} \, x_{\mbox{\scriptsize{p,dd}}}[k,l].
\end{eqnarray}where $x_{\mbox{\scriptsize{p,dd}}}[k,l] $ is given by (\ref{eqn_imp_pilot}). Note that $x_{\mbox{\scriptsize{s,dd}}}[k,l]$
is quasi-periodic since twisted convolution preserves quasi-periodicity.

More generally, we develop the fundamentals of filtering in the discrete DD domain.
 We start from the theory of linear time invariant (LTI) systems. Here it is well known that linear convolution
of a discrete-time periodic signal with a discrete-time filter is equivalent to periodic linear convolution of the periodic signal with the periodic extension of the filter. Thus, two discrete-time filters with identical periodic extensions are equivalent. If the period is $L$, then the discrete-time filter is specified by a vector in ${\mathbb C}^L$. In the discrete DD domain, we recall that quasi-periodic signals are \emph{periodic} with period $MN$ along both delay and Doppler axes (see (\ref{eqn925316e})).
We now show that twisted convolution of a quasi-periodic discrete DD signal with a discrete filter $w_s[k,l]$ is equivalent to a MN-periodic twisted convolution of that quasi-periodic signal with
the MN-periodic extension of the discrete filter.
Thus, two discrete DD domain filters with identical periodic extensions are equivalent.
Since the period is $MN$ along both delay and Doppler axes, a discrete DD domain filter is specified by a vector in ${\mathbb C}^{M^2N^2}$.

Given an arbitrary discrete DD domain filter $a_s[k,l]$, we define its $MN$-periodic extension $a[k,l]$ by
\begin{eqnarray}
\label{eqn_periodicMN}
    a[k,l] & \Define & \sum\limits_{n,m \in {\mathbb Z}} a_s[k + nMN, l+mMN].
\end{eqnarray}Note that $a[k,l]$ is periodic with period $MN$
along the delay and Doppler axes.
\begin{figure}[h!]
    	\vspace{-3mm}
     \centering
    	\begin{subfigure}[b]{0.5\textwidth}
    	\includegraphics[width=0.9\textwidth]{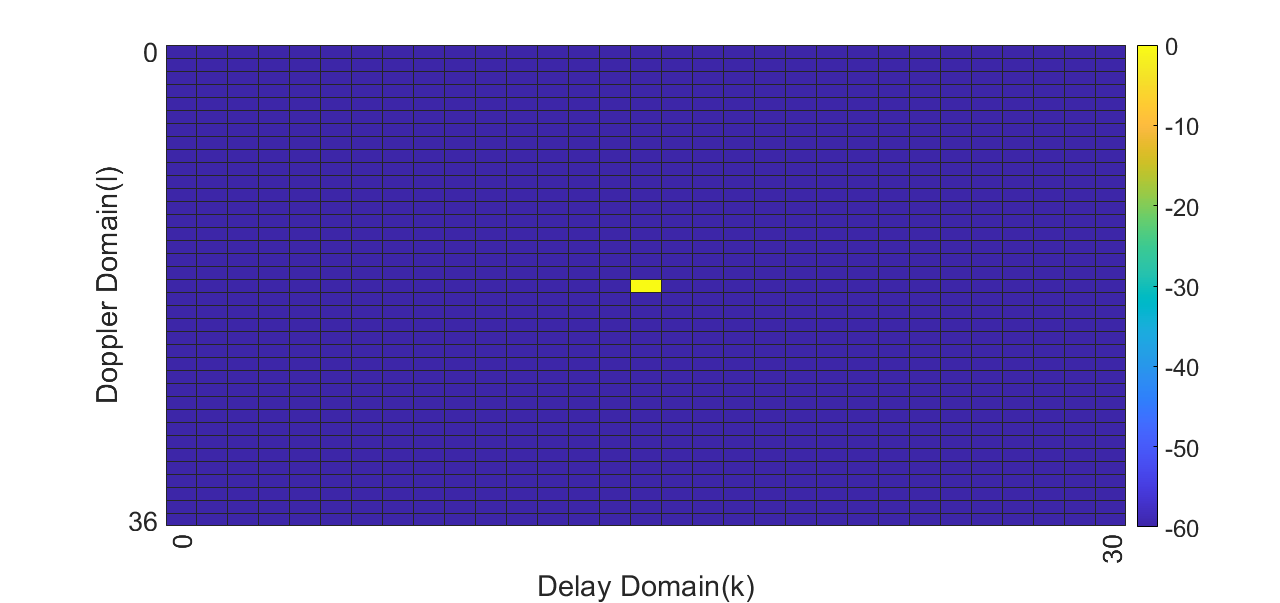}
    	\caption{{Point pilot at $(k_p,l_p) = \left( \frac{M+1}{2}, \frac{N+1}{2}\right)$, $M=31, N=37$.}}
    	\label{fig32}
    \end{subfigure}
    \centering
    \begin{subfigure}[b]{0.5\textwidth}
	\includegraphics[width=0.9\textwidth]{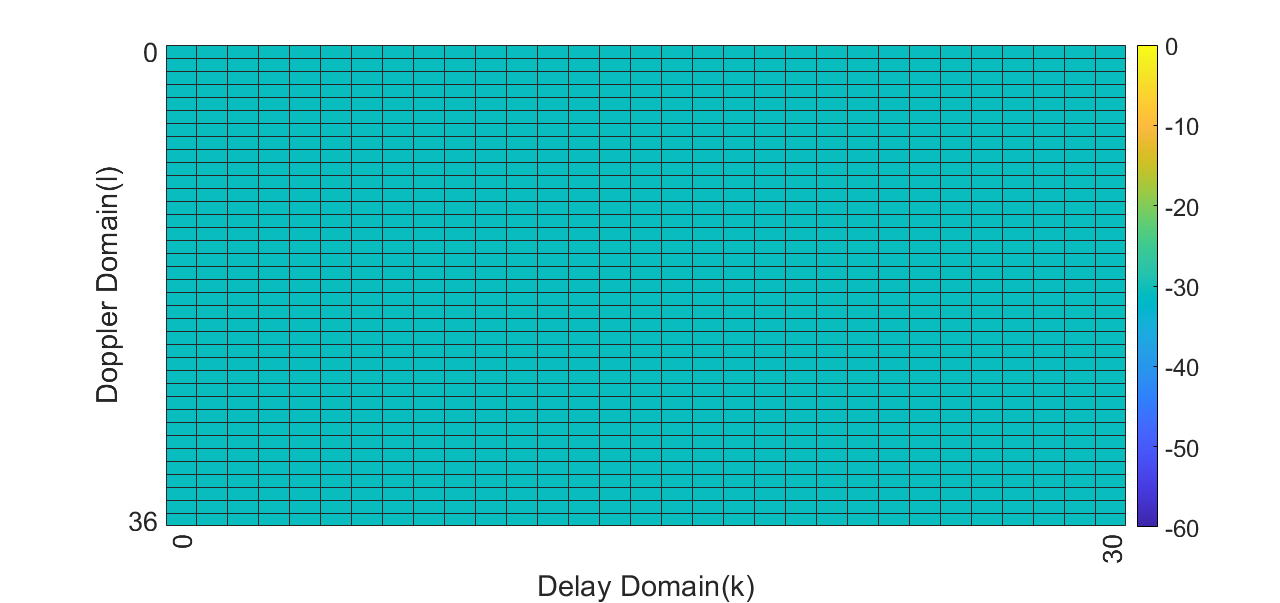}
	\caption{{Spread pilot. Chirp filter with $q = 3$, $M=31, N=37$. }}
	\label{fig33}
	\end{subfigure}
	\vspace{-2mm}
	\caption{Energy profiles of DD domain pilot signals. $M = 31, N = 37$ and the total DD domain energy is normalized to $1$ for both point pilot and spread pilot signals, that is, $ \sum\limits_{k=0}^{M-1} \sum\limits_{l=0}^{N-1}\vert x_{\mbox{\scriptsize{s,dd}}}[k,l] \vert^2 = \sum\limits_{k=0}^{M-1} \sum\limits_{l=0}^{N-1}\vert x_{\mbox{\scriptsize{p,dd}}}[k,l] \vert^2 = 1$. The point pilot takes the value $0$ at all locations except $(k_p,l_p)$ where the magnitude is $1$. The squared magnitude of each spread pilot symbol is about $ 1/(MN)$, i.e., about $10 \log_{10}(MN) = 30.6$ dB below the peak squared-magnitude of a point pilot DD signal.}
 \label{fig91}
\end{figure}
The extension $a[k,l]$ acts on a quasi-periodic signal
$b[k,l]$ by $MN$-periodic twisted convolution, denoted by $\circledast_{\sigma}$. More precisely
\begin{eqnarray}
\label{eqn_MNfilt}
    a[k,l] \, \circledast_{\sigma} \, b[k,l] \nonumber \\
    &  & \hspace{-33mm} = \sum\limits_{k'=0}^{MN-1} \sum\limits_{l'=0}^{MN-1} a[k',l'] \, b[k - k', l - l'] \, e^{j 2 \pi l' \frac{(k - k')}{MN}}.
\end{eqnarray}
\begin{theorem}
\label{thm1}
The twisted convolution of a discrete DD filter $a_s[k,l]$ with a
discrete quasi-periodic signal $b[k,l]$ coincides with the $MN$-periodic
twisted convolution of its $MN$-periodic extension $a[k,l]$
with $b[k,l]$, i.e.
\begin{eqnarray}
\label{eqn_stmt_thm1}
    a_s[k,l] \, *_{\sigma} b[k,l] & = & a[k,l] \circledast_{\sigma} \, b[k,l].
\end{eqnarray}
\end{theorem}
\begin{IEEEproof}
We start from the definition of twisted convolution, then view delay and Doppler indices modulo $MN$, then simplify using the definition of an $MN$-periodic extension (\ref{eqn_periodicMN}). The detailed derivation of (\ref{eqn_stmt_thm1}) is given by (\ref{eqn_deriv_MNfilt}) (see top of next page).  
\begin{figure*}
\vspace{-7mm}
\begin{eqnarray}
\label{eqn_deriv_MNfilt}
   a_s[k,l] \, *_{\sigma} \, b[k,l]  & \hspace{-3mm} = &  \hspace{-3mm} \sum\limits_{k', l' \in {\mathbb Z}} a_s[k', l'] \, b[k -k',l-l']  \, e^{j 2 \pi \frac{l' (k - k')}{MN}} \nonumber \\
    &  & \hspace{-9mm} = \sum\limits_{k'=0}^{MN-1} \sum\limits_{l'=0}^{MN-1} \sum\limits_{n,m \in {\mathbb Z}} a_s[k' + nMN, l' + mMN] \, b[k -k'-nMN,l-l'-mMN] \, e^{j 2 \pi \frac{(l' + m MN) (k - k' - nMN)}{MN}}  \nonumber \\
     &  & \hspace{-9mm} = \sum\limits_{k'=0}^{MN-1} \sum\limits_{l'=0}^{MN-1} \underbrace{\sum\limits_{n,m \in {\mathbb Z}} a_s[k' + nMN, l' + mMN]}_{= a[k',l'], \, \mbox{\small{see}} \, (\ref{eqn_periodicMN}) }\,  b[k -k',l-l']  \, e^{j 2 \pi \frac{l' (k - k')}{MN}} \nonumber \\
      &  & \hspace{-9mm} = \sum\limits_{k'=0}^{MN-1} \sum\limits_{l'=0}^{MN-1} a[k',l'] \,  b[k -k',l-l']  \, e^{j 2 \pi \frac{l' (k - k')}{MN}}   \nonumber \\
      & & \hspace{-9mm} = a[k,l] \, \circledast_{\sigma}  \, b[k,l].
\end{eqnarray}
\begin{eqnarray*}
    \hline
\end{eqnarray*}
\end{figure*}
\end{IEEEproof}

Returning to (\ref{eqn_20}), we apply Theorem \ref{thm1}
to obtain
\begin{eqnarray}
\label{eqn_periodtwistconv}
    x_{\mbox{\scriptsize{s,dd}}}[k,l] & \hspace{-3mm} \Define &  \hspace{-3mm} w_s[k,l] \, *_{\sigma} \, x_{\mbox{\scriptsize{p,dd}}}[k,l] \nonumber \\
    & = & w[k,l] \, \circledast_{\sigma}  \, x_{\mbox{\scriptsize{p,dd}}}[k,l] 
\end{eqnarray}where
\begin{eqnarray}
    w[k,l] & \Define &  \sum\limits_{n,m \in {\mathbb Z}} w_s[k + nMN, l + mMN]  
\end{eqnarray}is the DD domain $MN$-periodic extension of $w_s[k,l]$.
\begin{figure*}
\begin{eqnarray}
\label{eqn_spreadpilotexpr}
x_{\mbox{\scriptsize{s,dd}}}[k,l] & \hspace{-3mm} = &  \hspace{-3mm}  \sum\limits_{k'=0}^{MN-1} \sum\limits_{l' =0 }^{MN-1} w[k',l'] \, \sum\limits_{n,m \in {\mathbb Z}} e^{j 2 \pi n \frac{(l - l')}{N}} \, \delta[k - k' - k_p -nM] \, \delta[l - l' - l_p - mN] \, e^{j 2 \pi l' \frac{(k - k')}{MN}} \nonumber \\
    & & \hspace{-9mm} =   \sum\limits_{n=0}^{N-1} \sum\limits_{m=0}^{M-1} e^{j 2 \pi n \frac{(l_p + mN)}{N}} \,  e^{j 2 \pi \frac{(l - l_p -mN) (k_p + nM)}{MN}} \, w[k - k_p -nM, l - l_p - mN]
\end{eqnarray}
\begin{eqnarray*}
    \hline
\end{eqnarray*}
\end{figure*}We substitute for $x_{\mbox{\scriptsize{p,dd}}}[k,l]$ in (\ref{eqn_periodtwistconv}) using (\ref{eqn_imp_pilot}) to obtain the spread DD pilot signal expression given by (\ref{eqn_spreadpilotexpr}) (see top of next page).


\emph{$MN$-periodic discrete chirp filter $w[k,l]$:}
We define
\begin{eqnarray}
\label{eqnwpchirp}
    w[k,l] & = &  \frac{1}{MN} \, e^{j 2 \pi \frac{q (k^2 + l^2)}{MN}}, 
\end{eqnarray}for $k,l \in {\mathbb Z}$, and we refer to $q \in {\mathbb Z}$ as the \emph{slope parameter}.
In Section \ref{zakotfsspread} we will show that the constant
factor $1/(MN)$ in (\ref{eqnwpchirp}) results in a spread pilot signal $x_{\mbox{\scriptsize{s,dd}}}[k,l]$ with unit energy. We will observe that $\vert x_{\mbox{\scriptsize{s,dd}}}[k,l] \vert$ is almost constant
(say $\vert x_{\mbox{\scriptsize{s,dd}}}[k,l] \vert \approx \alpha$ for all $k,l$) when $M$ and $N$ are odd primes, and when $q$ is relatively prime to both $M$ and $N$. Since the total energy of the spread pilot is $1 = \sum\limits_{k=0}^{M-1} \sum\limits_{l=0}^{N-1} \vert x_{\mbox{\scriptsize{s,dd}}}[k,l] \vert^2 \approx MN \alpha^2$, we conclude $\alpha \approx 1/\sqrt{MN}$. Hence the peak amplitude
of the point pilot signal ($1$ at location $(k,l) = (k_p, l_p)$) is about $\sqrt{MN}$ times higher than that of the spread pilot signal. 
Fig.~\ref{fig91} compares the energy profiles of point and spread pilot signals.

\section{Zak-OTFS with spread sensing pulsone}
\label{zakotfsspread}
\subsection{Reducing PAPR}
\label{spreadpulsonepapr1}
\begin{figure}[h]
\centering
\includegraphics[width=9.5cm, height=5.2cm]{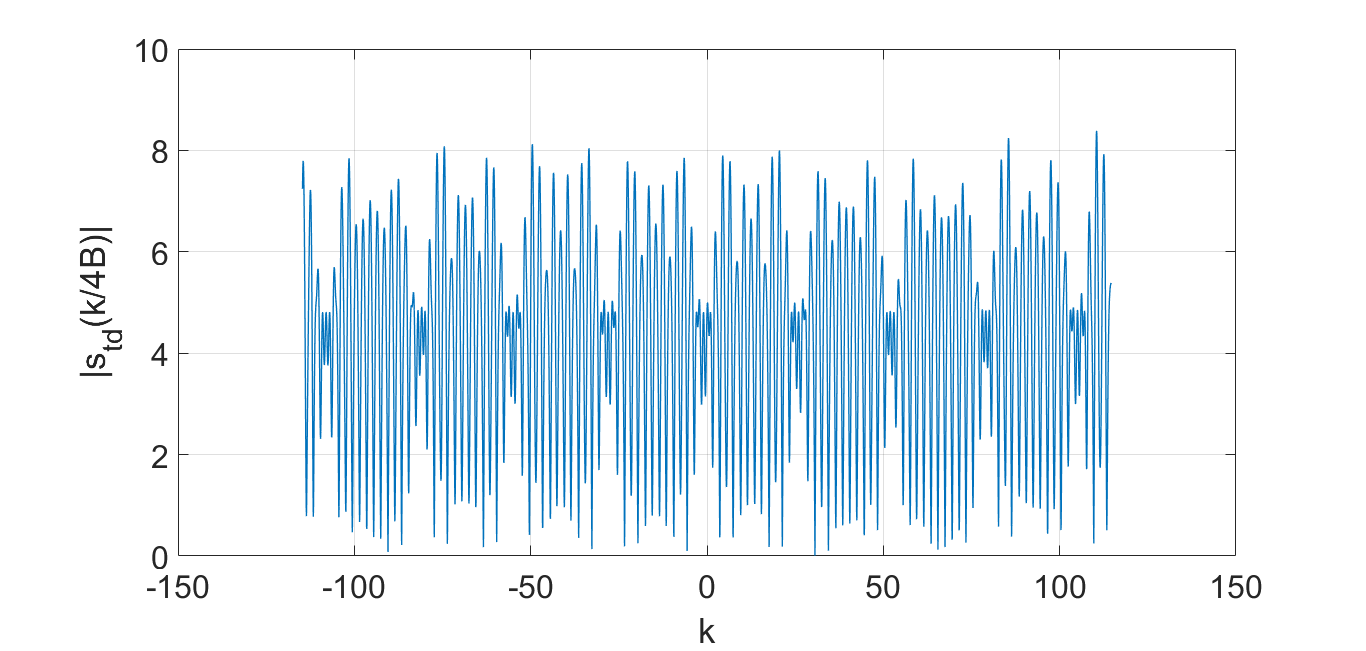}
\caption{TD realization $s_{_{td}}(t = k/(4 B)), k \in {\mathbb Z}$ of a spread pilot signal with Doppler period $\nu_p = 30$ KHz, $M = 31, N=37$. Discrete chirp filter with $q=3$. 
Only part of the entire TD realization is shown here, with samples taken every $\frac{1}{4 B}$ seconds where $B = M \nu_p = 930$ KHz. RRC pulse $w_{tx}(\tau, \nu)$ with roll-off factors, $\beta_{\tau} = \beta_{\nu} = 0.6$. Observe that the TD realization is not as peaky as that of the point pulsone shown in Fig.~\ref{fig3}.
} 
\label{fig4zakotfsspreadtd}
\end{figure}

In Section \ref{secfiltering}, we constructed a spread
pulsone using a discrete chirp filter, and we compared the energy profiles of point and spread pulsones in the discrete DD domain.
Recall that the TD realization of the point pulsone located at $(k_p,l_p)$ consists of narrow TD pulses at time instances $t = {\Big (} n \tau_p + k_p \frac{\tau_p}{M} {\Big )}, n \in {\mathbb Z}$ where the spread of each pulse is approximately $\tau_p/M$, which is the inverse bandwidth $1/B$ (see Fig.~\ref{fig3}). Fig.~\ref{fig4zakotfsspreadtd} shows the TD realization of the spread pulsone constructed in Section \ref{secfiltering}. This TD realization is the sum of all $MN$ pulsones located on the information grid
and the constant term $1/(MN)$ appearing in (\ref{eqnwpchirp}) reduces the amplitude of each point pulsone in the sum.
The $N$ pulsones located at a given $k_p$ interfere destructively, and the point pulsones at different locations $k_p$ result in trains of TD pulses at different locations.
This explains why the TD realization of the spread pulsone shown in Fig.~\ref{fig4zakotfsspreadtd} is less peaky than the TD realization of the point pulsone shown in Fig.~\ref{fig3}.     

\begin{figure}[h]
\centering
\includegraphics[width=9.5cm, height=5.2cm]{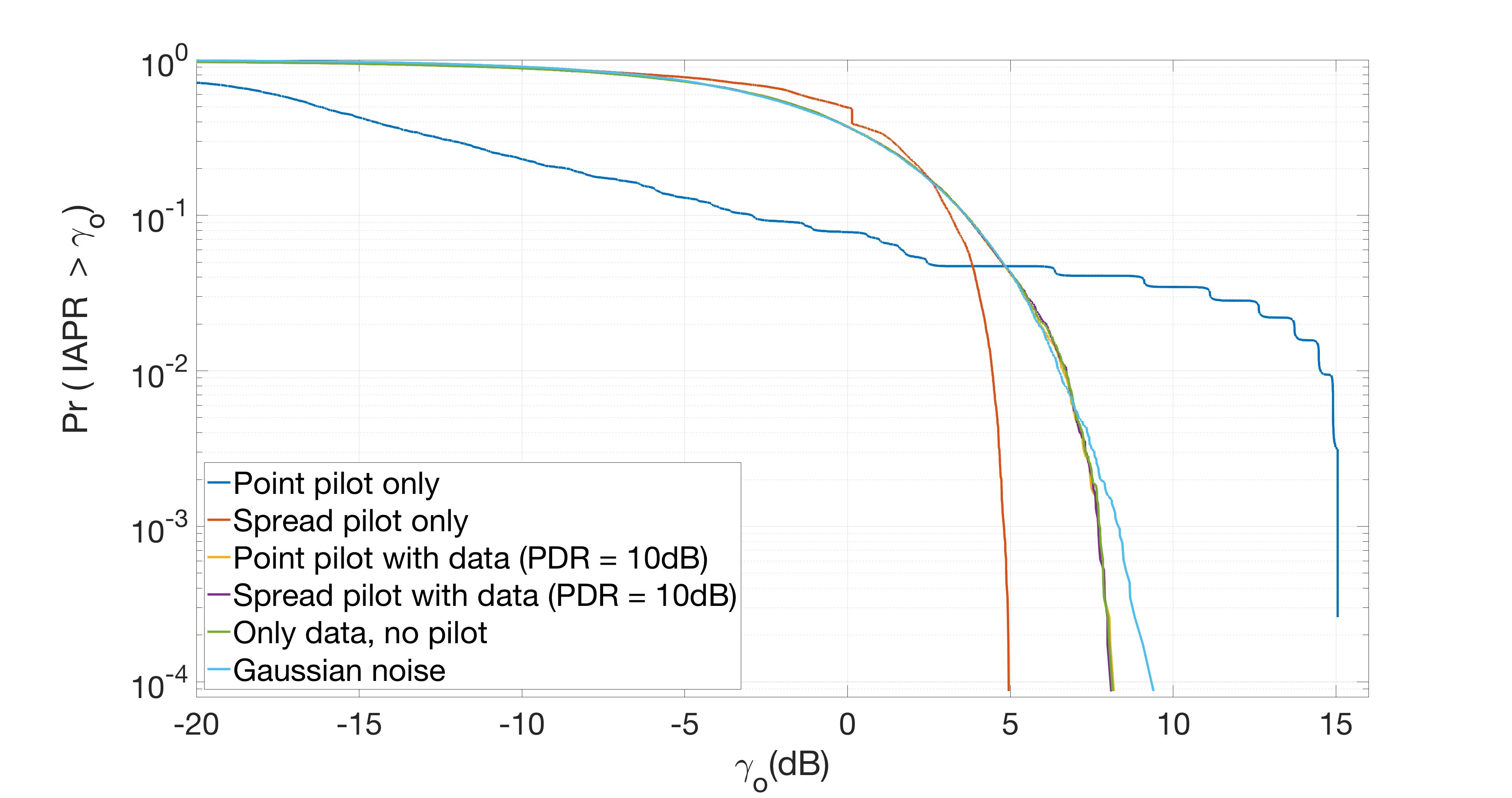}
\caption{Complementary CDF (CCDF) plot of IAPR. $M = 31, N = 37$. RRC pulse $w_{tx}(\tau, \nu)$ with roll-off factors, $\beta_{\tau} = \beta_{\nu} = 0.6$. Discrete chirp filter with $q=3$. PDR $\rho_p/\rho_d = 10$ dB.}
\label{fig55}
\end{figure}

Fig.~\ref{fig55} displays the complementary CDF of the instantaneous-to-average-power ratio (IAPR) for the spread and point pilots. In the absence of data, the IAPR of the spread pulsone does not exceed $5$ dB, whereas the IAPR of the point pilot is almost $15$ dB. With data alone, the IAPR is similar to that of Gaussian noise. In Section \ref{integdatasense} we choose a pilot to data power ratio (PDR) of $10$ dB to integrate sensing and communication
within a single Zak-OTFS subframe. For both spread pilot with data and point pilot with data, the IAPR rarely exceeds $7$ dB. However, for a higher PDR of $25$ dB, while the IAPR of spread pilot with data rarely exceeds $9$ dB,
the IAPR of point pilot with data can be as high as $12$ dB.
We have shown that spreading reduces the PAPR of the transmitted signal.

\subsection{Achieving predictability}
\label{secpredictspreadpilot}
In this Section we show that the self-ambiguity function of the spread pulsone is supported on a lattice $\Lambda_q$ that is obtained by applying a linear transformation to the period lattice $\Lambda_p$ that depends on the slope $q$ of the discrete chirp. When the effective channel filter satisfies
the crystallization condition with respect to the rotated lattice $\Lambda_q$, we will describe how to estimate the taps of $h_{\mbox{\scriptsize{eff}}}[k,l]$ from the cross-ambiguity function of the received pilot and the transmitted pilot. In this Section we will focus on sensing in the
absence of data and noise, but in Section \ref{integdatasense} we will describe methods for channel sensing in the presence of data and noise.

\textbf{Remark:} In the continuous DD domain, radar engineers construct waveform libraries for tracking applications consisting of a fixed waveform that has been
linearly frequency modulated or \emph{chirped} at various rates.
Chirping transforms the self-ambiguity surface, making
it possible to track a greater diversity of radar targets (see \cite{Howard2004} for more details). We plan to explore in a subsequent paper the connections between waveform design  in the discrete and continuous DD domains.

The spread pilot signal $x_{\mbox{\scriptsize{s,dd}}}[k,l] = w[k,l] \circledast_{\sigma} x_{\mbox{\scriptsize{p,dd}}}[k,l]$ is given by (\ref{eqn_periodtwistconv}).
Here $w[k,l]$ is the spreading filter and $x_{\mbox{\scriptsize{p,dd}}}[k,l]$
is the point pilot signal. In the absence of data, the received pilot signal is given by
\begin{eqnarray}
\label{eqn025448}
y_{\mbox{\scriptsize{s,dd}}}[k,l] & = & h_{\mbox{\scriptsize{eff}}}[k,l] *_{\sigma}  x_{\mbox{\scriptsize{s,dd}}}[k,l] \, + \, n_{\mbox{\scriptsize{dd}}}[k,l].
\end{eqnarray}The ML estimate of $\widehat{h}_{\mbox{\scriptsize{eff}}}[k,l]$ is then given by the cross-ambiguity between the received spread pilot and the transmitted spread pilot. Appendix \ref{app_prf_mlest} derives this ML estimate from first principles. It follows from (\ref{crossambig_1}) in Appendix \ref{app_prop_ambig} that
\begin{eqnarray}
\label{eqnhestambigauto}
{\widehat h}_{\mbox{\scriptsize{eff}}}[k,l] & = & A_{y_s, x_s}[k,l] \nonumber \\
& & \hspace{-22mm}  = \sum\limits_{k'=0}^{M-1} \sum\limits_{l'=0}^{N-1} y_{\mbox{\scriptsize{s,dd}}}[k',l'] \, x_{\mbox{\scriptsize{s,dd}}}^*[k' - k,l' - l] \, e^{-j 2 \pi \frac{l(k' - k)}{MN}},
\end{eqnarray}for $(k,l) \in {\mathcal S}$.  We apply Theorem \ref{thm_crossambig2} from Appendix \ref{app_prop_ambig} to obtain (in the absence of data and noise)
\begin{eqnarray}
\label{eqnaysays}
    A_{y_s, x_s}[k,l] & = &  {h}_{\mbox{\scriptsize{eff}}}[k,l] *_{\sigma}  A_{x_s, x_s}[k,l],
\end{eqnarray}where $A_{x_s, x_s}[k,l]$ is the self-ambiguity function of the spread pilot $x_{\mbox{\scriptsize{s,dd}}}[k,l]$ (cf. (\ref{ayxpeqn2})).

Henceforth, we restrict our attention to the spread pulsone
\begin{eqnarray}
    x_{\mbox{\scriptsize{s,dd}}}[k,l] & = & w[k,l] \, \circledast_{\sigma}  x_{\mbox{\scriptsize{p,dd}}}[k,l]
\end{eqnarray}arising from the discrete chirp filter
$w[k,l] = \frac{1}{MN} e^{j 2 \pi q \frac{(k^2 + l^2)}{MN}}$.
Since ambiguity functions are periodic along both delay and Doppler axes with period $MN$, i.e., $A_{x_s, x_s}[k+p_1MN,l+p_2MN] = A_{x_s, x_s}[k,l]$, for all $p_1,p_2 \in {\mathbb Z}$, it suffices to characterize the support set of
$A_{x_s, x_s}[k,l]$ modulo $MN$ along both axes.
Theorem \ref{thm_askl1} below shows that the self-ambiguity function
$A_{x_s, x_s}[k,l]$ is supported on a lattice.
\begin{theorem}
\label{thm_askl1}
Given odd primes $N, M$ and $q$
relatively prime to both $M$ and $N$
\begin{eqnarray}
\label{eqnaskl4}
\vert A_{x_s,x_s}[k,l] \vert & = \begin{cases}
    1 &,  \, [2qk - l]_{M} = 0 \\
    &,\, [k - l( (2q)^{-1} \,  - 2q )]_{N} = 0\\
    0 &, \mbox{\small{otherwise}} \\
\end{cases},
\end{eqnarray}
where $(2q)^{-1}$ is the unique inverse of $2q$ modulo $MN$.\footnote{\footnotesize{For any integer $a$ and positive integer $M$, $[ a ]_{M}$ denotes the unique smallest non-negative integer which is congruent to $a$ modulo $M$.}}
\end{theorem}
\begin{IEEEproof}
See Appendix \ref{prf_thm_askl1}.
\end{IEEEproof}
Theorem \ref{thm_askl1} implies that the self-ambiguity function $A_{x_s,x_s}[k,l]$
is non-zero if and only if there exist integers $n,m$ such that
\begin{eqnarray}
\label{eqn8747}
l & = & nM + 2qk \,,\, \nonumber \\
k & = & \theta l \, + \, mN, \nonumber \\
\theta & \Define & [(2q)^{-1} \,  - 2q]_{MN}.
\end{eqnarray}We rewrite
(\ref{eqn8747}) as
\begin{eqnarray}
\label{eqn826649}
    (1 - 2q \theta) k & = & \theta n M  \, + \, mN \nonumber \\
    (1 - 2q \theta) l & = & nM \, + \, 2q m N.
\end{eqnarray}It follows that the self-ambiguity function $A_{x_s,x_s}[k,l]$
is supported on a lattice $\Lambda_q$
given by\footnote{\footnotesize{In (\ref{eqn873412}), arithmetic is interpreted as modulo $MN$.}}
\begin{eqnarray}
\label{eqn873412}
    (1 - 2 q \theta) \, \Lambda_q & = & 
\begin{bmatrix}  \theta & 1 \\ 1 & 2 q \\ \end{bmatrix} \, \Lambda_p.
\end{eqnarray}We show that the lattices
$\Lambda_p, \Lambda_q$ have the same fundamental volume by showing that they have the same density. We show the densities are the same by showing that the number of $\Lambda_q$ points in the rectangle $\Omega$ bounded by $(0,0)$,
$(MN, 0)$, $(0, MN)$ and $(MN, MN)$ is the same as the number 
of $\Lambda_p$ points.

Since $\theta  =   [(2q)^{-1} \,  - 2q]_{MN}$, $(1 - 2q \theta) \equiv 4q^2$ modulo $MN$. Since $M$ and $N$ are odd primes and $q$ is relatively prime to both $M$ and $N$, $4q^2$ is relatively prime to $MN$. Therefore, $(1 - 2q \theta)$ is relatively prime to $MN$, i.e., $(1 - 2q \theta) \not \equiv 0$ modulo $MN$ and hence, the linear transformation $ \begin{bmatrix}  \theta & 1 \\ 1 & 2 q \\ \end{bmatrix}$ is non-singular modulo $MN$.
Hence the number of $\Lambda_q$ points in $\Omega$ equals the number of $\Lambda_p$ points in $\Omega$.

From (\ref{crossambig_1}) in Appendix \ref{app_prop_ambig} we know that
the total energy of the spread pulsone is simply the value of its cross-ambiguity at the origin, i.e.,
\begin{eqnarray}
\label{eqnspilotenergy}
    A_{x_s,x_s}[0,0] & = & \sum\limits_{k=0}^{M-1} \sum\limits_{l=0}^{N-1} \vert x_{\mbox{\scriptsize{s,dd}}}[k,l] \vert^2 \, = \, 1. 
\end{eqnarray}The last equality follows from the proof of Theorem \ref{thm_askl1} in Appendix \ref{prf_thm_askl1}.

From (\ref{eqnaskl4}) it follows that the self-ambiguity function $A_{x_s,x_s}[k,l]$ is given by
\begin{eqnarray}
\label{eqnlatticelq}
    A_{x_s,x_s}[k,l] & \hspace{-3mm} = & \hspace{-3mm}  \sum\limits_{(k_i, l_i) \in \Lambda_q } \hspace{-2mm} e^{j \theta_i} \, \delta[k - k_i] \, \delta[l - l_i].
\end{eqnarray}where $(k_i, l_i)$ are the lattice points of $\Lambda_q$.
Using (\ref{eqnlatticelq}) in (\ref{eqnaysays}) gives
\begin{eqnarray}
\label{eqn026445}
     \, h_{\mbox{\scriptsize{eff}}}[k,l] \, *_{\sigma} \, {\mathcal A}_{x_s,x_s}[k,l]  &  & \nonumber \\
    & & \hspace{-38mm} \mya   \sum\limits_{(k_i, l_i) \in \Lambda_q }   \hspace{-3mm} h_{\mbox{\scriptsize{eff}}}[k,l] \, *_{\sigma} \, \left( \, e^{j \theta_i} \, \delta[k - k_i] \delta[l - l_i] \right) \nonumber \\
    & & \hspace{-38mm} =  \hspace{-3mm} \sum\limits_{(k_i, l_i) \in \Lambda_q }  \hspace{-3mm} e^{j \theta_i} \, h_{\mbox{\scriptsize{eff}}}[k - k_i,l - l_i] \, e^{j 2 \pi \frac{(l - l_i) k_i}{MN}}  \nonumber \\
    & & \hspace{-40mm} =  h_{\mbox{\scriptsize{eff}}}[k ,l] \, +    \hspace{-12mm} \sum\limits_{(k_i, l_i) \in \Lambda_q , (k_i, l_i) \ne (0,0)}  \hspace{-13mm} e^{j \theta_i} \, h_{\mbox{\scriptsize{eff}}}[k - k_i,l - l_i] \, e^{j 2 \pi \frac{(l - l_i) k_i}{MN}}.  \nonumber \\
\end{eqnarray}The first term in the R.H.S. of the last equation above is the discrete channel filter $h_{\mbox{\scriptsize{eff}}}[k,l]$ itself. The other terms in the R.H.S. correspond to the other
lattice points not at the origin. Therefore, accurate estimation of the taps of $h_{\mbox{\scriptsize{eff}}}[k,l]$ is not possible if for any non-zero lattice point $(k_i, l_i) \ne (0,0)$ the support set of the corresponding term $ e^{j \theta_i} \,  h_{\mbox{\scriptsize{eff}}}[k - k_i,l - l_i] \, e^{j 2 \pi \frac{(l - l_i) k_i}{MN}}$ overlaps with the support set of $h_{\mbox{\scriptsize{eff}}}[k,l]$. We now show that if the channel satisfies the crystallization condition with respect to $\Lambda_q$, then we can accurately estimate the taps of $h_{\mbox{\scriptsize{eff}}}[k,l]$. 

\begin{figure}[h]
\centering
\includegraphics[width=7.5cm, height=5.2cm]{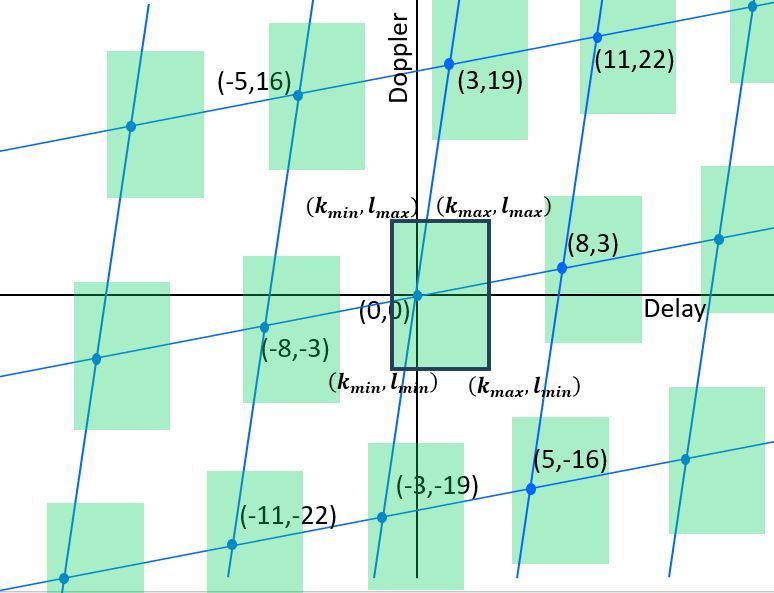}
\caption{Blue dots mark the support of the self-ambiguity function of the spread pulsone with $M=11, N= 13, q=5$, $(k_p, l_p) = (0,0)$. The support of $h_{\mbox{\scriptsize{eff}}}[k,l]$ is limited to a rectangular region with delay spread $(k_{max} - k_{min})$ and Doppler spread $(l_{max} - l_{min})$
shown as the green rectangle with the black border.}
\label{fig0284}
\end{figure}

\emph{\underline{Example 1}:}
The blue dots in Fig.~\ref{fig0284} mark the points of the lattice $\Lambda_q$ i.e., support of the self-ambiguity function of the spread pulsone with $M = 11, N= 13$, $q=5$, $(k_p , l_p) = (0,0)$. The fundamental period/region of $\Lambda_q$ is a parallelogram with area $143 = 11 \times 13 = M N$ bounded by the points $(0,0)$, $(3,19)$, $(11,22)$ and $(8,3)$. The lattice $\Lambda_q$ is generated by $(3,19)$ and $(8,3)$.
The cross-ambiguity between the received spread pulsone and the transmitted spread pulsone is supported on the union of the green rectangles (these rectangles correspond to the terms in the RHS of (\ref{eqn026445})). Since the rectangles do not overlap, there is no DD domain aliasing (i.e., the channel satisfies the crystallization condition with respect to $\Lambda_q$), and we can accurately estimate $h_{\mbox{\scriptsize{eff}}}[k,l]$ from the response received within the green rectangle with the black border.

\begin{figure}[h]
\hspace{-2mm}
\includegraphics[width=9.0cm, height=4.0cm]{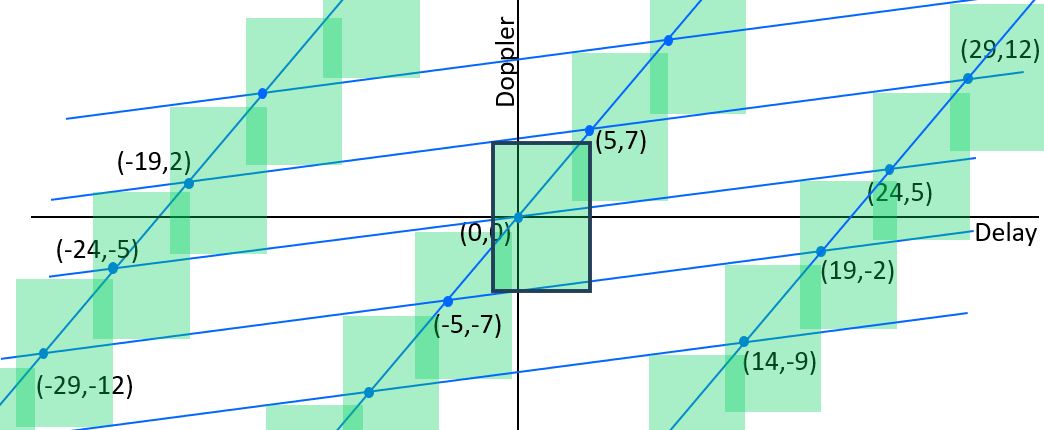}
\caption{
Blue dots mark the lattice $\Lambda_q$ supporting the
self-ambiguity function of the spread pulsone with $M = 11, N = 13, q=4$, $(k_p, l_p) = (0,0)$.
The lattice $\Lambda_q$ is generated by the vectors $(24,5)$ and $(5,7)$. The channel spread is the same as that in Fig.~\ref{fig0284}. The support of $h_{\mbox{\scriptsize{eff}}}[k,l]$ is limited to the green rectangle with the black border.}
\label{fig0288}
\end{figure}

\emph{\underline{Example 2:}}
The blue dots in Fig.~\ref{fig0288} mark the support
of the self-ambiguity function of the spread pulsone
with $M = 11, N= 13, q=4$, $(k_p, l_p) = (0,0)$.
The image of the fundamental period/region is a parallelogram bounded by $(0,0)$, $(24, 5)$, $(5,7)$ and $(29,12)$. The support of $h_{\mbox{\scriptsize{eff}}}[k,l]$ is the same as that in Fig.~\ref{fig0284} and is again shown as the green rectangle with the black border.\footnote{\footnotesize{Note that the size and shape of the green rectangle does not change with $q$ since it depends only on the delay Doppler spreading function $h_{\mbox{\scriptsize{phy}}}(\tau, \nu)$ of the underlying physical channel, the pulse shaping filters $w_{tx}(\tau, \nu)$, $w_{rx}(\tau, \nu)$, and the information grid ${\Lambda}_{dd}$ (which in turn depends only on $T$ and $B$, and is independent of $(\tau_p, \nu_p)$).}} This rectangle overlaps the green rectangles located at $(5,7)$ and $(-5, -7)$ and taps of $h_{\mbox{\scriptsize{eff}}}[k,l]$ located in the overlapped regions cannot be estimated accurately. By contrast, taps of $h_{\mbox{\scriptsize{eff}}}[k,l]$ located in the non-overlapped region can be estimated accurately.   

In \cite{zakotfs2} we have emphasized that the (strong) crystallization condition is satisfied when the channel delay spread is less than
the delay period and the channel Doppler spread is less than the Doppler period. We have also described a weaker mathematical condition that eliminates DD domain aliasing (\cite{zakotfs2}, Section II-D). When we use discrete chirp filters to construct spread pulsones, the self-ambiguity function is supported on a transformed lattice $\Lambda_q$. The significance of the weaker mathematical condition is that there are channels for which aliasing-free acquisition of $h_{\mbox{\scriptsize{eff}}}[k ,l ]$ is not possible with the period lattice $\Lambda_p$, but becomes possible with the transformed lattice $\Lambda_q$.

Consider a spread pulsone where the self-ambiguity function $A_{x_s,x_s}[k,l]$ is supported on a lattice $\Lambda_q$. Define the support ${\mathcal S}_{(k_i, l_i)}$ of $h_{\mbox{\scriptsize{eff}}}[k - k_i ,l - l_i ]$ by
\begin{eqnarray}
\label{eqndefskili}
    {\mathcal S}_{(k_i,l_i)} & \Define & \{ (k,l) \, | \, h_{\mbox{\scriptsize{eff}}}[k - k_i ,l - l_i ] \ne 0 \}.
\end{eqnarray}The weaker mathematical condition that eliminates DD domain aliasing is
\begin{eqnarray}
\label{gencondition3}
{\mathcal S}_{(0,0)} & \cap  &  {\Bigg (} \bigcup_{\substack{(k_i, l_i) \in \Lambda_q, \\  (k_i, l_i) \ne (0,0)}}  \, {\mathcal S}_{(k_i,l_i)} {\Bigg )} \, \,   = \, \,  \varnothing.
\end{eqnarray}Observe that the set ${\mathcal S}_{(k_i,l_i)}$ is obtained by translating
${\mathcal S}_{(0,0)}$ by the lattice point $(k_i, l_i) \in \Lambda_q$. Note also that when (\ref{gencondition3}) is satisfied, two distinct taps in ${\mathcal S}_{(0,0)}$ cannot differ by a
lattice point in $\Lambda_q$.

\begin{lemma}
\label{eqnlemma1}
    Let $(k'_2, l'_2) \ne (k'_1, l'_1)$,
    be two distinct taps in ${\mathcal S}_{(0,0)}$.
    If the weaker crystallization condition (\ref{gencondition3}) is satisfied then  
    \begin{eqnarray}
    A_{x_s,x_s}[k'_2 - k'_1, l'_2 - l'_1] & = & 0.
    \end{eqnarray}
\end{lemma}
\begin{IEEEproof}The self-ambiguity
function $A_{s,s}[k,l]$ is supported on the lattice
$\Lambda_q$ and $(k'_2 - k'_1, l'_2 - l'_1) \notin \Lambda_q$ when (\ref{gencondition3}) is satisfied. This is because, if $(k'_2 - k'_1, l'_2 - l'_1)$ were to be some $(k_i, l_i) \in \Lambda_q$ ($(k_i, l_i) \ne (0,0)$), then $(k_i + k_1', l_i + l_1') \in {\mathcal S}_{(k_i, l_i)}$ which is not possible since $(k_i + k_1', l_i + l_1') = (k_2' , l_2') \in {\mathcal S}_{(0,0)}$ and this point cannot lie both in ${\mathcal S}_{(0,0)}$ and ${\mathcal S}_{(k_i, l_i)}$ as (\ref{gencondition3}) is satisfied.
\end{IEEEproof}


\begin{figure}[h]
\hspace{-4mm}
\includegraphics[width=9.4cm, height=4.0cm]{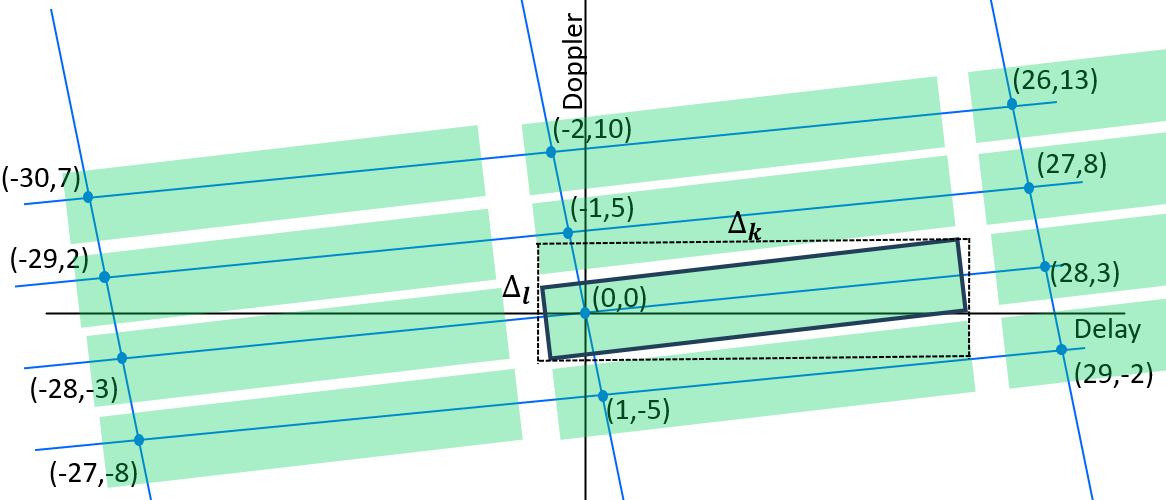}
\caption{Blue dots mark the lattice $\Lambda_q$ supporting the self-ambiguity function of the spread pulsone
with $M=11, N=13, q=3$, $(k_p, l_p) = (0,0)$.
The support ${\mathcal S}_{(0,0)}$ of the effective channel taps is shown as the green rectangle with the black border. The dotted rectangle is the smallest rectangle with axes parallel to the delay and Doppler axes that circumscribes ${\mathcal S}_{(0,0)}$.}
\label{fig0295}
\end{figure}

\emph{\underline{Example 3:}}
We consider a channel with support set ${\mathcal S}_{(0,0)}$ for which the delay spread $\Delta_k = (\max_{(k,l) \in {\mathcal S}_{(0,0)}} k  \, - \, \min_{(k,l) \in {\mathcal S}_{(0,0)}} k )$ and the Doppler spread $\Delta_l \Define (\max_{(k,l) \in {\mathcal S}_{(0,0)}} l  \, - \, \min_{(k,l) \in {\mathcal S}_{(0,0)}} l )$ satisfy $\Delta_k \, \Delta_l > MN = BT$ ($MN = (B \tau_p) (T \nu_p) = B T$). For a given $B$ and $T$, it is not possible to choose periods $M$ and $N$ such that $\Delta_k < M$ and $\Delta_l < N$, so the channel does not satisfy the (strong) crystallization condition with respect to the period lattice $\Lambda_p$. Here we show that aliasing-free acquisition of $h_{\mbox{\scriptsize{eff}}}[k,l]$ may still be possible if the channel is underspread ($\left\vert  {\mathcal S}_{(0,0)}\right\vert < MN$). The support set of the effective channel filter is shown as the green rectangle with the black border in Fig.~\ref{fig0295}. The dotted rectangle in Fig.~\ref{fig0295} is the smallest rectangle with axes parallel to the delay and Doppler axes that circumscribes ${\mathcal S}_{(0,0)}$. The crystallization condition is not satisfied with respect to the period lattice $\Lambda_p$ since the area of the dotted rectangle $\Delta_k \Delta_l $ is greater than $MN$. However the (weaker) crystallization condition (\ref{gencondition3}) is satisfied with respect to $\Lambda_q$ because the green rectangles are disjoint. Hence, we can accurately estimate the effective channel taps $h_{\mbox{\scriptsize{eff}}}[k,l]$.   

\begin{figure}[h!]
    	\vspace{-3mm}
     \centering
    	\begin{subfigure}[b]{0.5\textwidth}
    	\includegraphics[width=\textwidth]{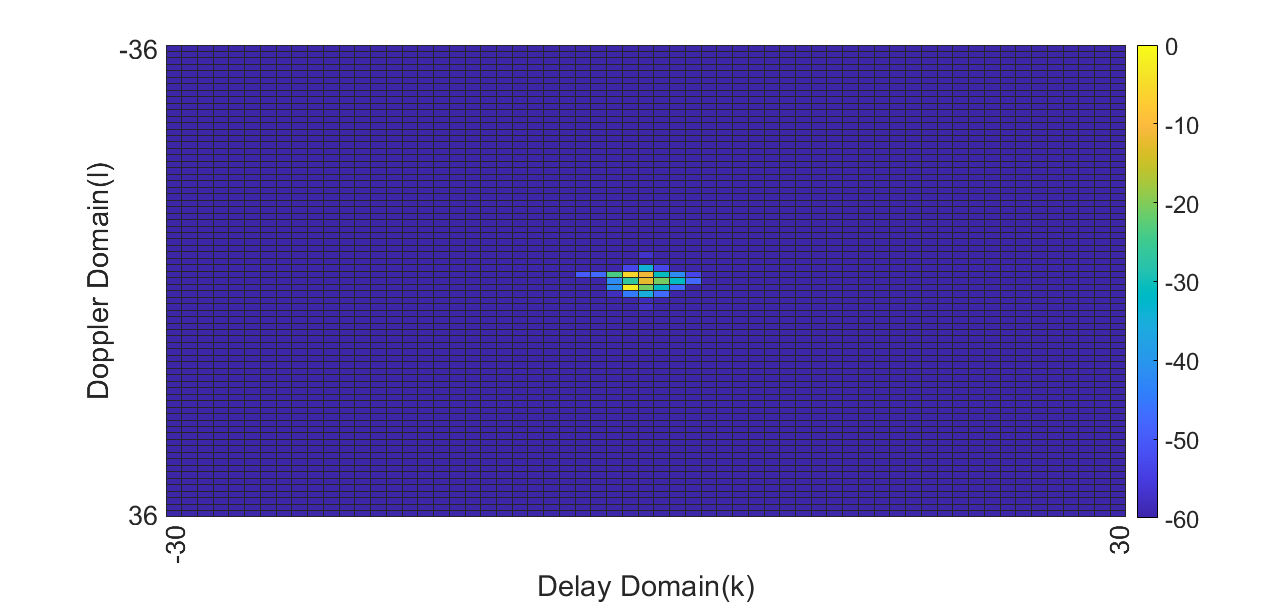}
    	\caption{{$\nu_{max} = 815$ Hz}}
    	\label{fig917523}
    \end{subfigure}
    \begin{subfigure}[b]{0.5\textwidth}
	\includegraphics[width=\textwidth]{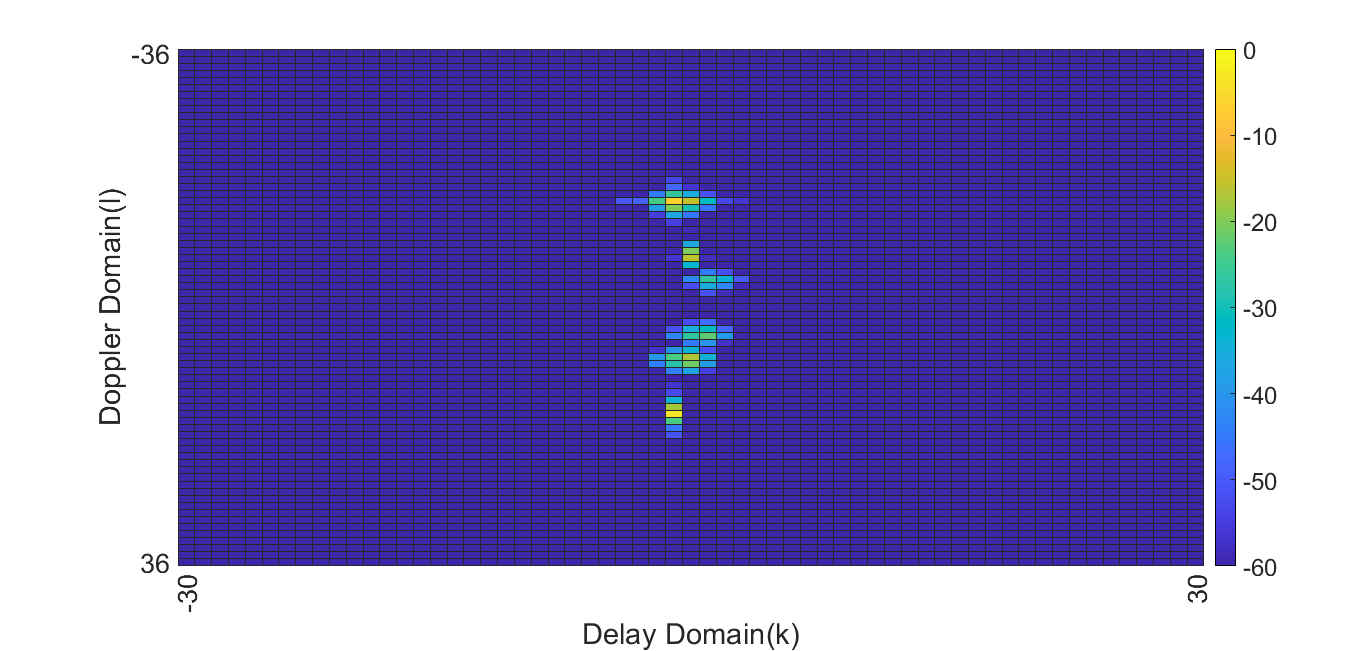}
	\caption{{$\nu_{max} = 12$ KHz}}
	\label{fig917524}
	\end{subfigure}
	\vspace{-2mm}
	\caption{Heat map of $\vert h_{\mbox{\scriptsize{eff}}}[k,l] \vert$. Veh-A channel, RRC pulse shaping filter ($\beta_{\tau} = \beta_{\nu} = 0.6$), Doppler period $\nu_p = 30$ KHz,  $M = 31, N=37$.}
 \label{fig9298}
\end{figure}

\emph{\underline {Example 4:}}
We consider the Veh-A channel with the power delay profile given in Table-\ref{tab1_paper2} of Section \ref{zakotfspointpilot} and with path Doppler shifts $\nu_1 = \nu_{max}$, $\nu_2 = - \nu_{max}$, $\nu_3 = \nu_{max}/2$, $\nu_4 = -\nu_{max}/2$, $\nu_5 = \nu_{max}/4$, $\nu_6 = -\nu_{max}/4$. We now illustrate the importance of filter design by comparing heat maps of the
cross-ambiguity $A_{y_s,x_s}[k,l]$ (see (\ref{eqnaysays}))
between the received spread pulsone and the transmitted spread pulsone, for spread pulsones defined by chirp filters with $q=3$ and $q=36$. Before we study these comparisons, through Fig.~\ref{fig9298} we illustrate the heat map for the effective discrete DD domain channel filter $h_{\mbox{\scriptsize{eff}}}[k,l]$. It is observed that the effective channel Doppler spread is higher for $\nu_{max} = 12$ KHz than for $\nu_{max} = 815$ Hz.

\begin{figure}[h!]
    	\vspace{-3mm}
     \centering
    	\begin{subfigure}[b]{0.5\textwidth}
    	\includegraphics[width=\textwidth]{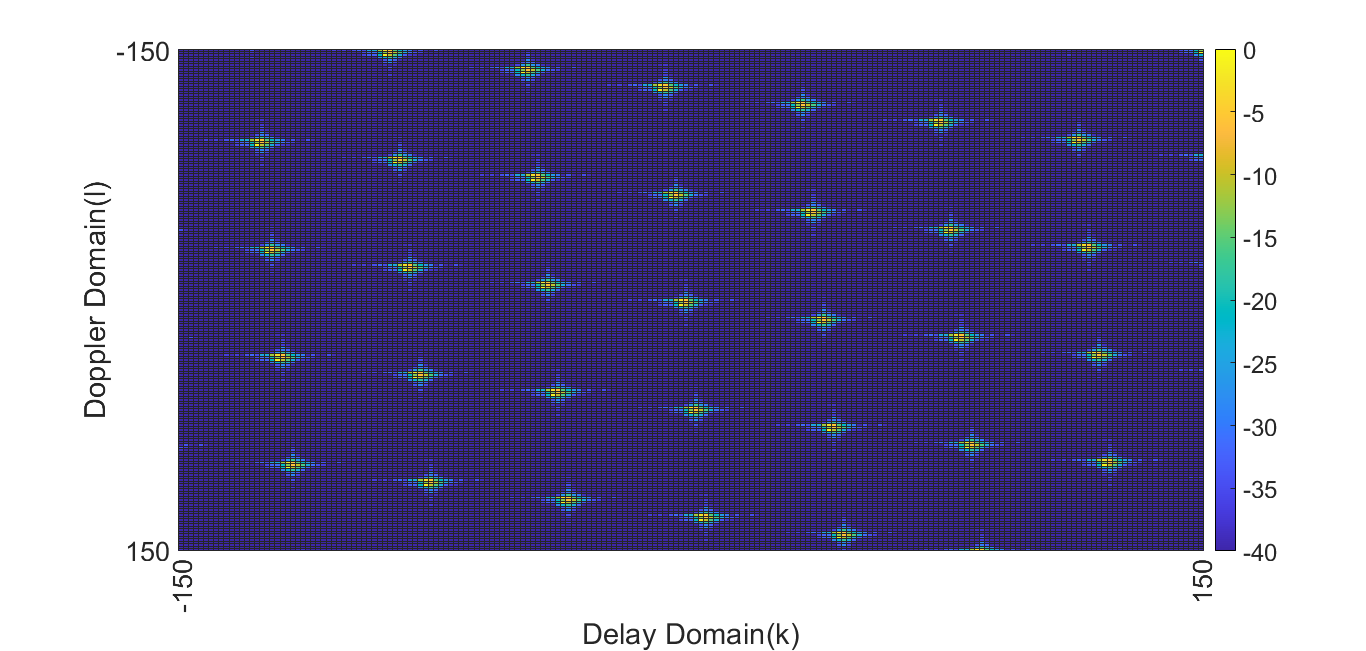}
    	\caption{{$\nu_{max} = 815$ Hz}}
    	\label{fig917223}
    \end{subfigure}
    \begin{subfigure}[b]{0.5\textwidth}
	\includegraphics[width=\textwidth]{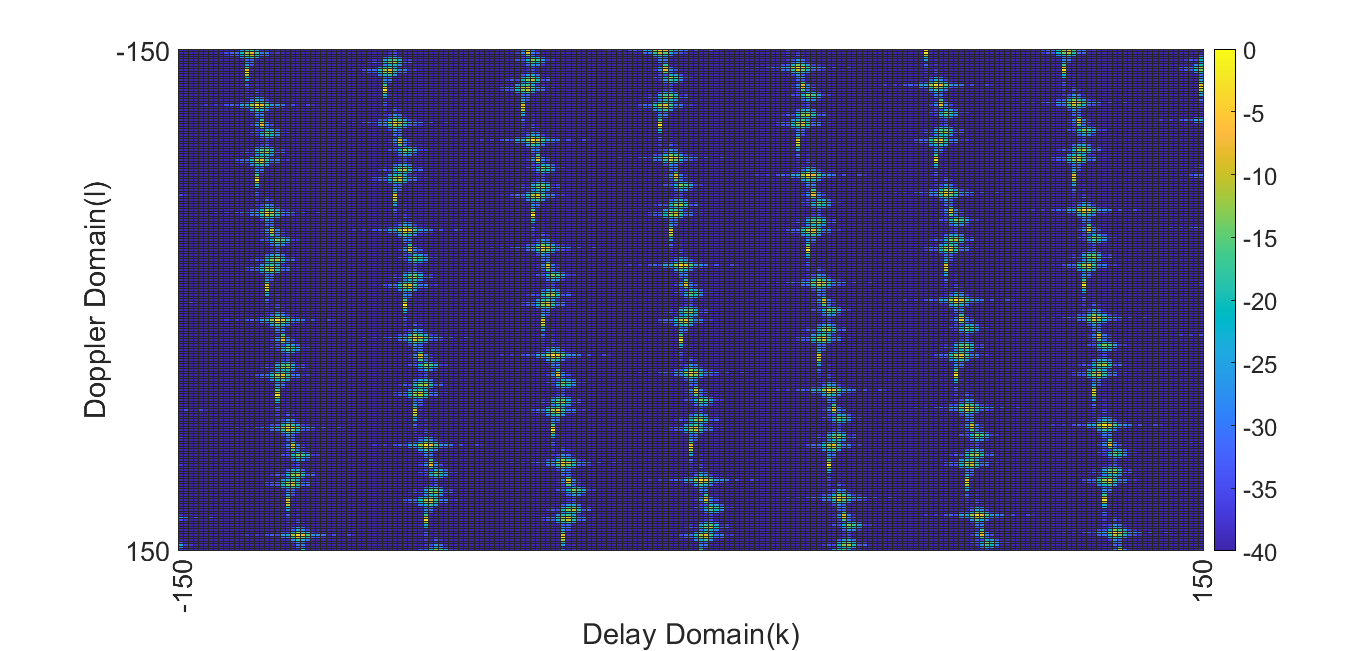}
	\caption{{$\nu_{max} = 12$ KHz}}
	\label{fig967524}
	\end{subfigure}
	\vspace{-2mm}
	\caption{Heat map of $\vert A_{y_s,x_s}[k,l] \vert$ for chirp filter with $q=3$. Veh-A channel described in Fig.~\ref{fig9298}.}
 \label{fig4298}
\end{figure}

We first consider the spread pulsone defined by the chirp filter with $q=3$. Fig.~\ref{fig4298} shows the heat map of $\vert A_{y_s,x_s}[k,l] \vert$ for $\nu_{max} = 815$ Hz and $\nu_{max} = 12$ KHz. In both cases, the support sets ${\mathcal S}_{k_i,l_i}$, $(k_i, l_i) \in {\Lambda}_q$
of $h_{\mbox{\scriptsize{eff}}}[k - k_i,l - l_i]$
do not overlap and the (weaker) crystallization condition is satisfied.

\begin{figure}[h!]
    	\vspace{-3mm}
     \centering
    	\begin{subfigure}[b]{0.5\textwidth}
    	\includegraphics[width=\textwidth]{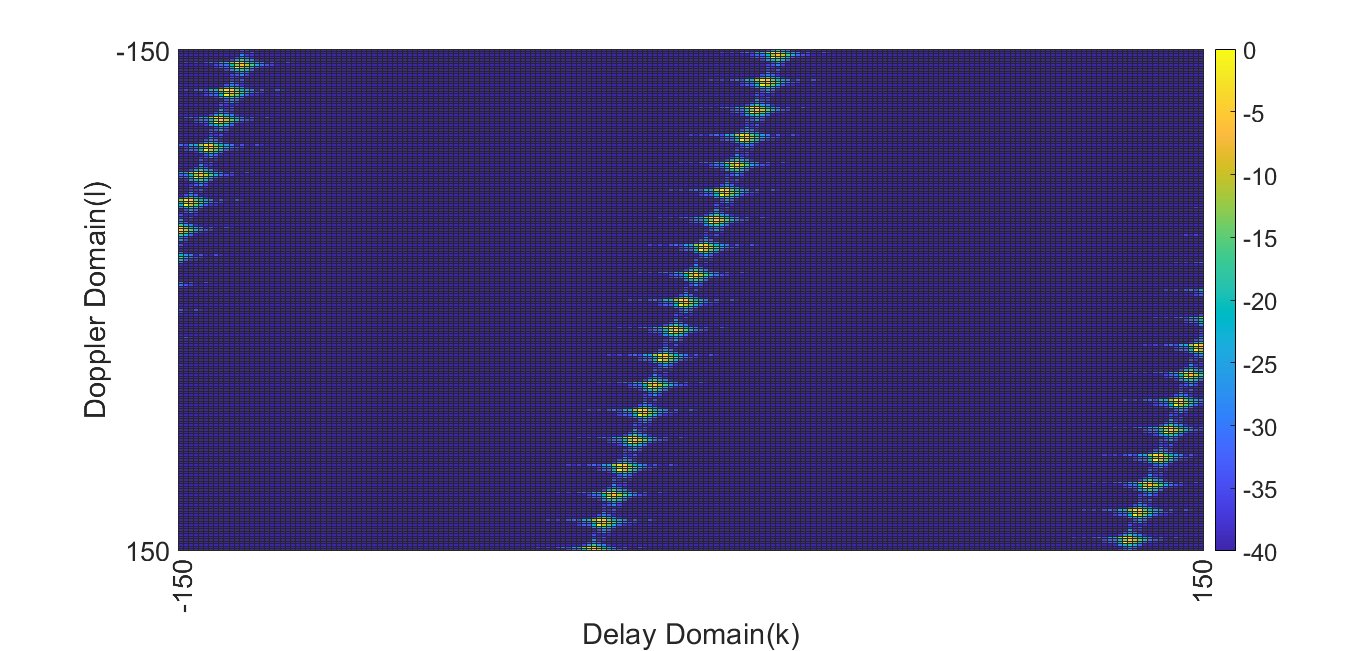}
    	\caption{{$\nu_{max} = 815$ Hz}}
    	\label{fig117223}
    \end{subfigure}
    \begin{subfigure}[b]{0.5\textwidth}
	\includegraphics[width=\textwidth]{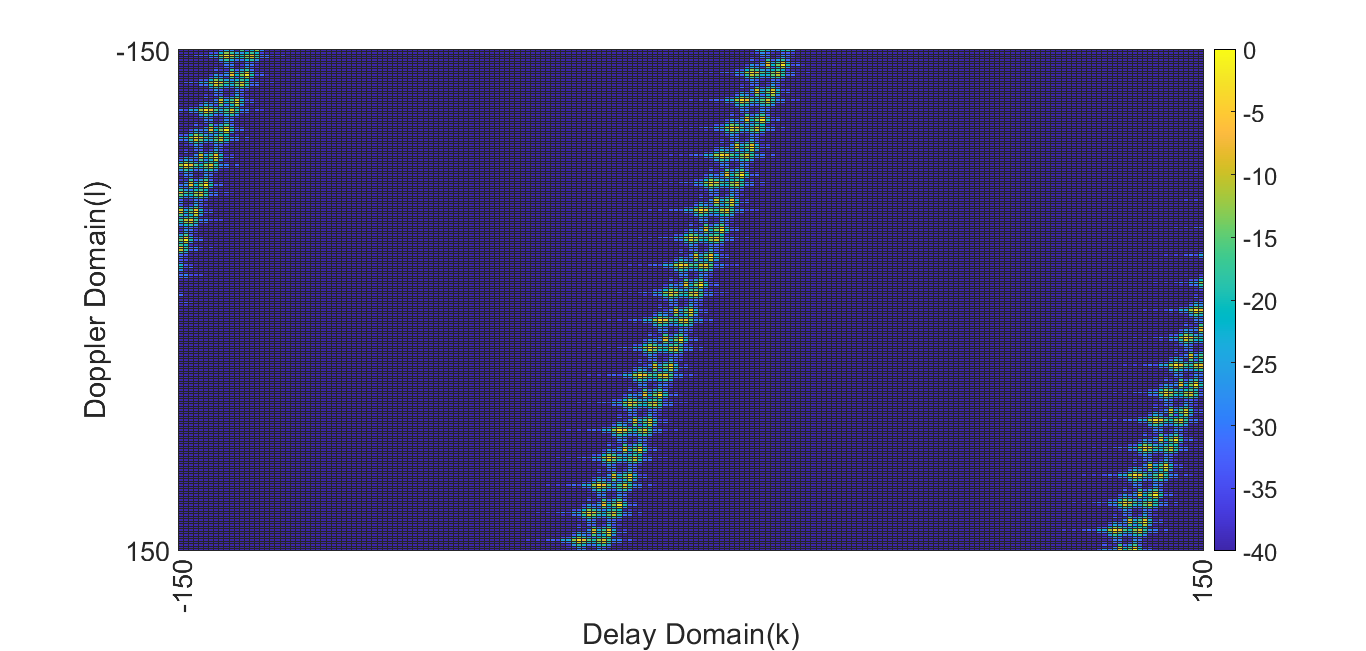}
	\caption{{$\nu_{max} = 12$ KHz}}
	\label{fig867524}
	\end{subfigure}
	\vspace{-2mm}
	\caption{Heat map of $\vert A_{y_s,x_s}[k,l] \vert$ for chirp filter with $q=36$. Veh-A channel described in Fig.~\ref{fig9298}.}
 \label{fig5298}
\end{figure}

Next we consider the spread pulsone defined by the
chirp filter with $q=36$. Fig.~\ref{fig5298} shows the heat map of $\vert A_{y_s,x_s}[k,l] \vert$ for $\nu_{max} = 815$ Hz and $\nu_{max} = 12$ KHz. When $\nu_{max} = 815$ Hz, the support sets do not overlap. When $\nu_{max} = 12$ KHz, the support sets do overlap and the (weaker) crystallization condition is not satisfied. This compromises the accuracy of channel estimation, which in turn degrades BER performance (see Section \ref{integdatasense} for more details).

\section{Integrated sensing and communication (ISAC)}
\label{integdatasense}

\subsection{ISAC with spread sensing pulsone}
\label{spreadjcas}
When we integrate channel sensing and data transmission in the same Zak-OTFS subframe, the spread pulsone used for channel sensing interferes with the point pulsones used for data transmission. In this Section, we describe how to design the spread pulsone so that interference is noise-like. This property eliminates the need for a guard band at the cost of reducing the SNR for channel sensing. Noise-like interference translates to a spread pulsone with energy that is almost uniformly distributed over the discrete DD domain. This explains why the PAPR of the spread pulsone is significantly lower than that of the point pulsone.

The discrete DD domain transmit signal $x_{\mbox{\scriptsize{dd}}}[k, l]$ is given by
\begin{eqnarray}
\label{eqn30_1}
    x_{\mbox{\scriptsize{dd}}}[k,l] & = & \sqrt{E_d} \, x_{\mbox{\scriptsize{d,dd}}}[k,l] \, + \, \sqrt{E_p} \, x_{\mbox{\scriptsize{s,dd}}}[k,l]
\end{eqnarray}is the sum of a data term and a sensing term. The quasi-periodic data signal $x_{\mbox{\scriptsize{d,dd}}}[k,l]$, $k,l \in {\mathbb Z}$ is given by
\begin{eqnarray}
\label{eqnxddklinf}
    x_{\mbox{\scriptsize{d,dd}}}[k,l] & = & \frac{1}{\sqrt{MN}} \, x[k \, \mbox{\footnotesize{mod}} \, M \, , \, l \, \mbox{\footnotesize{mod}} \, N] \, e^{j2 \pi \frac{\left\lfloor \frac{k}{M} \right\rfloor l }{N} }.
\end{eqnarray}The terms $x[k,l]$, $k=0,\cdots, M-1$, $l=0,1,\cdots, N-1$ are the $MN$ information symbols, each with unit average energy. The average energy of the data signal, i.e., $\sum\limits_{k=0}^{M-1} \sum\limits_{l=0}^{N-1}  {\mathbb E} \left[ \left\vert x_{\mbox{\scriptsize{d,dd}}}[k,l] \right\vert^2 \right] $ is
\begin{eqnarray}
\label{norm_infsymb}
\frac{E_d}{MN} \sum\limits_{k=0}^{M-1} \sum\limits_{l=0}^{N-1}  {\mathbb E} \left[ \left\vert x[k,l] \right\vert^2 \right] 
  &  =  & E_d.
\end{eqnarray}It follows from (\ref{eqnspilotenergy}) that the
average energy of the sensing signal is
\begin{eqnarray}
    E_p \left( \sum\limits_{k=0}^{M-1} \sum\limits_{l=0}^{N-1} \vert x_{\mbox{\scriptsize{s,dd}}}[k,l] \vert^2 \right) = E_p.
\end{eqnarray}We define PDR $= E_p/E_d$ to be the ratio of pilot power to data power. It follows from 
(\ref{eqn_io_relation}) that the received discrete DD domain signal is then given by
\begin{eqnarray}
\label{rxjointdatasense}
y_{\mbox{\scriptsize{dd}}}[k,l] & = &  h_{\mbox{\scriptsize{eff}}}[k,l] \, *_{\sigma} \, x_{\mbox{\scriptsize{dd}}}[k,l]  \, + \, n_{\mbox{\scriptsize{dd}}}[k,l] \nonumber \\
&  & \hspace{-16mm} = \sqrt{E_d} \underbrace{\left( h_{\mbox{\scriptsize{eff}}}[k,l] \, *_{\sigma} \, x_{\mbox{\scriptsize{d,dd}}}[k,l] \right)}_{\mbox{\scriptsize{Received data signal}}} \nonumber \\
&  & \hspace{-12mm} \, + \, \sqrt{E_p} \, \underbrace{\left( h_{\mbox{\scriptsize{eff}}}[k,l] \, *_{\sigma} \, x_{\mbox{\scriptsize{s,dd}}}[k,l] \right)}_{\mbox{\scriptsize{Received sensing signal}}} \, + \, n_{\mbox{\scriptsize{dd}}}[k,l].
\end{eqnarray}where $h_{\mbox{\scriptsize{eff}}}[k,l]$ is the effective discrete DD domain channel filter.

\textbf{Channel sensing:}
Here we suppose that the (weaker) crystallization condition (\ref{gencondition3}) holds, and we describe how to estimate the
effective channel filter $h_{\mbox{\scriptsize{eff}}}[k,l]$ from the received signal $y_{\mbox{\scriptsize{dd}}}[k,l]$. Recall from (\ref{eqnhestambigauto}), that in the absence of data, the ML estimate 
${\widehat h}_{\mbox{\scriptsize{eff}}}[k,l]$ is given by the cross-ambiguity function between the received signal and the transmitted spread pulsone.

\begin{theorem}
\label{coro1}
In the presence of data, the cross-ambiguity function
${ A}_{y,x_s}[k,l]$ between $y_{\mbox{\scriptsize{dd}}}[k,l]$ and $x_{\mbox{\scriptsize{s,dd}}}[k,l]$ is given by

{\vspace{-4mm}
\small
\begin{eqnarray}
\label{eqn352_ays}
    { A}_{y,x_s}[k,l]  & \hspace{-3mm} = &  \hspace{-4mm}  \sqrt{E_p} \, h_{\mbox{\scriptsize{eff}}}[k,l] \, *_{\sigma} \, { A}_{x_s,x_s}[k,l] \nonumber \\
    & &  \hspace{-5mm} +  \sqrt{E_d} \,  h_{\mbox{\scriptsize{eff}}}[k,l] \, *_{\sigma} \, {A}_{x_d,x_s}[k,l] \nonumber \\
    & & \hspace{-6mm}  + \underbrace{\sum\limits_{k'=0}^{M-1} \sum\limits_{l'=0}^{N-1}  n_{\mbox{\scriptsize{dd}}}[k',l'] \, x_{\mbox{\scriptsize{s,dd}}}^*[k' - k, l' - l] \, e^{-j 2 \pi \frac{l (k' - k)}{MN}}}_{\Define { A}_{n,x_s}[k,l]}, \nonumber \\
\end{eqnarray}\normalsize}where ${ A}_{x_s,x_s}[k,l]$ is the self-ambiguity function of the spread pulsone and

{\vspace{-4mm}
\small
\begin{eqnarray}
\label{expradskl}
    { A}_{x_d,x_s}[k,l]  & \hspace{-3mm} = & \hspace{-4mm} \sum\limits_{k'=0}^{M-1} \sum\limits_{l'=0}^{N-1} x_{\mbox{\scriptsize{d,dd}}}[k',l'] \, x_{\mbox{\scriptsize{s,dd}}}^*[k' - k, l' - l] \, e^{-j 2 \pi \frac{l (k' - k)}{MN}} \nonumber \\
\end{eqnarray}\normalsize}is the cross-ambiguity between the data signal and the sensing signal.
\end{theorem}
\begin{IEEEproof}
See Appendix \ref{app_prf_corrambigthm2}.
\end{IEEEproof}
When the crystallization condition (\ref{gencondition3}) holds, we can obtain $h_{\mbox{\scriptsize{eff}}}[k,l]$ by evaluating the first term on the RHS of (\ref{eqn352_ays}) inside ${\mathcal S}_{(0,0)}$ (see (\ref{eqndefskili})). For $(k,l) \in {\mathcal S}_{(0,0)}$, the cross-ambiguity function ${ A}_{y,x_s}[k,l]$ now reduces to 

{\vspace{-4mm}
\small
\begin{eqnarray}
\label{eqnaysfinal}
{ A}_{y,x_s}[k,l]  & \hspace{-3mm} = & \hspace{-3mm} \sqrt{E_p} \, h_{\mbox{\scriptsize{eff}}}[k,l] \,  +  \, \sqrt{E_d} \,  h_{\mbox{\scriptsize{eff}}}[k,l] \, *_{\sigma} \, {A}_{x_d,x_s}[k,l]  \nonumber \\
& & \, + \, A_{n,x_s}[k,l].
\end{eqnarray}\normalsize}Our estimate of $h_{\mbox{\scriptsize{eff}}}[k,l]$ is
\begin{eqnarray}
\label{chesteq1jointdatasens}
    {\widehat h}_{\mbox{\scriptsize{eff}}}[k,l] & = & \frac{{ A}_{y,x_s}[k,l]}{\sqrt{E_p}}, \,\, \mbox{\small{for}} \,\, (k,l) \in {\mathcal S}_{(0,0)}.
\end{eqnarray}
Theorem \ref{thm_927482} provides an estimate for the mean squared error contributed by the second term in (\ref{eqnaysfinal})
which measures interference to sensing from data. This error term is given by
\begin{eqnarray}
\sqrt{E_d} \, h_{\mbox{\scriptsize{eff}}}[k,l] *_{\sigma} A_{x_d,x_s}[k,l] &  & \nonumber \\
& & \hspace{-44mm} = \sqrt{E_d} \hspace{-2mm} \sum\limits_{(k',l') \in {\mathcal S}_{(0,0)}}  \hspace{-5mm} h_{\mbox{\scriptsize{eff}}}[k',l'] \, A_{x_d,x_s}[k - k', l - l'] \, e^{j 2 \pi  \frac{l'(k - k')}{MN}}. \nonumber \\
\end{eqnarray}

\begin{theorem}
\label{thm_927482}
    When the (weaker) crystallization condition (\ref{gencondition3}) holds, interference from data to sensing contributes
    \begin{eqnarray}
    \label{dataintrfjointdatasense}
     {\mathbb E} \left[ \left\vert  \sqrt{E_d}  \,  h_{\mbox{\scriptsize{eff}}}[k,l] *_{\sigma} A_{x_d,x_s}[k,l] \right\vert^2 \right] &  & \nonumber \\
     & & \hspace{-30mm} = \frac{E_d}{MN} \, \sum\limits_{(k,l) \in {\mathcal S}_{(0,0)}} \left\vert h_{\mbox{\scriptsize{eff}}}[k,l] \right\vert^2
    \end{eqnarray}to the mean squared error of the estimate ${\widehat h}_{\mbox{\scriptsize{eff}}}[k,l]$ given in (\ref{chesteq1jointdatasens}).
\end{theorem}
\begin{IEEEproof}
See Appendix \ref{prf_app_thm_927482}.
\end{IEEEproof}
Fig.~\ref{fig0287} shows the heat map for the quantities
$\vert A_{x_d,x_s}[k,l] \vert$ which measure interference to sensing from data. We observe that the values are roughly of the order $10 \log_{10}\left( 1/\sqrt{MN}\right) \approx -15$ dB, in other words $\vert A_{x_d,x_s}[k,l] \vert \approx 1/\sqrt{MN}$.  

The factor of $1/(MN)$ is due to the fact that $\vert A_{x_d,x_s}[k,l] \vert$ is roughly of order $1/\sqrt{MN}$.
In Fig.~\ref{fig0287} we have plotted the heat map for $\vert A_{x_d,x_s}[k,l] \vert$ with $M = 31, N=37, q=3$ and for a random realization of the information symbols. It is observed that the values are roughly of the order $10 \log_{10}(1/\sqrt{MN}) \approx -15$ dB.
\begin{figure}[h]
\centering
\includegraphics[width=9.5cm, height=5.2cm]{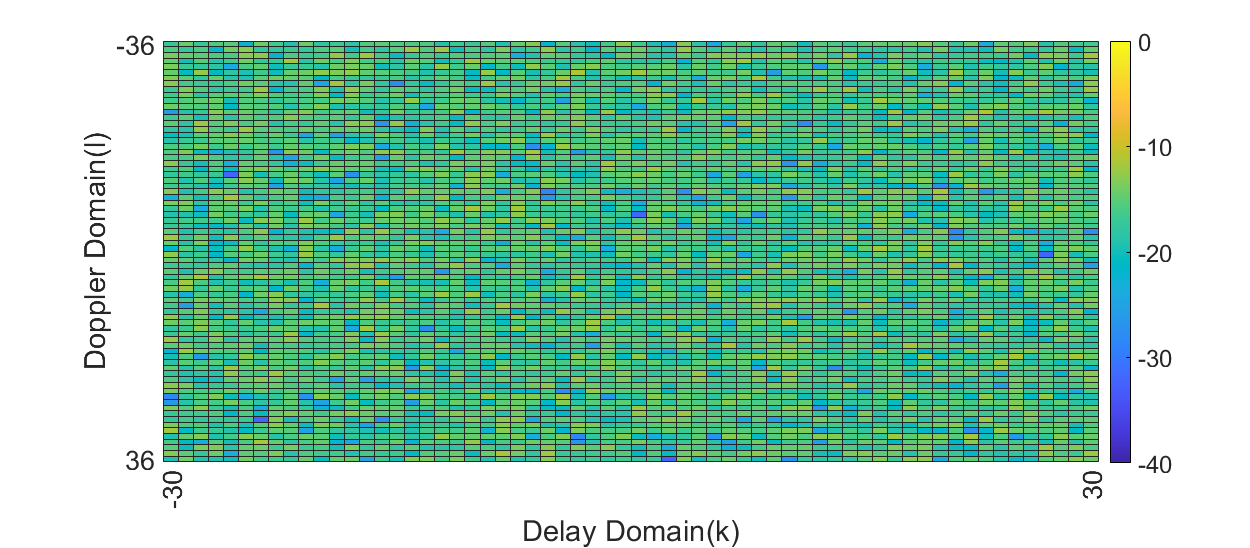}
\caption{Heat map of $\left\vert A_{x_d,x_s}[k,l] \right\vert$. PDR $= E_p/E_d = 0$ dB. $M = 31, N = 37$,
spread pulsone defined by chirp filter with $q=3$.
}
\label{fig0287}
\end{figure}

The last term in (\ref{eqnaysfinal}) represents the
contribution of noise to channel estimation error. Since the received DD domain noise samples are i.i.d. ${\mathcal C} {\mathcal N}(0, N_0)$ distributed, the samples of
$A_{n,x_s}[k,l]$ are zero mean. Since $\sum\limits_{k=0}^{M-1} \sum\limits_{l=0}^{N-1} \left\vert x_{\mbox{\scriptsize{s,dd}}}[k,l] \right\vert^2 = 1$, it follows from (\ref{eqn352_ays}) that the
variance is given by

{\vspace{-4mm}
\small
\begin{eqnarray}
\label{eqn9728}
    {\mathbb E}\left[ \left\vert A_{n, x_s}[k,l] \right\vert^2 \right] &  =  & N_0.
\end{eqnarray}\normalsize}
It follows from (\ref{chesteq1jointdatasens}) that
the normalized mean squared estimation error (NMSE) is given by
\begin{eqnarray}
    \label{chesterroreqn}
    \frac{{\mathbb E}\left[\left\vert h_{\mbox{\scriptsize{eff}}}[k,l] \, - \, {\widehat h}_{\mbox{\scriptsize{eff}}}[k,l] \right\vert^2 \right]}{\sum\limits_{(k,l) \in {\mathcal S}_{(0,0)}} \hspace{-5mm} \left\vert  h_{\mbox{\scriptsize{eff}}}[k,l]  \right\vert^2 }  & = & \frac{1}{MN} \left( \frac{1 + \rho_d}{\rho_p} \right) 
\end{eqnarray}where we have used the expression for $A_{y,x_s}[k,l]$ in the R.H.S. of (\ref{eqnaysfinal}), and the mean squared values of the data interference and noise terms in (\ref{eqnaysfinal}) from (\ref{dataintrfjointdatasense}) and (\ref{eqn9728}) respectively. In (\ref{chesterroreqn}), $\rho_d$ refers to data SNR (see (\ref{eqnrhoddef})) and $\rho_p$ refers to the pilot SNR (see (\ref{rhopdef})).\footnote{\footnotesize{In practical implementations, while reading the estimate of $h_{\mbox{\scriptsize{eff}}}[k,l]$ from
$A_{y,x_s}[k,l]$, $(k,l) \in {\mathcal S}_{(0,0)}$, all taps of $A_{y,x_s}[k,l]$ in ${\mathcal S}_{(0,0)}$ may not be genuine taps of $h_{\mbox{\scriptsize{eff}}}[k,l]$ primarily due to the randomness of noise and data interference. Therefore, in practice, only those taps of $A_{y,x_s}[k,l]$, $(k,l) \in {\mathcal S}_{(0,0)}$, are considered whose magnitude exceeds a pre-determined threshold. This threshold could be $2$ or $3$ times the standard deviation of the estimation error $(h_{\mbox{\scriptsize{eff}}}[k,l]  - {\widehat h}_{\mbox{\scriptsize{eff}}}[k,l])$ (which can be calculated from (\ref{chesterroreqn})).}}   


When the crystallization condition is satisfied, the factor $1/(MN)$ in (\ref{chesterroreqn}) significantly reduces the impact of data and noise on the accuracy of channel estimation. The factor $1/(MN)$ is present because
interference between the spread pulsone used for sensing and the point pulsones used for data transmission is noise-like.

\begin{figure}[h]
\centering
\includegraphics[width=9.3cm, height=7.2cm]{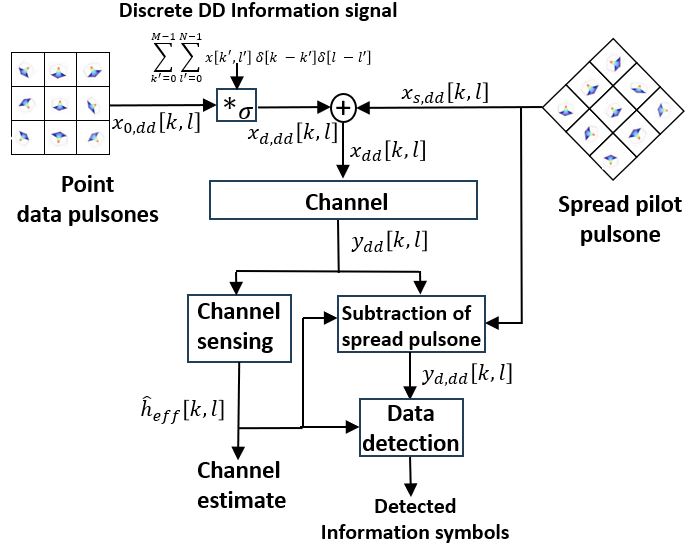}
\caption{Architecture of integrated sensing and communication with a spread pilot pulsone.
} 
\label{figjsac}
\end{figure}

\textbf{Cancellation of spread pilot followed by data detection}
Fig.~\ref{figjsac} provides a flow chart of the method
we propose for integrated sensing and communication (ISAC). We first estimate the received spread pilot pulsone as
\begin{eqnarray}
    \sqrt{E_p} \, {\widehat h}_{\mbox{\scriptsize{eff}}}[k,l] *_{\sigma} x_{\mbox{\scriptsize{s,dd}}}[k,l].
\end{eqnarray}using the estimate ${\widehat h}_{\mbox{\scriptsize{eff}}}[k,l]$ derived in (\ref{chesteq1jointdatasens}).
It follows from (\ref{chesterroreqn}) that this estimate
is accurate when the (weaker) crystallization condition
is satisfied with respect to the lattice $\Lambda_q$.
The estimate becomes more accurate for larger $M,N$, that is
\begin{eqnarray}
{\widehat h}_{\mbox{\scriptsize{eff}}}[k,l] *_{\sigma} x_{\mbox{\scriptsize{s,dd}}}[k,l] & \approx & h_{\mbox{\scriptsize{eff}}}[k,l] *_{\sigma} x_{\mbox{\scriptsize{s,dd}}}[k,l].
\end{eqnarray}We then cancel this estimate from the received signal to obtain
\begin{eqnarray}
\label{pilotfreeeqn}
    y_{\mbox{\scriptsize{d,dd}}}[k,l] & = & y_{\mbox{\scriptsize{dd}}}[k,l] -  \sqrt{E_p} \, {\widehat h}_{\mbox{\scriptsize{eff}}}[k,l] *_{\sigma} x_{\mbox{\scriptsize{s,dd}}}[k,l].
\end{eqnarray}When the (weaker) crystallization
condition is satisfied with respect to the period lattice $\Lambda_p$, we can reliably recover data from the almost sensing pulsone free received signal $y_{\mbox{\scriptsize{d,dd}}}[k,l]$ in (\ref{pilotfreeeqn}). See \cite{zakotfs2} for more details.

Our approach increases effective throughput by eliminating the need to \emph{divide} physical resources between sensing and communication (see Section \ref{simsec} for numerical results).



The flow chart of the proposed spread pilot pulsone based joint sensing and communication is depicted in Fig.~\ref{figjsac}.


\subsection{Numerical simulations}
\label{simsec}
The main finding of this paper is that Zak-OTFS allows for ISAC with spread sensing pulsone where there is no sensing overhead and the transmit signal has low PAPR. In this section we support this finding with numerical simulations of the BER performance of ISAC using point pulsones and spread sensing pulsones.

\begin{figure}[h]
\hspace{-6mm}
\includegraphics[width=9.7cm, height=6.2cm]{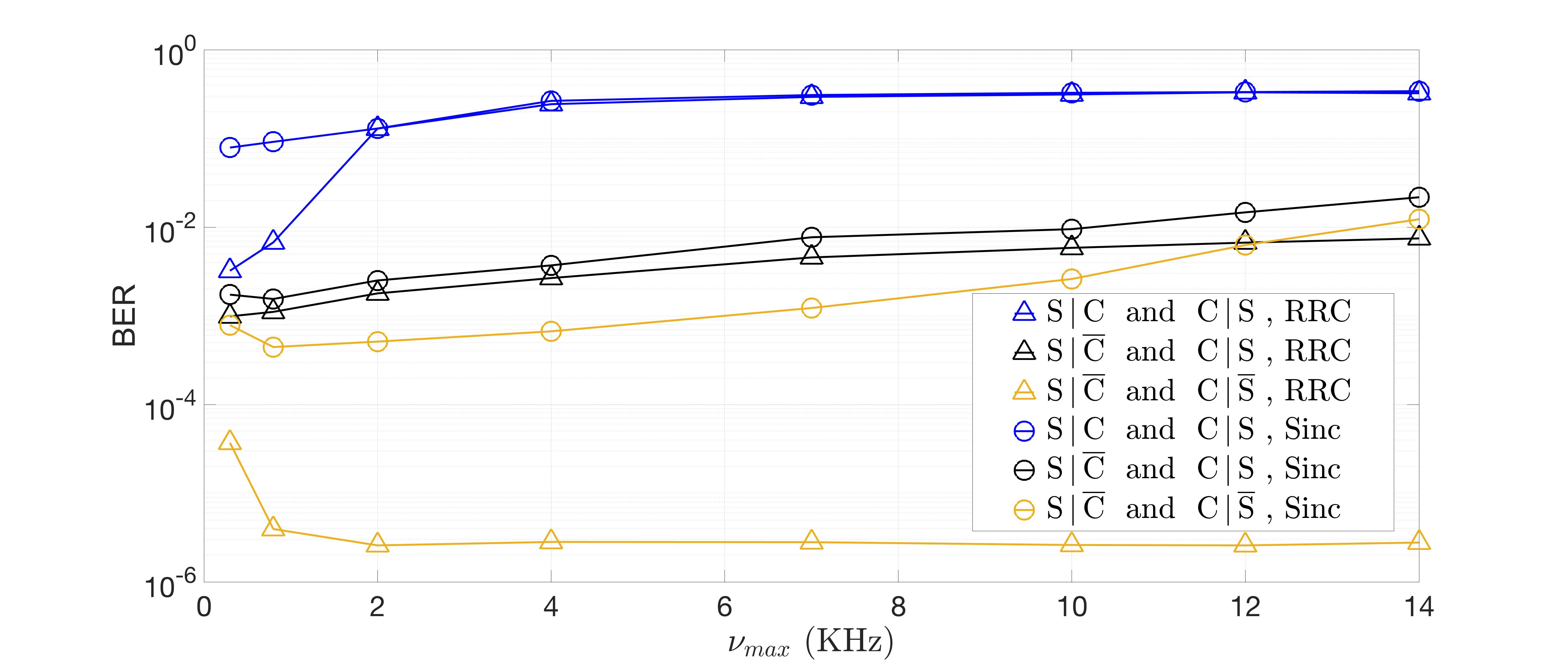}
\caption{Uncoded $4$-QAM BER vs $\nu_{max}$ for ISAC with point sensing pulsone. Veh-A channel. Data SNR and PDR are fixed at $\rho_d = 25$ dB and $\frac{\rho_p}{\rho_d} = 10$ dB respectively. $\nu_p = 30$ KHz, $M=31, N = 37$ and fixed guard region ${\mathcal G}$ ($7 \times 7$ rectangle around $(k_p, l_p) = ((M+1)/2, (N+1)/2)$). RRC pulse shaping filter ($\beta_{\tau} = \beta_{\nu} = 0.6$). Data detection with MMSE equalization.
} 
\label{fig71}
\end{figure}
Fig.~\ref{fig71} plots BER of uncoded $4$-QAM 
as a function of increasing Doppler spread $(2 \nu_{max})$
for the Veh-A channel described in Section \ref{zakotfspointpilot} (see Table-\ref{tab1_paper2} and (\ref{paper2_eqn13})).
Point pulsones are used for both channel sensing and data transmission.

Here, and throughout this Section, we consider channel sensing in the presence of interference from communication data (\textbf{S$\vert $C}) and in the absence of interference from data (\textbf{S$\vert$$\overline{\mbox{C}}$}). Similarly, we consider data equalization in the presence of interference from sensing
(\textbf{C$\vert$S}) and in the absence of interference
from sensing (\textbf{C$\vert$$\overline{\mbox{S}}$}).

\textbf{S$\vert $$\overline{\mbox{C}}$} and \textbf{C$\vert$$\overline{\mbox{S}}$}: This is the baseline (yellow curves) where we dedicate separate Zak-OTFS subframes to channel sensing and to data transmission (see \cite{zakotfs2}). With a sinc filter (yellow circles), Doppler domain aliasing increases with $\nu_{max}$, and BER increases as the accuracy of channel sensing degrades. We can reduce Doppler domain aliasing by substituting an RRC filter (yellow triangles) for a sinc filter.

\textbf{S$\vert $${\mbox{C}}$} and \textbf{C$\vert$${\mbox{S}}$}:
When we integrate channel sensing and data transmission
in the same Zak-OTFS subframe (blue curves), the point pulsone used for channel sensing interferes with the point pulsones used for data transmission. This interference severely compromises BER performance when the Doppler spread exceeds $2$ KHz ($\nu_{max} > 1$ KHz) because the channel response to a point pulsone extends beyond the guard band, regardless of the choice of filter. When the Doppler spread is less than $1$ kHz, the choice of filter makes a significant difference. When we introduce a guard band, we
\emph{divide} DD domain resources between sensing and communication. By increasing the size of the guard band (\emph{sacrificing transmission rate}) we can extend the range of reliable operation to higher Doppler spreads.

\textbf{S$\vert $${\mbox{C}}$} and \textbf{C$\vert$$\overline{\mbox{S}}$}:
We integrate sensing and data transmission in the same Zak-OTFS subframe so the point pulsones used to transmit data interfere with the point pulsone used to sense the channel (S$\vert $${\mbox{C}}$). We ask what BER performance
would be, if when recovering the data, there was no interference from the sensing pulsone (C$\vert$$\overline{\mbox{S}}$). We estimate the channel within the guard band of the ISAC subframe (S$\vert $${\mbox{C}}$), then generate a data-only subframe carrying the same data as the ISAC subframe, then equalize the data using the channel estimate (C$\vert$$\overline{\mbox{S}}$). We observed little difference in BER performance (compared to the blue curves) and have not shown these curves in Fig.~\ref{fig71}. 

\textbf{S$\vert $$\overline{\mbox{C}}$} and \textbf{C$\vert$${\mbox{S}}$}:
 We ask what BER performance would be if channel estimation were not subject to interference from data.
Fig.~\ref{fig71} illustrates BER performance when we use a separate Zak-OTFS subframe for channel sensing (black curves). The improvement in BER performance is significant when compared to the blue curves. 


\begin{figure}[h]
\centering
\includegraphics[width=9.5cm, height=6.2cm]{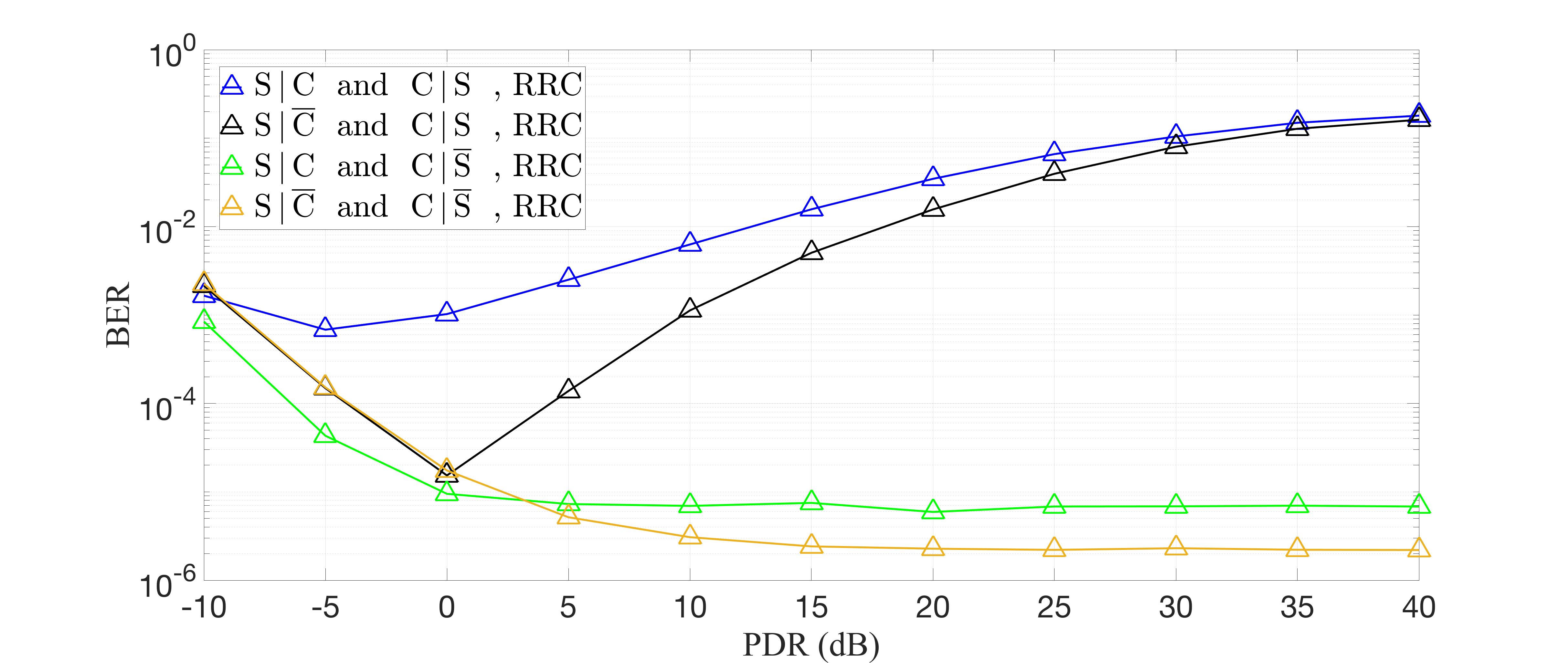}
\caption{
Uncoded $4$-QAM bit error rate (BER) as a function of the ratio of
the point sensing pulsone power to the total power of data pulsones $ \frac{\rho_p}{\rho_d}$ (PDR) for ISAC with point sensing pulsones. Veh-A channel. RRC pulse shaping filter ($\beta_{\tau} = \beta_{\nu} = 0.6$). Data SNR fixed at $\rho_d = 25$ dB, Doppler period $\nu_p = 30$ KHz, $M=31, N = 37$ and the guard region is a $7 \times 7$ rectangle. Note the characteristic ``U" shaped curves for integrated sensing and communication.}
\label{figpointbervspdr}
\end{figure}

Fig.~\ref{figpointbervspdr} plots BER of uncoded $4$-QAM as a function of increasing pilot to data power ratio (PDR) for the Veh-A channel described in Section \ref{zakotfspointpilot}. We fix the data SNR $\rho_d = 25$ dB, and we fix $\nu_{max} = 815$ Hz. By limiting the Doppler spread to $1.63$ KHz, we limit the fraction of received pulsone energy that falls outside the guard band. Point pulsones are used for both channel sensing and data transmission, and we only consider RRC pulse-shaping filters. 

\textbf{S$\vert$$\overline{\mbox{C}}$} and \textbf{C$\vert$$\overline{\mbox{S}}$}: This is the baseline (yellow curve) where we dedicate separate Zak-OTFS subframes to channel estimation and data transmission. The accuracy of channel sensing increases with PDR, but DD domain aliasing in the data subframe limits BER, resulting in an error floor.

\textbf{S$\vert $${\mbox{C}}$} and \textbf{C$\vert$${\mbox{S}}$}: The point pulsone used for channel sensing interferes with the point pulsones used for data transmission. When PDR is small, the interference is small, and BER improves as the channel estimate becomes more accurate. When the power of the pilot pulsone exceeds the total power of all data pulsones (PDR $ > 0$ dB), interference becomes more significant than noise, and BER degrades with increasing PDR. This explains the characteristic ``U" shape of the blue curve.

\textbf{S$\vert $${\mbox{C}}$} and \textbf{C$\vert$$\overline{\mbox{S}}$}:
There is no interference from the pilot pulsone when recovering the data, and therefore at high PDR, BER is almost independent of PDR. This explains the high PDR flatness of the green curve.
For small PDR, increase in PDR improves the BER due to improvement in channel estimation accuracy.

\textbf{S$\vert $$\overline{\mbox{C}}$} and \textbf{C$\vert$${\mbox{S}}$}: When there is a dedicated sensing subframe, the accuracy of the channel estimate improves with increasing PDR. When PDR $> 0$, interference from the pilot pulsone becomes more significant than noise, and BER degrades with increasing PDR. This explains why the black curve has the same characteristic ``U" shape as the blue curve. We observed only a small improvement in BER when we replaced the estimated channel by perfect CSI and we have not shown this curve in Fig.~\ref{figpointbervspdr}.

\begin{figure}[h]
\centering
\includegraphics[width=9.5cm, height=6.2cm]{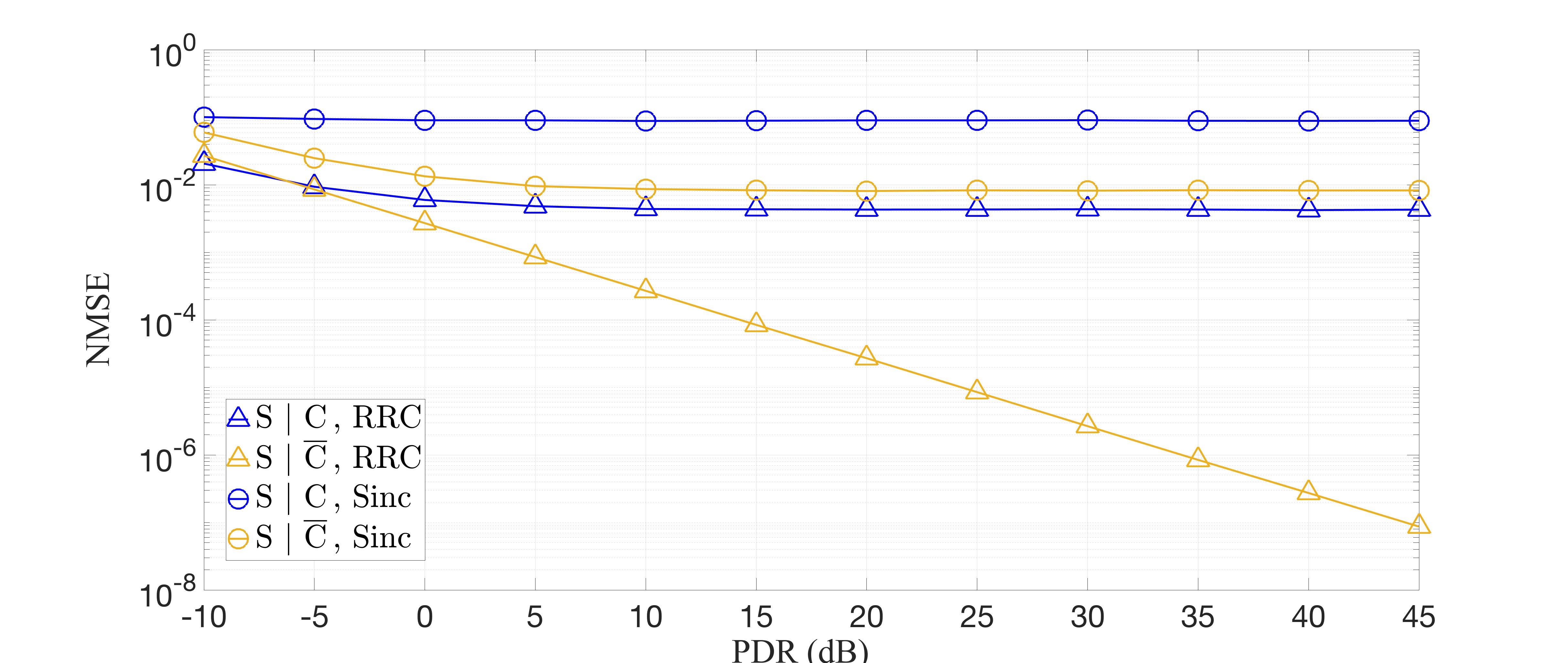}
\caption{NMSE vs. PDR for ISAC with point sensing pulsone. NMSE of estimation of the taps of the effective channel filter $h_{\mbox{\scriptsize{eff}}}[k,l]$ as a function of increasing PDR for a fixed data SNR of $\rho_d = 25$ dB. RRC filter ($\beta_{\tau} = \beta_{\nu} = 0.6$). $\nu_p = 30$ KHz.
} 
\label{fig2p1}
\vspace{-1em}
\end{figure}

The normalized mean squared error (NMSE) of the channel estimate is given by
\begin{eqnarray}
    \frac{\sum\limits_{(k,l) \in {\mathcal S}}\left\vert { h}_{\mbox{\scriptsize{eff}}}[k,l]  -  {\widehat h}_{\mbox{\scriptsize{eff}}}[k,l] \right\vert^2} {\sum\limits_{(k,l) \in {\mathcal S}} \left\vert { h}_{\mbox{\scriptsize{eff}}}[k,l] \right\vert^2 }.
\end{eqnarray}Fig.~\ref{fig2p1} plots NMSE as a function of PDR for the Veh-A channel considered
in Fig.~\ref{fig71}. We consider sinc and RRC pulse shaping filters to understand the significance of the fraction of received sensing pulsone energy that
falls outside the guard band.

\textbf{S$\vert $$\overline{\mbox{C}}$}:
When the received pilot pulsone is confined within the guard band, NMSE decreases linearly with increasing PDR (yellow triangles). Otherwise NMSE
decreases to a floor that depends on the fraction of the energy of the received pulsone that lies outside the guard band (yellow circles).

\textbf{S$\vert $${\mbox{C}}$}:
The limiting behavior is the same, but the floors are different (blue triangles and blue circles).

\begin{figure}[h]
\centering
\includegraphics[width=9.5cm, height=6.2cm]{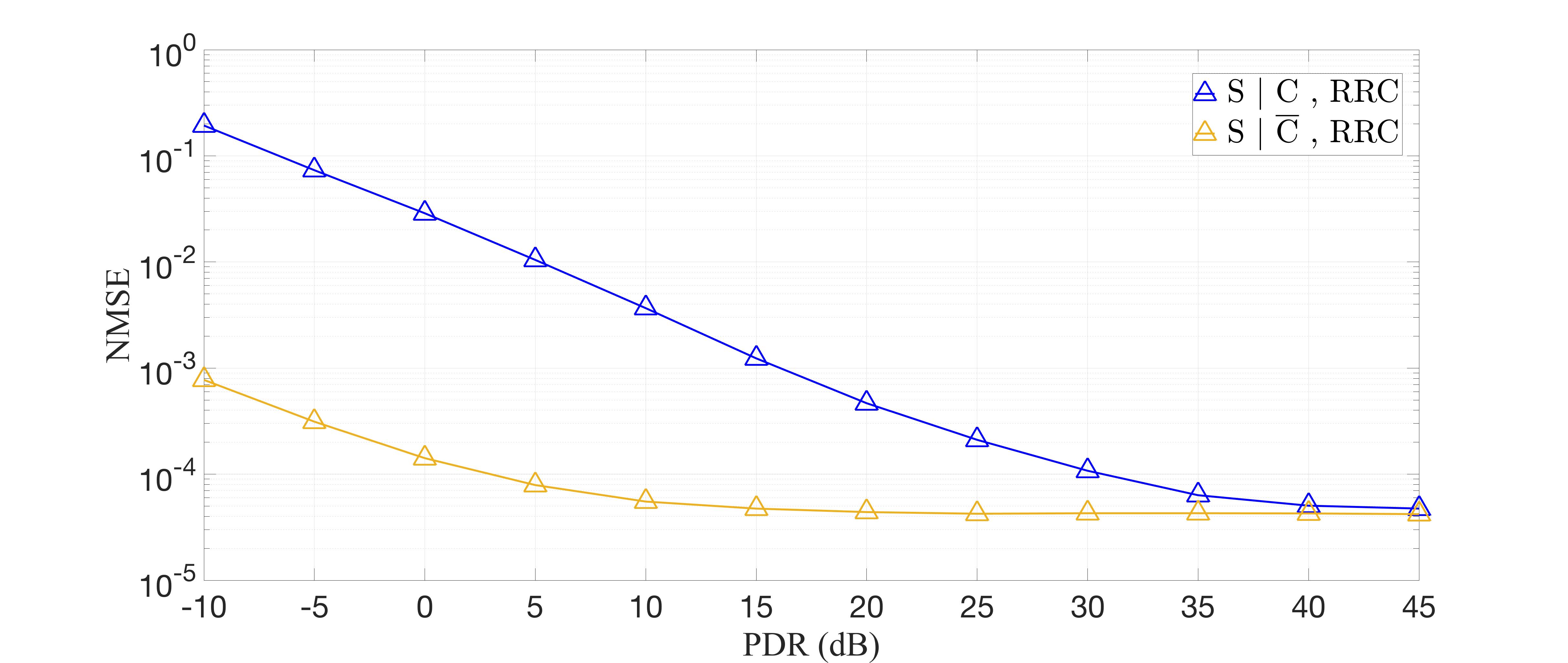}
\caption{NMSE vs. PDR for a spread sensing pulsone with $q=3$. Veh-A channel, RRC
pulse shaping filter ($\beta_{\tau}= \beta_{\nu} = 0.6$), data SNR $\rho_d = 25$ dB, $\nu_{max} = 815$ Hz, $\nu_p = 30$ KHz, $M = 31, N = 37$.
} 
\label{fig13}
\end{figure}Fig.~\ref{fig13} plots NMSE as a function of increasing PDR for a spread sensing pulsone ($q=3$) on the Veh-A channel considered in Fig.~\ref{fig71}. The Doppler spread $2 \nu_{max} = 1.63$ KHz is significantly less than the Doppler period $\nu_p = 30$ KHz, but even with RRC pulse shaping filters, there is residual DD domain aliasing. 

\textbf{S$\vert $$\overline{\mbox{C}}$}: 
This is the baseline (yellow curve) where we dedicate separate Zak-OTFS subframes to channel estimation and data transmission. At high PDR, the pilot power to noise power ratio (PNR) is high.
NMSE saturates at
high PNR due to DD domain aliasing.

\textbf{S$\vert $${\mbox{C}}$}: 
As PDR increases, the spread pilot becomes stronger than the combination of data interference and noise. DD domain aliasing again causes NMSE to saturate at high PDR.

\begin{figure}[h]
\centering
\includegraphics[width=9.5cm, height=6.2cm]{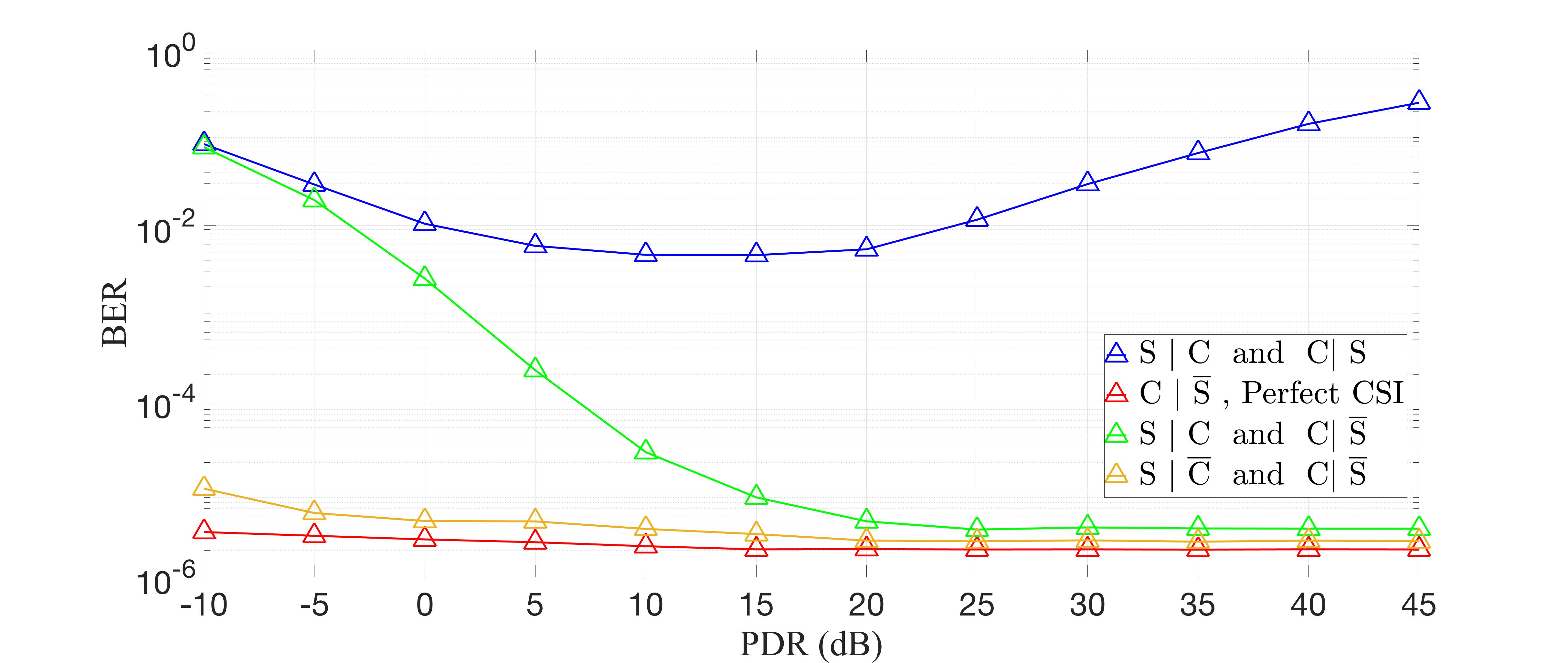}
\caption{BER vs. PDR for a spread sensing pulsone with $q=3$. Veh-A channel. RRC
pulse shaping filter ($\beta_{\tau}= \beta_{\nu} = 0.6$), data SNR $\rho_d = 25$ dB, $\nu_{max} = 815$ Hz, $\nu_p = 30$ KHz, $M = 31, N = 37$.}
\label{fig10}
\end{figure}
Fig.~\ref{fig10} plots BER of uncoded $4$-QAM as a function of increasing PDR. Note that with perfect CSI, the BER performance is independent of PDR (red curve).

\textbf{S$\vert $$\overline{\mbox{C}}$} and \textbf{C$\vert$$\overline{\mbox{S}}$}: This is the baseline (yellow curve) where we dedicate separate Zak-OTFS subframes to channel estimation and data transmission. BER performance is similar to NMSE performance in Fig.~\ref{fig13}, improving with increasing PDR (for low PDR less than $0$ dB) and saturating at high PDR. The green curve illustrates the case \textbf{S  $\vert $${\mbox{C}}$} \& \textbf{C$\vert$$\overline{\mbox{S}}$} where we estimate the channel in the presence of interference from data. Clearly, at low PDR there is hardly any interference from the pilot to data and therefore performance is same as that of the blue curve. However, with increasing PDR the BER performance improves due to a stronger pilot and no pilot interference to data, saturating at high PDR due to a floor in the NMSE performance resulting from DD domain aliasing (see blue curve in Fig.~\ref{fig13}). 

\textbf{S$\vert $${\mbox{C}}$} and \textbf{C$\vert$${\mbox{S}}$}: The spread pulsone used for channel sensing interferes with the point pulsones used for data transmission. When PDR $< 10$ dB, BER improves with increasing PDR. Beyond $10$ dB, interference from the residual pilot pulsone (after cancellation) degrades BER with increasing PDR. This explains the characteristic ``U" shape of the blue curve. In this case, there is an optimal PDR which minimizes BER. When we dedicate a separate Zak-OTFS subframe to sensing (\textbf{S$\vert $$\overline{\mbox{C}}$} \& \textbf{C$\vert$${\mbox{S}}$}) we observe
only a small improvement in BER at high PDR, and we have not shown this curve in Fig.~\ref{fig10}.

\begin{figure}[h]
\centering
\includegraphics[width=9.5cm, height=6.2cm]{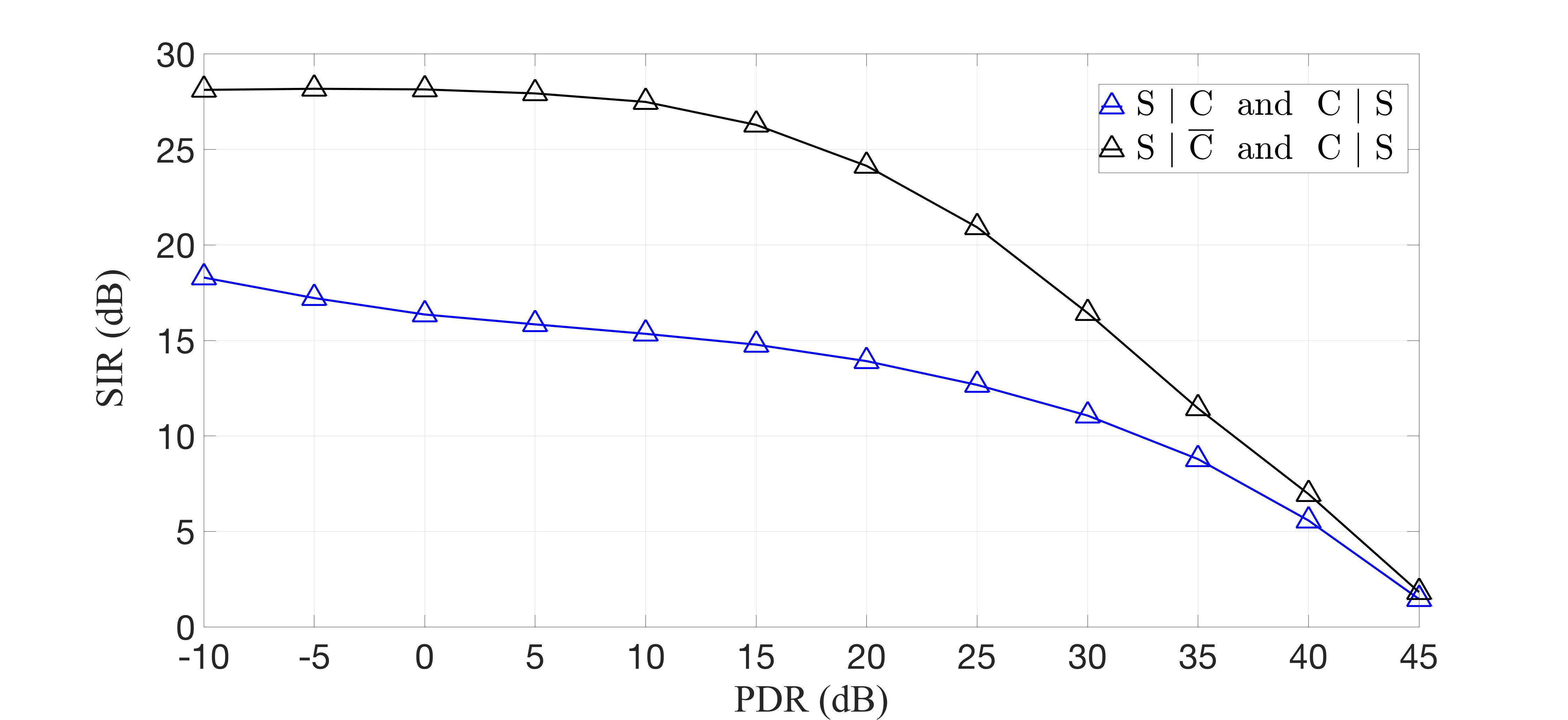}
\caption{Signal to Interference ratio (SIR) of data power to the power of the residual spread sensing pulsone (after cancellation) as a function of increasing PDR (for fixed data SNR $\rho_d = 25$ dB). RRC pulse shaping filter ($\beta_{\tau} = \beta_{\nu} = 0.6$). $\nu_p = 30$ KHz. Chirp filter with $q=3$. Veh-A channel is same as that considered in Figs.~\ref{fig13} and \ref{fig10}.
} 
\label{fig12}
\end{figure}

We use SIR to represent the ratio of the average power of the data pulsones to the average power of the residual sensing pulsones after cancellation. Fig.~\ref{fig12} plots SIR as a function of increasing PDR for the Veh-A channel considered in Figs.~\ref{fig13} and \ref{fig10}. 

\textbf{S$\vert $$\overline{\mbox{C}}$} and \textbf{C$\vert$${\mbox{S}}$}: Sensing is performed on a separate Zak-OTFS subframe (black curve). Estimation accuracy improves with increasing PDR, but the energy of the residual pilot pulsone increases proportionally. This is because, there are always some taps of the effective channel filter that are not estimated. We sense taps within a support set, and outside this set, taps are noisy because they are subject to DD domain aliasing. This explains why SIR decreases with increasing PDR.

\textbf{S  $\vert $${\mbox{C}}$} and \textbf{C$\vert$${\mbox{S}}$}: Here the channel estimate
is inferior to the estimate obtained from a dedicated sensing subframe. Hence the SIR ratio is smaller. The effect diminishes as PDR increases because the effect of data interference on sensing diminishes.

\begin{figure}[h]
\centering
\includegraphics[width=9.5cm, height=6.2cm]{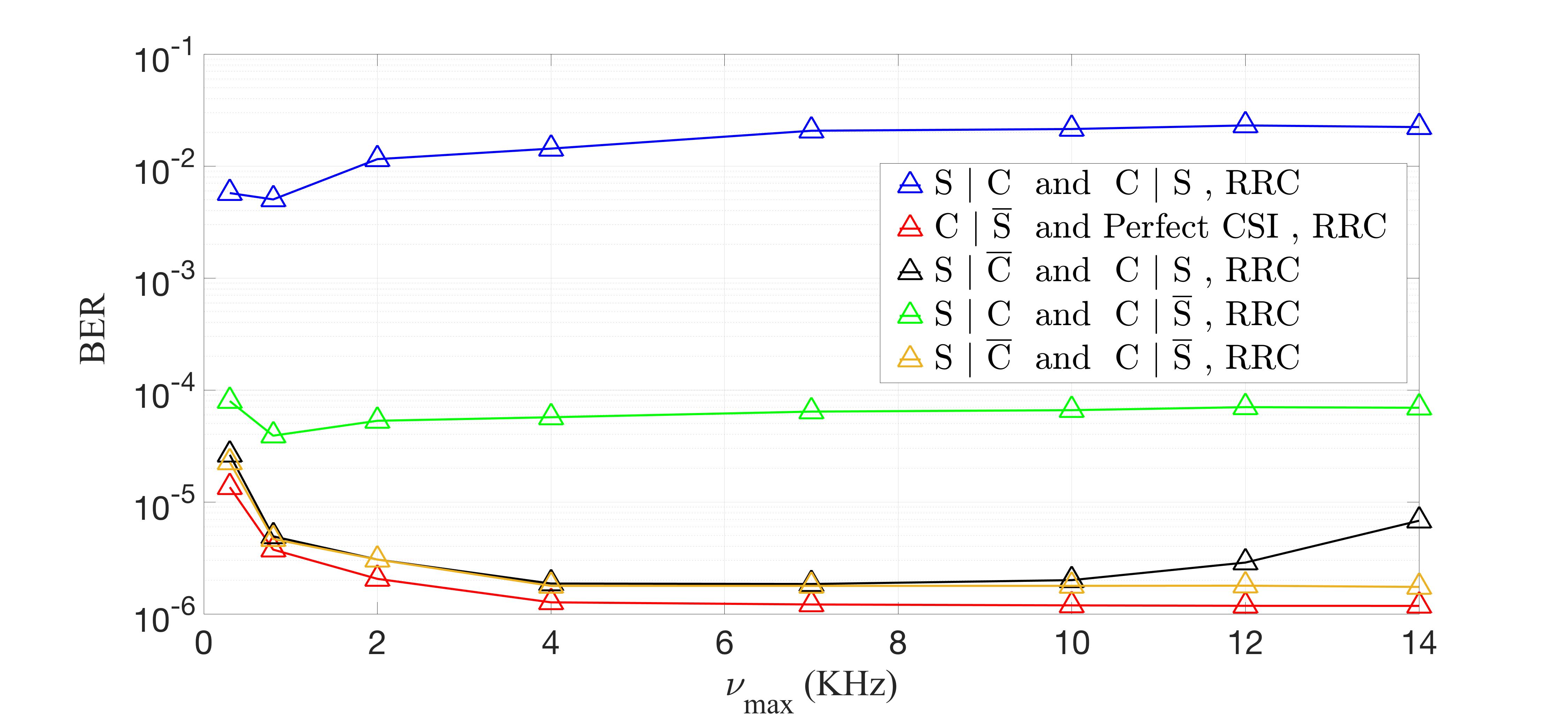}
\caption{BER vs $\nu_{max}$ for spread sensing pulsone ($q=3$). Veh-A channel model, RRC pulse shaping filter ($\beta_{\tau} = \beta_{\nu} = 0.6$), Doppler period $\nu_p = 30$ KHz, PDR $\frac{\rho_p}{\rho_d} = 10$ dB, data SNR $\rho_d = 25$ dB. 
} 
\label{fig11p1}
\end{figure}
Fig.~\ref{fig11p1} plots BER as a function of increasing $\nu_{max}$.

\textbf{S$\vert $$\overline{\mbox{C}}$} and \textbf{C$\vert$$\overline{\mbox{S}}$}: This is the baseline (yellow curve) where we dedicate separate subframes to sensing and data transmission.
There is no residual pilot pulsone to interfere with the data pulsones. Channel estimation is accurate even when $\nu_{max}$ is large. This is because the lattice $\Lambda_q$ supporting the self-ambiguity function of the spread sensing pulsone satisfies the crystallization condition (\ref{gencondition3}), even for large $\nu_{max}$. This explains why BER is excellent and almost independent of $\nu_{max}$. The green curve (\textbf{S$\vert $${\mbox{C}}$} \& \textbf{C$\vert$$\overline{\mbox{S}}$}) also exhibits a BER performance almost invariant of $\nu_{max}$ since channel sensing is accurate due to high PDR $= 10$ dB and the crystallization condition in (\ref{gencondition3}) is satisfied even for $\nu_{max} = 14$ KHz. Compared to the yellow baseline curve, the green curve has a higher BER since sensing in the presence of data pulsones is not as accurate as that in their absence. The red curve
(\textbf{C$\vert$$\overline{\mbox{S}}$}, perfect CSI) has almost the same performance (only slightly better) than the baseline yellow curve which again confirms that the impact of DD domain aliasing on the accuracy of channel sensing is small even for $\nu_{max} = 14$ KHz (since the crystallization condition is satisfied).

\textbf{S$\vert $${\mbox{C}}$} and \textbf{C$\vert$${\mbox{S}}$}: Here the spread pulsone used for channel sensing interferes with the point pulsones used for data transmission. Nevertheless, the variation in BER is small over a wide range of Doppler spreads. This is very different from BER performance for point sensing pulsones illustrated in Fig.~\ref{fig71} (BER $= 5 \times 10^{-3}$ at $\nu_{max} = 300$ Hz and BER $= 2 \times 10^{-2}$
at $\nu_{max} = 14$ KHz in Fig.~\ref{fig11p1} versus BER $= 3 \times 10^{-3}$ at $\nu_{max} = 300$ Hz and BER $= 0.3$ at $\nu_{max} = 14$ KHz in Fig.~\ref{fig71}). 
We conclude that spread sensing pulsones improve upon point sensing pulsones by extending the range of reliable operation to a wider range of Doppler spreads.

\begin{figure}[h]
\centering
\includegraphics[width=9.5cm, height=6.2cm]{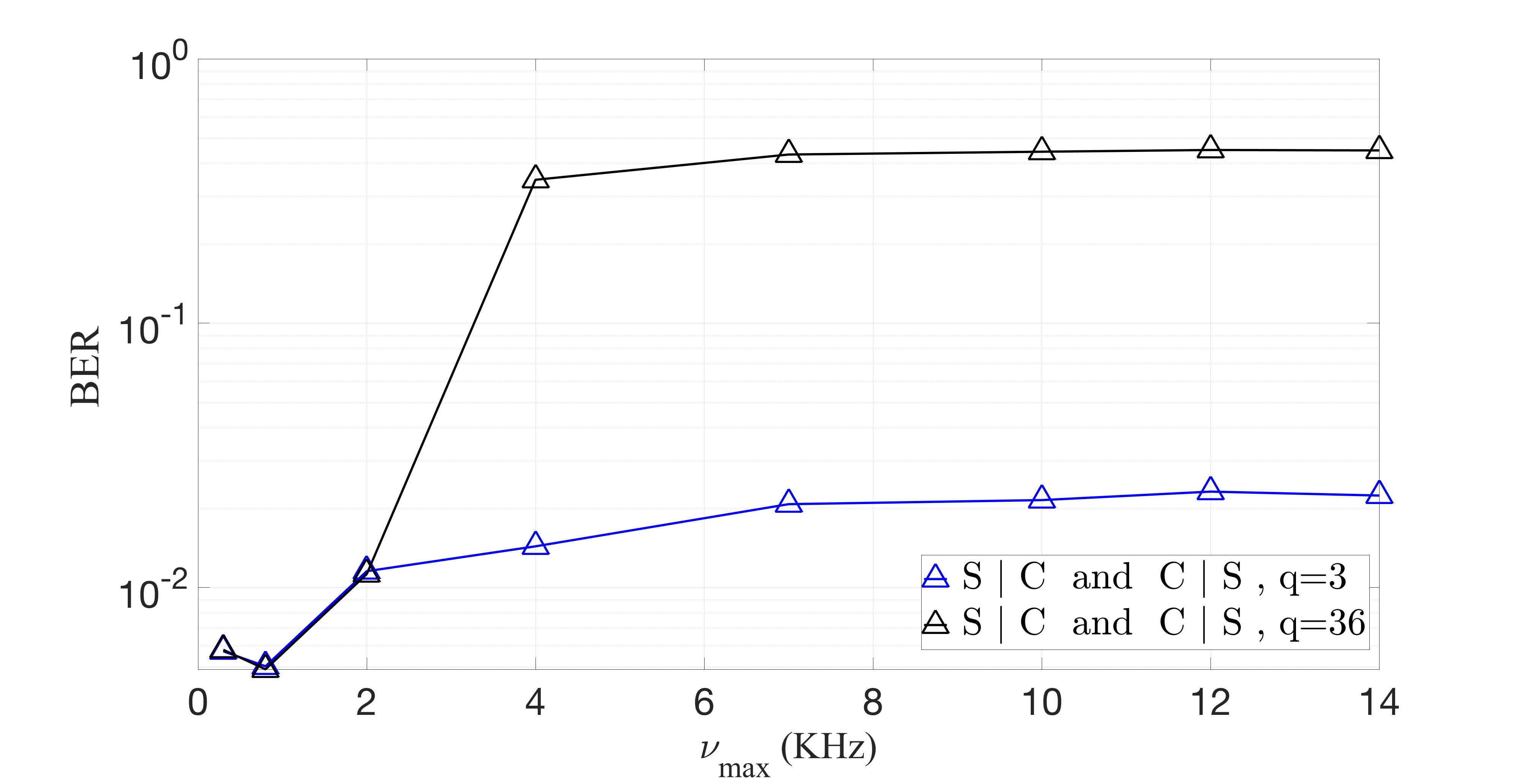}
\caption{BER vs $\nu_{max}$ for spread sensing pulsones defined by chirp filters with $q=3$ and $q=36$. Veh-A channel described in Fig.~\ref{fig11p1}. $\nu_p = 30$ KHz. Fixed PDR $\frac{\rho_p}{\rho_d} = 10$ dB, data SNR $\rho_d = 25$ dB. 
} 
\label{fig11p78}
\end{figure} 
Fig.~\ref{fig11p78} illustrates the importance of filter design by comparing BER performance of chirp filters with $q=3$ and $q=36$. Accurate sensing requires crystallization with respect to the lattice $\Lambda_q$, and reliable data detection requires crystallization with respect to the period lattice $\Lambda_p$. The crystallization condition is satisfied for the period lattice $\Lambda_p$ since $\nu_{max} < 15$ KHz and the delay spread for the Veh-A channel is $2.5 \, \mu s$, which is less than $\tau_p = 1/\nu_p = 33.33 \, \mu s$. When $q=3$, crystallization conditions w.r.t. both $\Lambda_p$ and $\Lambda_q$ are satisfied and BER performance is almost constant for a wide range of Doppler shifts (blue curve). When $q=36$, the crystallization condition with respect to $\Lambda_q$ is not satisfied for $\nu_{max} > 2$ KHz and BER degrades sharply (black curve).

\begin{figure}[h]
\centering
\includegraphics[width=9.5cm, height=6.0cm]{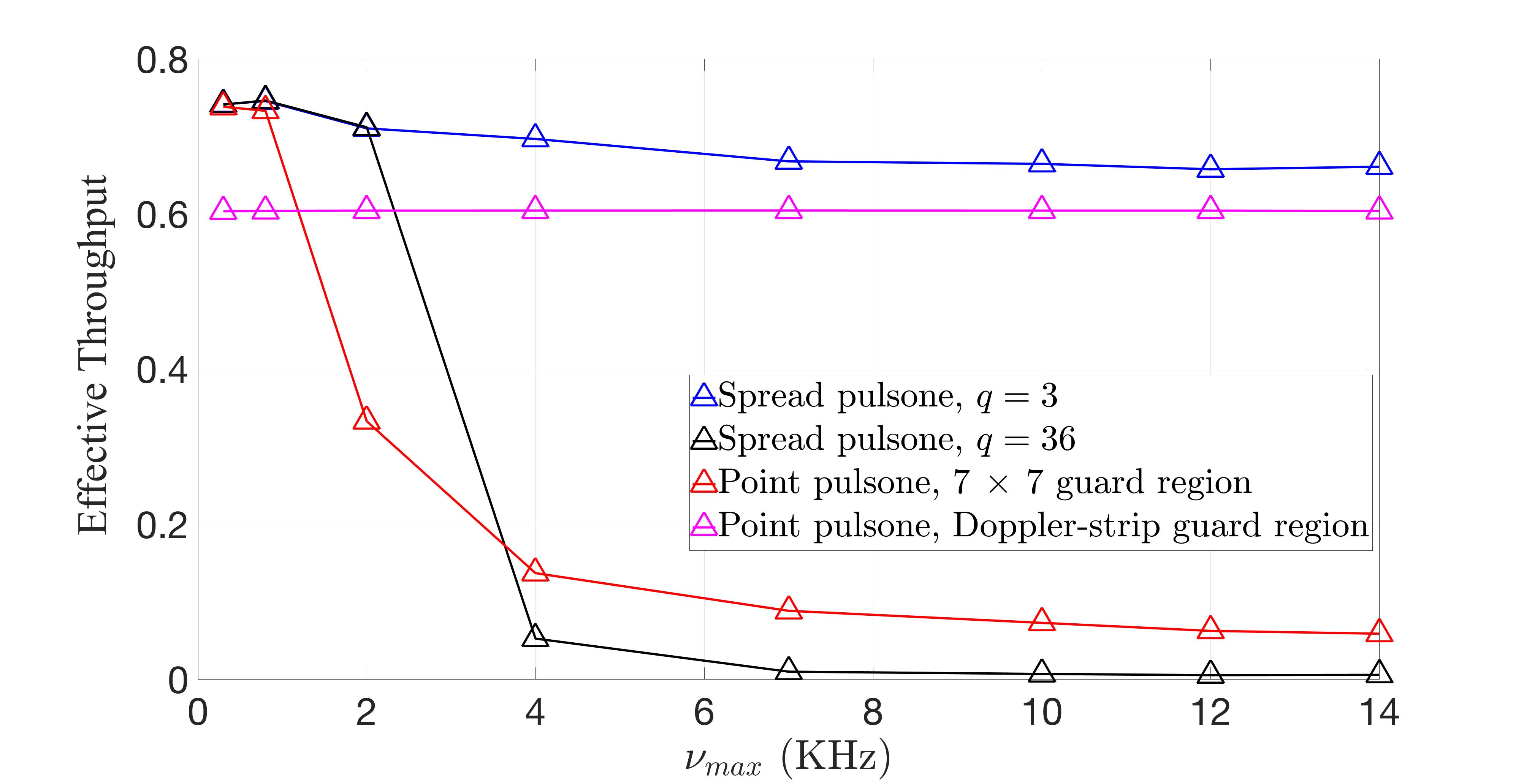}
\caption{Effective throughput (bits/sec/Hz) as a function of increasing $\nu_{max}$. Integrated sensing and communication (\textbf{S$\vert $ ${\mbox{C}}$} \& \textbf{C$\vert$${\mbox{S}}$}). $\nu_p = 30$ KHz. $M = 31, N=37$. RRC pulse shaping filter ($\beta_{\tau} = \beta_{\nu} = 0.6$). Veh-A channel as described in Fig.~\ref{fig11p1}.} 
\label{fig11}
\end{figure}
\textbf{Effective Throughput}: This is defined to be the ratio of the number of bits reliably communicated in each subframe to the available degrees of freedom. We only consider channel sensing and data transmission in the same Zak-OTFS subframe (\textbf{S$\vert$${\mbox{C}}$} \& \textbf{C$\vert$${\mbox{S}}$}). When we use a point pulsone for channel sensing we need to introduce a guard band, and this division of DD domain resources between sensing and communication decreases effective throughput. When we use a spread pulsone for channel sensing we avoid this overhead.

Assuming RRC pulse shaping filter with roll-off factors $\beta_{\tau}$ and $\beta_{\nu}$, the number of degrees of freedom is $BT (1 + \beta_{\tau})(1 + \beta_{\nu})$. Setting BER = $R$, the number of reliably communicated bits in each subframe is $2MN(1 - H(R))$ when using a spread sensing pulsone, compared with $2(MN - 49)(1 - H(R))$ when using a point sensing pulsone (assuming a $7 \times 7$ guard band) and $2(MN - 7N)(1 - H(R))$ (assuming the guard region to be a strip along the Doppler domain having width $7$ along the discrete delay axis). Here $H(\cdot)$ denotes the binary entropy function and we note that for small $R$ we have $H(R) \approx R/\ln2$.

Fig.~\ref{fig11} plots effective throughput as a function of increasing $\nu_{max}$. Recall from Fig.~\ref{fig11p78} that for a spread sensing pulsone defined by a chirp filter with $q=3$, BER performance is almost constant for a wide range of Doppler spreads. This is reflected in Fig.~\ref{fig11}. Recall from Fig.~\ref{fig11p78} that when $q=36$, the crystallization condition with respect to $\Lambda_q$ is not satisfied for $\nu_{max} > 2$ KHz, and the BER degrades sharply. This explains the difference in effective throughput for spread sensing pulsones defined by chirp filters with $q=3$ and $q=36$.

Recall from Fig.~\ref{fig71} that for $\nu_{max} > 1$ KHz, the channel response to a point sensing pulsone extends beyond the guard band regardless of the choice of pulse shaping filter. Interference between the point pulsone used for sensing and the point pulsones used for data transmission severely compromises BER performance. This explains the characteristics of the effective throughput curve for a point sensing pulsone (red curve).

Fig.~\ref{fig11} illustrates the gain in effective throughput that results from \emph{sharing} DD domain resources between sensing and communication.

\section{Conclusions}
We started from a parametric family of pulsone waveforms that can be matched to the delay and Doppler spreads of different propagation environments. We reviewed how the (point) pulsone signal in the time domain realizes a quasi-periodic localized function on the delay-Doppler (DD) domain. The characteristic structure of a pulsone is a train of pulses modulated by a tone, a signal with unattractive peak-to-average power ratio (PAPR). We reviewed system performance in the crystalline regime where the delay period of the pulsone is greater than the delay spread of the channel, and the Doppler period of the pulsone is greater than the Doppler spread of the channel. When channel sensing and data transmission take place in separate subframes, the point pulsone used to sense the channel does not interfere with the point pulsones used to transmit data. We have reviewed how the I/O relation of the sampled communication system can be read off from the response to the point pulsone used for channel sensing.
In this paper, we have introduced the notion of filtering in the discrete delay-Doppler domain. We have demonstrated that it is possible to construct spread waveforms with desirable characteristics by applying a chirp filter in the discrete DD domain to a point pulsone. One desirable characteristic is low PAPR, about $5$ dB for the exemplar spread pulsone, compared with about $15$ dB for the point pulsone. A second desirable characteristic is the ability to read off the I/O relation of the sampled communication system provided a second crystallization condition is satisfied. 
We have demonstrated how to integrate sensing and communication within a single Zak-OTFS subframe through the combination of a spread pulsone used for channel sensing and point pulsones for data transmission. The filter in the discrete DD domain enables coexistence by minimizing interference between sensing and data transmission. We have demonstrated that sharing DD domain resources in this way increases effective throughput compared with traditional approaches that use guard bands to divide DD domain resources between sensing and communication. 

\appendices
\section{Zak transform and quasi-periodicity}
\label{prelim_zak}
The definition of a DD domain pulse depends on a delay period $\tau_p$, and a Doppler period $\nu_p$, where the two periods are reciprocal, that is $\nu_p \, \tau_p = 1$. The Zak transform, denoted ${\mathcal Z}_t$,
provides a unitary equivalence between TD signals and a subclass of quasi-periodic DD domain signals. 
The Zak transform of a TD signal $x(t)$ is given by
\begin{eqnarray}
\label{prelim_1}
    x_{\mbox{\scriptsize{dd}}}(\tau, \nu) & = & {\mathcal Z_t} {\Big (} x(t) {\Big )} \nonumber \\
    & \Define & \sqrt{\tau_p} \, \sum\limits_{k \in {\mathbb Z}} x(\tau + k \tau_p) \, e^{-j 2 \pi k \nu \tau_p}.
\end{eqnarray}The quasi-periodicity condition is that, for any $n, m \in {\mathbb Z}$
\begin{eqnarray}
\label{prelim_2}
    x_{\mbox{\scriptsize{dd}}}(\tau + n \tau_p, \nu + m \nu_p)   &  =  &  e^{j 2 \pi n \nu \tau_p} \,  x_{\mbox{\scriptsize{dd}}}(\tau, \nu)
\end{eqnarray}and we refer the reader to equation $(5)$ in \cite{zakotfs1}, for a detailed derivation. We observe that $x_{\mbox{\scriptsize{dd}}}(\tau, \nu)$ is periodic along the Doppler axis with period $\nu_p$, and that it is quasi-periodic along the delay axis with period $\tau_p$.

The inverse Zak transform, denoted by ${\mathcal Z}_t^{-1}$, of a quasi-periodic DD domain signal $x_{\mbox{\scriptsize{dd}}}(\tau, \nu)$ is given by
\begin{eqnarray}
\label{prelim_3}
x(t) & = & {\mathcal Z}_t^{-1}{\Big (} x_{\mbox{\scriptsize{dd}}}(\tau, \nu) {\Big )}  \, = \,  \sqrt{\tau_p} \, \int\limits_{0}^{\nu_p} x_{\mbox{\scriptsize{dd}}}(t, \nu) \, d\nu.
\end{eqnarray}We refer the reader to equation $(7)$ in \cite{zakotfs1} for more details.

\section{TD pulsones: Carrier waveforms for Zak-OTFS}
\label{app_pulsones}
The following theorem describes the Zak-OTFS carrier waveforms (pulsones).
\begin{theorem}
\label{thm0}
The transmit TD signal $s_{\mbox{\scriptsize{td}}}(t)$ is given by
\begin{eqnarray}
    s_{\mbox{\scriptsize{td}}}(t) & = & {\mathcal Z_t}^{-1}{\Big (}  x_{\mbox{\scriptsize{dd}}}^{w_{tx}}(\tau , \nu ) {\Big )} \nonumber \\
    & = & \sum\limits_{k=0}^{M-1} \sum\limits_{l=0}^{N-1} x[k,l] \,  s_{\mbox{\scriptsize{td},k,l}}(t)
\end{eqnarray}where $s_{\mbox{\scriptsize{td}},k,l}(t)$ is the carrier waveform associated to the $(k,l)$-th information symbol $x[k,l]$, given by
\begin{eqnarray}
\label{eqn_18}
   s_{\mbox{\scriptsize{td}}, k,l}(t) &  \hspace{-3mm} =  &  \hspace{-3.5mm} \iint \hspace{-1mm} w_{tx}(\tau, \nu) \, x_{k,l}(t - \tau )  \, e^{j 2 \pi \nu (t - \tau)} \, d\tau \, d\nu, \nonumber \\
   x_{k,l}(t) & \Define & \sqrt{\tau_p} \,  \sum\limits_{n \in {\mathbb Z}} e^{j 2 \pi \frac{n l}{N} } \, \delta\left(t - n \tau_p - \frac{k \tau_p}{M}\right) \nonumber \\
   &  &  \hspace{-13mm} = e^{j 2 \pi \frac{l \nu_p}{N} \, \left(t - \frac{k \tau_p}{M} \right)} \, \sqrt{\tau_p}  \sum\limits_{n \in {\mathbb Z}} \delta\left(t - n \tau_p - \frac{k \tau_p}{M}\right).
\end{eqnarray}The FD representation (i.e., Fourier transform) of the TD carrier waveform $s_{\mbox{\scriptsize{td}}, k,l}(t)$ is given by
\begin{eqnarray}
\label{eqn_sfd1}
    s_{\mbox{\scriptsize{fd}},k,l}(f) & \hspace{-3mm} \Define & \hspace{-3mm} \int s_{\mbox{\scriptsize{td}},k,l}(t) \, e^{-j 2 \pi f t} \, dt \nonumber \\
    & & \hspace{-23mm} =  \hspace{-1mm} \iint \hspace{-1mm} w_{tx}(\tau, \nu)  \, e^{-j 2 \pi f \tau} \,   X_{k,l}(f - \nu ) \,  d\tau \, d\nu
    \end{eqnarray}where
    \begin{eqnarray}
\label{eqn_fdpulsone}
X_{k,l}(f) & \Define & \int x_{k,l}(t) \, e^{-j 2 \pi f t} \, dt \nonumber \\
&  & \hspace{-10mm} = \sqrt{\nu_p} \, \sum\limits_{m \in {\mathbb Z}} \delta\left( f - m \nu_p - \frac{l \nu_p}{N} \right) \, e^{-j 2 \pi f \frac{k \tau_p}{M}}
\end{eqnarray} is the Fourier transform of $x_{k,l}(t)$.
The DD domain representation of the $(k,l)$-th carrier waveform is
\begin{eqnarray}
    \label{eqn_DDrepresentation}
     s_{\mbox{\scriptsize{dd}},k,l}(\tau, \nu) & \Define & {\mathcal Z_t}  {\Big (} s_{\mbox{\scriptsize{td}},k,l}(t) {\Big )} \nonumber \\
     & = & w_{tx}(\tau, \nu) \, *_{\sigma} \, x_{\mbox{\scriptsize{dd}},k,l}(\tau, \nu ) 
\end{eqnarray}where
{\vspace{-2mm}
\small
\begin{eqnarray}
\label{eqn_xtaunukl}
     x_{\mbox{\scriptsize{dd}},k,l}(\tau, \nu ) & \hspace{-3.5mm} \Define & \hspace{-6mm} \sum\limits_{n,m \in {\mathbb Z}}  \hspace{-1.5mm}  e^{j 2 \pi \frac{n l}{N}} \, \delta\left(\tau - n\tau_p - \frac{k \tau_p}{M}\right) \delta\left(\nu - m \nu_p - \frac{l \nu_p}{N}\right), \nonumber \\
\end{eqnarray}\normalsize} is a quasi-periodic Dirac-delta impulse at $\left( \frac{k \tau_p}{M} \,,\, \frac{l \nu_p}{N}\right)$.
\end{theorem} (see Proof after the following). 
$\hfill\blacksquare$

\textbf{Remarks:} From (\ref{eqn_DDrepresentation}) and (\ref{eqn_xtaunukl}) it follows that the DD representation of the Zak-OTFS carrier waveform is given by

{\vspace{-3mm}
\small
\begin{eqnarray}
 s_{\mbox{\scriptsize{dd}},k,l}(\tau, \nu) & \hspace{-3mm} = & \hspace{-3mm} \sum\limits_{n,m \in {\mathbb Z}} {\Bigg [} w_{tx}\left( \tau - n \tau_p - \frac{k \tau_p}{M} \,,\, \nu - m \nu_p - \frac{l \nu_p}{N} \right) \nonumber \\
 & & \hspace{7mm} e^{j 2 \pi \frac{n l}{N}}  \, e^{j 2 \pi \left( \nu - m \nu_p - \frac{l \nu_p}{N} \right) \left( n \tau_p + \frac{k \tau_p}{M} \right)}{\Bigg ]}
\end{eqnarray}\normalsize}which is a quasi-periodic pulse in the DD domain. The representative pulse within the fundamental period ${\mathcal D}_0$, is located at $(k \tau_p/M, l \nu_p/N)$ (see Fig.~\ref{fig4_paper1}).

We now make fundamental observations about the TD and FD realization of the Zak-OTFS carrier waveform.\\

    

\textbf{Carrier waveforms having unlimited time and bandwidth:} For $w_{tx}(\tau, \nu) = \delta(\tau) \, \delta(\nu)$,
it follows from (\ref{eqn_18}) that
\begin{eqnarray}
    s_{\mbox{\scriptsize{td}},k,l}(t) & = & x_{k,l}(t).
\end{eqnarray}For $(k,l) = (0,0)$, the TD carrier waveform is therefore
\begin{eqnarray}
 s_{\mbox{\scriptsize{td}},0,0}(t) & = & \sqrt{\tau_p} \sum\limits_{n \in {\mathbb Z}} \delta\left(t - n \tau_p\right)
\end{eqnarray}which is simply an infinite train of Dirac-delta impulses where the $n$-th impulse is located at $t = n \tau_p$. The $(k,l)$-th carrier waveform $s_{\mbox{\scriptsize{td}},k,l}(t)  =  x_{k,l}(t)$ is nothing but $s_{\mbox{\scriptsize{td}},0,0}(t)$ with a Doppler shift of $l \nu_p/N$ and a delay shift of $k \tau_p/M$, since it can be checked that
\begin{eqnarray}
    x_{k,l}(t) & \hspace{-3mm} = & \hspace{-3mm} x_{0,0}(t - \underbrace{k \tau_p/M}_{\mbox{\tiny{delay shift}}} ) \, \underbrace{e^{j 2 \pi \frac{l \nu_p}{N} \, (t - k \tau_p/M)}}_{\mbox{\tiny{Doppler shift}}}.
\end{eqnarray}Note that the $(k,l)$-th carrier waveform is the TD realization of a quasi-periodic DD domain impulse at $\left( \frac{k \tau_p}{M} , \frac{l \nu_p}{N}\right)$ (see RHS of (\ref{eqn_xtaunukl})). The $(k,l)$-th carrier waveform , which we refer to as a TD pulsone, is simply a train of impulses where the $n$-th impulse is located at $t = n \tau_p + k \tau_p/M$, modulated by a complex tone/sinusoid $e^{j 2 \pi \frac{l \nu_p}{N} \, (t - k \tau_p/M)}$ (see (\ref{eqn_18})). 

It is clear from (\ref{eqn_fdpulsone}) that the FD representation of the $(k,l)$-th TD pulsone is an FD impulse train (with $m$-th impulse at $f = m \nu_p + l\nu_p/N$) modulated by a FD tone $e^{-j 2 \pi f k \tau_p/M}$. We refer to this waveform as a FD pulsone.
When $w_{tx}(\tau, \nu) = \delta(\tau) \, \delta(\nu)$, it follows from (\ref{eqn_18}) and (\ref{eqn_fdpulsone}) that the corresponding carrier waveforms have infinite time duration and bandwidth.\\

\textbf{Carrier waveforms with limited time and bandwidth:} The carrier waveforms can be limited in time duration and bandwidth by choosing an appropriate $w_{tx}(\tau, \nu)$. 
Appendix \ref{app_timebandlimited} starts
from (\ref{eqn_18}) and shows that if the delay spread of $w_{tx}(\tau, \nu)$ is approximately $1/B$, and the Doppler spread is approximately $1/T$, then the TD pulsones have time duration $T$ and bandwidth $B$.
It follows from (\ref{eqn_18}) that
$s_{\mbox{\scriptsize{td}},k,l}(t)$ is simply $x_{k,l}(t)$ spread by $1/B$ in the time domain. Given the structure of $x_{k,l}(t)$, it is clear that the TD pulsone $s_{\mbox{\scriptsize{td}},k,l}(t)$ consists of a train of narrow pulses modulated by a tone, with each narrow pulse having spread $1/B$ and adjacent pulses separated by $\tau_p$ seconds.
Similarly, from
the integral expression in (\ref{eqn_sfd1}) it follows that the FD representation of the $(k,l)$-th carrier waveform, i.e., $s_{\mbox{\scriptsize{fd}},k,l}(f)$, is simply $X_{k,l}(f)$ spread by $1/T$ in the frequency domain, since the Doppler domain spread of $w_{tx}(\tau, \nu)$ is $1/T$.
Given the structure of $X_{k,l}(f)$ in (\ref{eqn_fdpulsone}), it follows from the integral expression in (\ref{eqn_sfd1}) that the FD representation of the carrier waveform i.e., $s_{\mbox{\scriptsize{fd}},k,l}(f)$ consists of a train of narrow FD pulses modulated by a FD tone, with each narrow pulse having spread $1/T$  and adjacent pulses separated by $\nu_p$ Hz. Therefore, $s_{\mbox{\scriptsize{td}},k,l}(t)$ also has a pulsone structure in the FD, and is referred to as a FD pulsone. Fig.~\ref{fig4_paper1} illustrates the TD and FD representations of the Zak-OTFS carrier waveform.\\

\textbf{Proof of Theorem \ref{thm0}:} From (\ref{eqn_discreteinformationsig}) and (\ref{eqn_qp}) it follows that
\begin{eqnarray}
    x_{\mbox{\scriptsize{dd}}}(\tau, \nu) & \Define & \sum\limits_{k=0}^{M-1} \sum\limits_{l=0}^{N-1} x[k,l] \, x_{\mbox{\scriptsize{dd}},k,l}(\tau, \nu)
\end{eqnarray}where

{\vspace{-4mm}
\small
\begin{eqnarray}
     x_{\mbox{\scriptsize{dd}},k,l}(\tau, \nu ) & \hspace{-3.5mm} \Define & \hspace{-6mm} \sum\limits_{n,m \in {\mathbb Z}}  \hspace{-1.5mm}  e^{j 2 \pi \frac{n l}{N}} \, \delta\left(\tau - n\tau_p - \frac{k \tau_p}{M}\right) \delta\left(\nu - m \nu_p - \frac{l \nu_p}{N}\right), \nonumber \\
\end{eqnarray}\normalsize}$k=0,1,\cdots, M-1$, $l=0,1,\cdots,N-1$, is the quasi-periodic DD pulse localized at $\left(\frac{k \tau_p}{M}, \frac{l \nu_p}{N}\right)$ in the continuous DD domain.
Since twisted convolution is a linear operation
\begin{eqnarray}
\label{eqn_xddwtxkl}
    x_{\mbox{\scriptsize{dd}}}^{w_{tx}}(\tau, \nu) & = & w_{tx}(\tau, \nu) \, *_{\sigma}  \,  x_{\mbox{\scriptsize{dd}}}(\tau, \nu) \nonumber \\
    & = & \sum\limits_{k=0}^{M-1} \sum\limits_{l=0}^{N-1} x[k,l] \, x_{\mbox{\scriptsize{dd}},k,l}^{w_{tx}}(\tau, \nu ) ,\nonumber \\
    x_{\mbox{\scriptsize{dd}},k,l}^{w_{tx}}(\tau, \nu ) & \Define & w_{tx}(\tau, \nu) \, *_{\sigma} \, x_{\mbox{\scriptsize{dd}},k,l}(\tau, \nu ).
\end{eqnarray}Therefore, the transmit TD signal in (\ref{eqn_stdt}) is given by
\begin{eqnarray}
    s_{\mbox{\scriptsize{td}}}(t) & = & {\mathcal Z_t}^{-1}{\Big (}  x_{\mbox{\scriptsize{dd}}}^{w_{tx}}(\tau , \nu ) {\Big )} \nonumber \\
    & = & \sum\limits_{k=0}^{M-1} \sum\limits_{l=0}^{N-1} x[k,l] \,  s_{\mbox{\scriptsize{td}},k,l}(t) \nonumber \\
    s_{\mbox{\scriptsize{td}},k,l}(t) & \Define & {\mathcal Z_t}^{-1}{\Big (}  x_{\mbox{\scriptsize{dd}},k,l}^{w_{tx}}(\tau , \nu ) {\Big )},
\end{eqnarray}which follows from the linearity of the Inverse Zak transform.
Note that, $s_{\mbox{\scriptsize{td}},k,l}(t)$ is the carrier waveform for the $(k,l)$-th information symbol $x[k,l]$.
Let
\begin{eqnarray}
\label{eqn_22_1}
x_{k,l}(t) & \hspace{-3mm} \Define &  \hspace{-3mm} {\mathcal Z}_t^{-1} {\Big (} x_{\mbox{\scriptsize{dd}},k,l}(\tau, \nu ) {\Big )} \nonumber \\
& &  \hspace{-6mm} = {\mathcal Z}_t^{-1} {\Bigg (} \sum\limits_{n,m \in {\mathbb Z}}  \hspace{-1.5mm}  e^{j 2 \pi \frac{n l}{N}} \, \delta\left(\tau - n\tau_p - \frac{k \tau_p}{M}\right) \nonumber \\
& &  \hspace{19mm} \delta\left(\nu - m \nu_p - \frac{l \nu_p}{N}\right) {\Bigg )}.
\end{eqnarray}
Recall that in the absence of AWGN, the Zak transform of the received TD signal $r_{\mbox{\scriptsize{td}}}(t)$ in (\ref{eqn_rtdt}) is
\begin{eqnarray}
{\mathcal Z}_t{\Big (} \iint h_{\mbox{\scriptsize{phy}}}(\tau, \nu) \, s_{\mbox{\scriptsize{td}}}(t - \tau)  \, e^{j 2 \pi \nu (t - \tau)} \, d\tau \, d\nu  {\Big )} & &  \nonumber \\
& & \hspace{-60mm} =  h_{\mbox{\scriptsize{phy}}}(\tau, \nu) \, *_{\sigma} \, {\mathcal Z}_t{\Big (} s_{\mbox{\scriptsize{td}}}(t)  {\Big )} \nonumber \\
&  &  \hspace{-60mm} =  h_{\mbox{\scriptsize{phy}}}(\tau, \nu) \, *_{\sigma} \, x_{\mbox{\scriptsize{dd}}}^{w_{tx}}(\tau, \nu).
\end{eqnarray}where the last step follows from (\ref{eqn_stdt}). We substitute $w_{tx}(\tau, \nu)$ for
$h_{\mbox{\scriptsize{phy}}}(\tau, \nu)$, and 
$x_{\mbox{\scriptsize{dd}},k,l}(\tau, \nu) = {\mathcal Z}_t\left(  x_{k,l}(t) \right)$
for ${\mathcal Z}_t{\Big (} s_{\mbox{\scriptsize{td}}}(t)  {\Big )} = x_{\mbox{\scriptsize{dd}}}^{w_{tx}}(\tau, \nu)$ to obtain
\begin{eqnarray}
\label{eqn_32}
w_{tx}(\tau, \nu) \, *_{\sigma} \, x_{\mbox{\scriptsize{dd}},k,l}(\tau, \nu) & & \nonumber \\
& & \hspace{-50mm} = {\mathcal Z}_t{\Big (} \iint w_{tx}(\tau, \nu) \, x_{k,l}(t - \tau)  \, e^{j 2 \pi \nu (t - \tau)} \, d\tau \, d\nu  {\Big )}.
\end{eqnarray}Since $x_{\mbox{\scriptsize{dd}},k,l}^{w_{tx}}(\tau , \nu ) = w_{tx}(\tau, \nu) \, *_{\sigma} \, x_{\mbox{\scriptsize{dd}},k,l}(\tau , \nu )$ (see (\ref{eqn_xddwtxkl})), it follows from (\ref{eqn_32}) that
\begin{eqnarray}
s_{\mbox{\scriptsize{td}}, k,l}(t) & \hspace{-3mm} = &  \hspace{-3mm}  {\mathcal Z_t}^{-1}{\Big (}  x_{\mbox{\scriptsize{dd}},k,l}^{w_{tx}}(\tau , \nu ) {\Big )} \nonumber \\
&  &  \hspace{-15mm} = {\mathcal Z_t}^{-1}{\Big (}  w_{tx}(\tau, \nu) \, *_{\sigma} \, x_{\mbox{\scriptsize{dd}},k,l}(\tau , \nu )   {\Big )} \nonumber \\
&  & \hspace{-15mm}  = \iint w_{tx}(\tau, \nu) \, x_{k,l}(t - \tau )  \, e^{j 2 \pi \nu (t - \tau)} \, d\tau \, d\nu.
\end{eqnarray}
We apply the integral expression for the inverse Zak transform (\ref{prelim_3}) to the DD domain signal in (\ref{eqn_22_1}) to obtain
\begin{eqnarray}
x_{k,l}(t) & \hspace{-3mm} = &  \hspace{-3mm} {\mathcal Z}_t^{-1} {\Big (} x_{\mbox{\scriptsize{dd}},k,l}(\tau, \nu ) {\Big )} \nonumber \\ 
&  &  \hspace{-20mm} = \sqrt{\tau_p} \, \int\limits_{0}^{\nu_p} x_{\mbox{\scriptsize{dd}},k,l}(t, \nu ) \, d\nu  \nonumber \\
& & \hspace{-20mm} =  \sqrt{\tau_p} \sum\limits_{n,m \in {\mathbb Z}} {\Bigg [} e^{j 2 \pi \frac{n l}{N}} \delta\left(t - n\tau_p - \frac{k \tau_p}{M}\right) \, \nonumber \\
& & \int\limits_{0}^{\nu_p} \delta\left(\nu - m \nu_p - \frac{l \nu_p}{N}\right) \, d\nu {\Bigg ]} \nonumber \\
&  &  \hspace{-20mm} = \sqrt{\tau_p} \, \sum\limits_{n \in {\mathbb Z} } e^{j 2 \pi \frac{n l }{N} } \, \delta\left(t - n\tau_p - \frac{k \tau_p}{M}\right),
\end{eqnarray}since
\begin{eqnarray}
    \int\limits_{0}^{\nu_p} \delta\left(\nu - m \nu_p - \frac{l \nu_p}{N}\right) \, d\nu  & = \begin{cases}
        0 &, m \ne 0 \\
        1 &, m = 0
    \end{cases}.
\end{eqnarray}The FD representation (i.e., Fourier transform) of $x_{k,l}(t)$ is
\begin{eqnarray}
X_{k,l}(f) & = & \int x_{k,l}(t) \, e^{-j 2 \pi f t} \, dt \nonumber \\
&  & \hspace{-10mm} = \sqrt{\nu_p} \, \sum\limits_{m \in {\mathbb Z}} \delta\left( f - m \nu_p - \frac{l \nu_p}{N} \right) \, e^{-j 2 \pi f \frac{k \tau_p}{M}}.
\end{eqnarray}This follows from the fact that the FD representation
of the TD pulse train $\sum\limits_{n \in {\mathbb Z}} \delta(t - n \tau_p)$ is the FD pulse train $(1/\tau_p)\sum\limits_{m \in {\mathbb Z}} \delta(f - m \nu_p)$.

The FD representation of $s_{\mbox{\scriptsize{td}},k,l}(t)$ is
obtained from (\ref{eqn_18}) by taking
the Fourier transform.
\begin{eqnarray}
    s_{\mbox{\scriptsize{fd}},k,l}(f) & \hspace{-3mm} \Define & \hspace{-3mm} \int s_{\mbox{\scriptsize{td}},k,l}(t) \, e^{-j 2 \pi f t} \, dt \nonumber \\
    & & \hspace{-23mm} =  \hspace{-1mm} \iint \hspace{-1mm} w_{tx}(\tau, \nu) {\Bigg [}  \hspace{-1mm} \int  \hspace{-1mm} x_{k,l}(t - \tau) \, e^{-j 2 \pi f t} \, e^{j 2 \pi \nu (t - \tau) } \, dt {\Bigg ]} d\tau \, d\nu \nonumber \\
    & & \hspace{-23mm} =  \hspace{-1mm} \iint \hspace{-1mm} w_{tx}(\tau, \nu)  e^{-j 2 \pi f \tau}  \underbrace{{\Bigg [}  \hspace{-1mm} \int  \hspace{-1mm} x_{k,l}(t - \tau ) \, e^{-j 2 \pi (f - \nu) (t - \tau)}  dt {\Bigg ]}}_{= X_{k,l}(f - \nu )} d\tau \, d\nu \nonumber \\
    & & \hspace{-23mm} =  \hspace{-1mm} \iint \hspace{-1mm} w_{tx}(\tau, \nu)  \, e^{-j 2 \pi f \tau} \,   X_{k,l}(f - \nu) \,  d\tau \, d\nu
\end{eqnarray}where we have used the fact that $X_{k,l}(f )$ in (\ref{eqn_fdpulsone}) is the Fourier transform of $x_{k,l}(t)$.

This completes the proof of Theorem \ref{thm0}.

\section{Time and band-limited pulsones}
\label{app_timebandlimited}
 For factorizable pulse shaping waveform $w_{tx}(\tau, \nu) = w_1(\tau) \, w_2(\nu)$,  it follows from (\ref{eqn_18}) that the $(k,l)$-th carrier waveform is given by
\begin{eqnarray}
\label{eqn_23}
    s_{\mbox{\scriptsize{td}},k,l}(t) & \hspace{-3.5mm} = & \hspace{-3.5mm} \iint w_1(\tau) \, w_2(\nu) \, x_{k,l}(t - \tau ) \, e^{j 2 \pi \nu (t - \tau) } \, d\tau \, d\nu \nonumber \\
    & & \hspace{-19mm} = \int w_1(\tau) x_{k,l}(t - \tau ) {\Bigg [} \int w_2(\nu) \, e^{j 2 \pi \nu (t - \tau)} d\nu {\Bigg ]} \, d\tau  \nonumber \\
    & & \hspace{-19mm} = \int w_1(\tau) \, x_{k,l}(t - \tau ) \, W_2(t - \tau) \, d\tau
\end{eqnarray}where
\begin{eqnarray}
\label{eqn_24}
    W_2(t) & \Define & \int w_2(\nu) \, e^{j 2 \pi \nu t} \, d\nu
\end{eqnarray}is the inverse Fourier transform of the factor $w_2(\nu)$ along the Doppler domain. Let
\begin{eqnarray}
\label{eqn_25}
    x_{2,k,l}(t ) & \Define & x_{k,l}(t ) \, W_2(t).
\end{eqnarray}Then, $x_{2,k,l}(t )$ has time duration limited by the duration of $W_2(t)$. Using (\ref{eqn_25}) in (\ref{eqn_23}) we obtain
\begin{eqnarray}
\label{eqn_27}
    s_{\mbox{\scriptsize{td}},k,l}(t) &  = & \int w_1(\tau) \, x_{2,k,l}(t - \tau) \, d\tau \nonumber \\
    & = & w_1(t) \, \star \, x_{2,k,l}(t )
\end{eqnarray}where $\star$ denotes the usual TD convolution. Hence, the Fourier transform of $s_{\mbox{\scriptsize{td}},k,l}(t)$ is
\begin{eqnarray}
\label{eqn_26}
    s_{\mbox{\scriptsize{fd}},k,l}(f) &  \Define & {\mathcal F}{\Big (} s_{\mbox{\scriptsize{td}},k,l}(t) {\Big )} \nonumber \\
    & & \hspace{-25mm} = \int s_{\mbox{\scriptsize{td}},k,l}(t) \, e^{-j 2 \pi f t} \, dt = \, W_1(f) \, X_{2,k,l}(f ) 
\end{eqnarray}where $W_1(f)$ and $X_{2,k,l}(f )$ are the Fourier transforms of $w_1(t)$ and $x_{2,k,l}(t )$, respectively, given by
\begin{eqnarray}
    W_1(f) &  \Define &  \int w_1(t) \, e^{-j 2 \pi f t} \, dt, \nonumber \\
    X_{2,k,l}(f ) & \Define & \int x_{2,k,l}(t ) \, e^{-j 2 \pi f t } \, dt.
\end{eqnarray}It is clear from (\ref{eqn_26}) that the carrier waveform will be bandwidth limited to the bandwidth of $w_1(\cdot)$. For example choosing $w_1(\tau) = \sqrt{B} \, sinc(B \tau)$ limits the bandwidth of the carrier waveforms to exactly $B$ Hz. In general, for a given bandwidth constraint $B$, $w_1(\tau)$ must have a spread of approximately $1/B$ along the delay domain. Similarly, from (\ref{eqn_25}) it follows that the duration of $x_{2,k,l}(t )$ can be limited to approximately $T$ seconds by choosing the factor pulse $w_2(\nu)$ to have a spread of approximately $1/T$ along the Doppler domain (for example, with $w_2(\nu) = \sqrt{T} sinc( \nu T)$, $x_{2,k,l}(t)$ is limited exactly to the TD interval $[-T/2 \,,\, T/2]$). Therefore, $x_{2,k,l}(t ) =  x_{k,l}(t ) \, W_2(t)$ contains only $T/\tau_p = N$ number of Dirac-delta impulses of the infinite impulse train $x_{k,l}(t )$. Also, due to the TD convolution in (\ref{eqn_27}), in the carrier waveform $s_{\mbox{\scriptsize{td}},k,l}(t)$ these $N$ impulses are spread over a duration of approximately $1/B$ seconds since the spread of $w_1(\tau)$ is $1/B$. Therefore, under a finite duration and bandwidth constraint, the $(k,l)$ carrier waveform consists of narrow pulses at $t = n \tau_p + k \tau_p/M$ where the width of each pulse is $1/B = \tau_p/M$ (since $M = B \tau_p$). Note that $\tau_p/M$ is also the time between the location of the $n$-th pulses of the $(k,l)$-th and the $(k+1, l)$-th carrier waveforms.

\section{The discrete ambiguity function}
\label{app_prop_ambig}
The discrete cross-ambiguity function ${A}_{a,b}[k,l]$ between two discrete quasi-periodic DD domain functions $a[k,l]$ and $b[k,l]$ is given by
\begin{eqnarray}
\label{crossambig_1}
{ A}_{a,b}[k,l] & \hspace{-3mm} = & \hspace{-4mm} \sum\limits_{k'=0}^{M-1} \sum\limits_{l'=0}^{N-1} a[k',l'] \, b^*[k' - k, l' - l] \, e^{-j 2 \pi \frac{l (k' - k)}{MN}}. \nonumber \\
\end{eqnarray}The sum in the R.H.S. may be taken over any discrete delay period $M$ and discrete Doppler period $N$ (see Lemma \ref{lem_crossambig1} in this Appendix).

\begin{theorem}
\label{thm_crossambig2}
Let $a[k,l]$ and $b[k,l]$ be quasi-periodic DD functions related by
\begin{eqnarray}
    a[k,l] = g[k,l] \, *_{\sigma} \, b[k,l].
\end{eqnarray}Then, the cross ambiguity between $a[k,l]$ and another
quasi-periodic DD function $c[k,l]$ is given by the twisted convolution between
$g[k,l]$ and the cross ambiguity between $b[k,l]$ and $c[k,l]$, i.e.
\begin{eqnarray}
\label{eqn_aac1}
    { A}_{a,c}[k,l] & = & \sum\limits_{k'=0}^{M-1} \sum\limits_{l'=0}^{N-1} a[k',l'] \, c^*[k' - k, l' - l] \, e^{-j 2 \pi \frac{l (k' - k)}{MN}} \nonumber \\
    & = & g[k,l] \, *_{\sigma} \, {\mathcal A}_{b,c}[k,l].
\end{eqnarray}
\end{theorem}
\begin{IEEEproof}
    See Appendix \ref{app_prf_thm1}.
\end{IEEEproof}

\begin{lemma}
\label{lem_crossambig1}
The discrete cross-ambiguity between any two quasi-periodic functions $a[k,l]$
and $b[k,l]$ is invariant w.r.t. the period over which the sum in the R.H.S. of (\ref{crossambig_1}) is computed, i.e., for any $k_0, l_0 \in {\mathbb Z}$
\begin{eqnarray}
{\mathcal A}_{a,b}[k,l] & \hspace{-3mm} = &  \hspace{-6mm} \sum\limits_{k'=k_0}^{k_0+M-1} \sum\limits_{l'= l_0}^{l_0 + N-1} a[k',l'] \, b^*[k' - k, l' - l] \, e^{-j 2 \pi \frac{l (k' - k)}{MN}} \nonumber \\
& &  \hspace{-9mm} =\sum\limits_{k'=0}^{M-1} \sum\limits_{l'=0}^{N-1} a[k',l'] \, b^*[k' - k, l' - l] \, e^{-j 2 \pi \frac{l (k' - k)}{MN}}.
\end{eqnarray}
\end{lemma}
\begin{IEEEproof}

{\vspace{-4mm}
\small
\begin{eqnarray}
\label{eqn_crossambiglemeqn1}
{\mathcal A}_{a,b}[k,l] & \hspace{-3mm} = & \hspace{-6mm} \sum\limits_{k'=k_0}^{k_0+M-1} \sum\limits_{l'= l_0}^{l_0 + N-1} a[k',l'] \, b^*[k' - k, l' - l] \, e^{-j 2 \pi \frac{l (k' - k)}{MN}} \nonumber \\
& & \hspace{-16mm} = \hspace{-3mm} \sum\limits_{k'=k_0}^{k_0+M-1} \sum\limits_{l'= l_0}^{l_0 + N-1} \hspace{-2mm} a\left[k' \hspace{-3mm} \mod M + \left\lfloor \frac{k'}{M} \right\rfloor M  \,,\,  l' \hspace{-3mm} \mod N + \left\lfloor \frac{l'}{N} \right\rfloor N  \right] \nonumber \\
& &  \hspace{-7mm} b^*\left[k' \hspace{-3mm} \mod M - k + \left\lfloor \frac{k'}{M} \right\rfloor M , l' \hspace{-3mm} \mod N  - l + \left\lfloor \frac{l'}{N} \right\rfloor N \right]  \nonumber \\
& &    e^{-j 2 \pi \frac{l \left(k' \hspace{-3mm} \mod M  - k + \left\lfloor \frac{k'}{M} \right\rfloor M \right)}{MN}}. 
\end{eqnarray}\normalsize}Since both $a[k,l]$ and $b[k,l]$ are quasi-periodic DD functions, it follows from (\ref{eqn_01}) that
\begin{eqnarray}
\label{eqn9834513}
    a\left[k' \hspace{-3mm} \mod M + \left\lfloor \frac{k'}{M} \right\rfloor M  \,,\,  l' \hspace{-3mm} \mod N + \left\lfloor \frac{l'}{N} \right\rfloor N  \right] & & \nonumber \\
    & & \hspace{-63mm} =  a\left[k' \hspace{-3mm} \mod M  \,,\,  l' \hspace{-3mm} \mod N  \right] \, e^{j 2 \pi \left\lfloor \frac{k'}{M} \right\rfloor \frac{l' \hspace{-3mm} \mod N}{N} }
\end{eqnarray}and
\begin{eqnarray}
\label{eqn00037t52}
b\left[k' \hspace{-3mm} \mod M - k + \left\lfloor \frac{k'}{M} \right\rfloor M , l' \hspace{-3mm} \mod N  - l + \left\lfloor \frac{l'}{N} \right\rfloor N \right]  & & \nonumber \\
    & & \hspace{-83mm} = b\left[k' \hspace{-3mm} \mod M - k , l' \hspace{-3mm} \mod N  - l  \right] \, e^{j 2 \pi \left\lfloor \frac{k'}{M} \right\rfloor \frac{l' \hspace{-3mm} \mod N - l }{N}  }.
\end{eqnarray}Substituting (\ref{eqn9834513}) and (\ref{eqn00037t52}) in (\ref{eqn_crossambiglemeqn1}) gives

{\vspace{-4mm}
\small
\begin{eqnarray}
    {\mathcal A}_{a,b}[k,l] & \hspace{-3mm} = & \hspace{-6mm} \sum\limits_{k'=k_0}^{k_0+M-1} \sum\limits_{l'= l_0}^{l_0 + N-1} \hspace{-1mm} {\Bigg (} a\left[k' \hspace{-3mm} \mod M  \,,\,  l' \hspace{-3mm} \mod N  \right] \nonumber \\
    & &  \hspace{17mm} b^*\left[k' \hspace{-3mm} \mod M - k , l' \hspace{-3mm} \mod N  - l  \right] \nonumber \\
    & &  \hspace{20mm}  e^{-j 2 \pi l \frac{\left(k' \hspace{-3mm} \mod M - k \right)}{MN}} {\Bigg )}.
\end{eqnarray}\normalsize}The R.H.S. above depends on $k'$ and $l'$ only through
$k' \hspace{-2mm} \mod M$ and $l' \hspace{-2mm} \mod N$. Since the summation variables $k'$ and $l'$ span a full delay and Doppler period respectively, it follows that as $k'$ varies from $k_0$ to $k_0 + M -1$, $k' \hspace{-1mm} \mod M$ takes all possible values
from $0$ to $M-1$. Similarly, as $l'$ varies from $l_0$ to $l_0 + N -1$, $l' \hspace{-1mm} \mod N$ takes all possible values
from $0$ to $N-1$. Therefore, replacing summation variable $k'$ and $l'$ with ${\Tilde k} = k' \hspace{-2mm} \mod M$ and ${\Tilde l} = l' \hspace{-2mm} \mod N$ respectively, we get
\begin{eqnarray}
    {\mathcal A}_{a,b}[k,l] & \hspace{-3mm} = & \hspace{-4mm} \sum\limits_{{\Tilde k}=0}^{M-1} \sum\limits_{{\Tilde l}=0}^{N-1} a[{\Tilde k},{\Tilde l}] \, b^*[{\Tilde k} - k, {\Tilde l} - l] \, e^{-j 2 \pi \frac{l ({\Tilde k} - k)}{MN}}
\end{eqnarray}which is same as the expression in (\ref{crossambig_1}).
\end{IEEEproof}

\begin{lemma}
\label{lem3ambig}
For any two quasi-periodic discrete DD domain functions $a[k,l]$ and $b[k,l]$
\begin{eqnarray}
{\mathcal A}_{b,a}[k,l] & = & {\mathcal A}_{a,b}^*[-k, -l] \, e^{j 2 \pi \frac{k l}{M N}}.
\end{eqnarray}
\end{lemma}
\begin{IEEEproof}
    From the definition of ${\mathcal A}_{a,b}[k,l]$ in (\ref{crossambig_1})
    it follows that
    \begin{eqnarray}
        {\mathcal A}_{a,b}^*[-k,-l] & \hspace{-3mm} = &  \hspace{-4mm} \sum\limits_{k'=0}^{M-1} \sum\limits_{l'=0}^{N-1} a^*[k',l'] \, b[k' + k, l'+l] \, e^{-j 2 \pi l \frac{(k' + k)}{MN}} \nonumber \\
        & & \hspace{-22mm} = e^{-j 2 \pi \frac{l k}{MN}}  \hspace{-2mm}  \underbrace{\sum\limits_{{\Tilde k}=k}^{k+M-1} \sum\limits_{{\Tilde l}=l}^{l+N-1} b[{\Tilde k}, {\Tilde l}] \, a^*[{\Tilde k} -k, {\Tilde l} -l] \, e^{-j 2 \pi l \frac{({\Tilde k} - k)}{MN}}}_{= {\mathcal A}_{b,a}[k,l]} \nonumber \\
        &  &  \hspace{-22mm} =  e^{-j 2 \pi \frac{l k}{MN}} {\mathcal A}_{b,a}[k,l]
    \end{eqnarray}where in the second step the summation variables $k', l',$ are changed to
    ${\Tilde k} = (k' + k)$ and ${\Tilde l} = (l' + l)$ respectively.
\end{IEEEproof}

\section{self-ambiguity function of the point pilot signal}
\label{app_ppambig}
It follows from the definition of the ambiguity function (\ref{crossambig_1}) that the self-ambiguity of the point pilot signal is given by (\ref{eqnautoambig}) (see top of next page). Here, step (a) results from substituting for $ x_{\mbox{\scriptsize{p,dd}}}[k,l]$ using (\ref{eqn_imp_pilot}).  
For a non-zero sum in step (a), from the location of the Dirac-delta impulses in the R.H.S., it follows that $k' \equiv k_p \, mod \, M$ and $(k' - k) \equiv k_p  \, mod \, M$. This is only possible if $k$ is an integer multiple of $M$. Similarly, $l' \equiv l_p \, mod \, N$ and $(l' - l) \equiv l_p  \, mod \, N$. This is only possible if $l$ is an integer multiple of $N$, in other words the self-ambiguity function $A_{x_p, x_p}[k,l]$ is non-zero for only the DD points $(nM, mN), n,m \in {\mathbb Z}$, which is simply the period lattice. Again, from step (a) it is also clear that, for $(k,l) = (nM, mN)$, the only non-zero contribution in the sum in the R.H.S. is from the term corresponding to $n_1 = (k_p \, mod \, M - k_p)/M$, $n_2 = n_1 - n$, $m_1 = (l_p \, mod \, N - l_p)/N$ and $m_2 = m_1 - m$. The value of $A_{x_p, x_p}[nM, mN]$ is simply $e^{j 2 \pi \frac{n_1 l' }{N} } \, e^{-j 2 \pi \frac{n_2 (l' - l)}{N} } \, e^{-j 2 \pi \frac{l (k' - k)}{MN}}$ evaluated for the values of $n_1, n_2, l', k'$ mentioned above.
This leads to the final expression of $A_{x_p, x_p}[k,l]$ in step (b). 

\begin{figure*}
{\small
\vspace{-6mm}
\begin{eqnarray}
\label{eqnautoambig}
A_{x_p, x_p}[k,l] & \hspace{-2.5mm} = & \hspace{-2.5mm} \sum\limits_{k' = 0}^{M-1} \sum\limits_{l' = 0}^{N-1} x_{\mbox{\scriptsize{p,dd}}}[k' ,l' ]  \, x_{\mbox{\scriptsize{p,dd}}}^*[k' - k,l' - l] \, e^{- j 2 \pi \frac{l (k' - k)}{M N}} \nonumber \\
& &  \hspace{-20mm} \mya \sum\limits_{k' = 0}^{M-1} \sum\limits_{l' = 0}^{N-1} \sum\limits_{n_1, m_1 \in {\mathbb Z}} \sum\limits_{n_2, m_2 \in {\mathbb Z}}  \hspace{-3mm} e^{j 2 \pi \frac{n_1 l' }{N} } \, e^{-j 2 \pi \frac{n_2 (l' - l)}{N} } \, e^{-j 2 \pi \frac{l (k' - k)}{MN}} \, \delta[k' - k_p - n_1 M] \, \delta[l' - l_p - m_1 N] \, \delta[k' - k - k_p - n_2 M] \, \delta[l' - l - l_p - m_2 N] \nonumber \\
& &  \hspace{-20mm} \myb  \sum\limits_{n,m \in {\mathbb Z}} e^{j 2 \pi \frac{(nM l_p - mN k_p)}{MN}} \, \delta[k - nM] \, \delta[l - mN].
\end{eqnarray}\normalsize}
\vspace{-3mm}
\begin{eqnarray*}
    \hline
\end{eqnarray*}
\end{figure*}

\section{symmetry of cross-ambiguity}
\label{app_prf_thm1}
First
\begin{eqnarray}
    a[k,l] & \hspace{-3mm} = &  \hspace{-4mm} \sum\limits_{{\Tilde k}, {\Tilde l} \in {\mathbb Z}} g[{\Tilde k}, {\Tilde l}] \, b[k - {\Tilde k}, l - {\Tilde l}] \, e^{j 2 \pi {\Tilde l} \frac{(k - {\Tilde k})}{MN}}.
\end{eqnarray}Using this expression for $a[k,l]$ in the R.H.S. of the first step in (\ref{eqn_aac1}) we get

{\vspace{-4mm}
\small
\begin{eqnarray}
\label{eqn231_crossambigthm}
    {\mathcal A}_{a,c}[k,l] & \hspace{-3mm} = &  \hspace{-5mm} \sum\limits_{k'=0}^{M-1} \sum\limits_{l'=0}^{N-1}  \, \sum\limits_{{\Tilde k}, {\Tilde l} \in {\mathbb Z}} g[{\Tilde k}, {\Tilde l}] \, b[k' - {\Tilde k}, l' - {\Tilde l}]  \nonumber \\
    & & c^*[k' - k, l' - l] \, e^{-j 2 \pi \frac{l (k' - k)}{MN}} \, e^{j 2 \pi {\Tilde l} \frac{(k' - {\Tilde k})}{MN}} \nonumber \\
    & & \hspace{-12mm} =  \sum\limits_{{\Tilde k}, {\Tilde l} \in {\mathbb Z}} g[{\Tilde k}, {\Tilde l}] \, {\Bigg (} \sum\limits_{k'=0}^{M-1} \sum\limits_{l'=0}^{N-1} b[k' - {\Tilde k}, l' - {\Tilde l}] \, c^*[k' - k, l' - l] \nonumber \\
    & & \hspace{18mm} e^{-j 2 \pi \frac{l (k' - k)}{MN}} \, e^{j 2 \pi {\Tilde l} \frac{(k' - {\Tilde k})}{MN}} {\Bigg )}.
\end{eqnarray}\normalsize}We change variable ${\Tilde k}$ to $k''  = k' - {\Tilde k}$, variable ${\Tilde l}$
to $l'' = l' - {\Tilde l}$, and apply Lemma \ref{lem_crossambig1} to obtain 
\begin{eqnarray}
\label{eqn232_crossambigthm}
\sum\limits_{k''= - {\Tilde k}}^{(M-1 - {\Tilde k})} \sum\limits_{l''= - {\Tilde l}}^{(N-1- {\Tilde l})} {\Bigg (} b[k'', l''] \, c^*[k'' - (k - {\Tilde k} ), l'' - (l - {\Tilde l} )]   \nonumber \\
 e^{-j 2 \pi (l - {\Tilde l}) \frac{(k'' - (k - {\Tilde k}))}{MN}} \, {\Bigg )}  \, e^{j 2 \pi {\Tilde l} \frac{(k - {\Tilde k})}{MN}} \nonumber \\
 =  {\mathcal A}_{b,c}[k - {\Tilde k}, l - {\Tilde l}] \, \, e^{j 2 \pi {\Tilde l} \frac{(k - {\Tilde k})}{MN}}
 \end{eqnarray}Substituting (\ref{eqn232_crossambigthm}) into (\ref{eqn231_crossambigthm}) we obtain
 \begin{eqnarray}
     {\mathcal A}_{a,c}[k,l] & \hspace{-3mm} = &  \hspace{-3mm} \sum\limits_{{\Tilde k}, {\Tilde l} \in {\mathbb Z}} g[{\Tilde k}, {\Tilde l}] \, {\mathcal A}_{b,c}[k - {\Tilde k}, l - {\Tilde l}] \, \, e^{j 2 \pi {\Tilde l} \frac{(k - {\Tilde k})}{MN}} \nonumber \\
     & = & g[k,l] \, *_{\sigma} \, {\mathcal A}_{b,c}[k,l].
 \end{eqnarray}
 
\section{Proof of Theorem \ref{coro1}}
\label{app_prf_corrambigthm2}
The expression of the received signal $y_{\mbox{\scriptsize{dd}}}[k,l]$ in (\ref{rxjointdatasense}) consists of the sum
of the received data signal, pilot signal and noise.
The cross-ambiguity of the received signal with the spread pilot signal $x_{\mbox{\scriptsize{s,dd}}}[k,l]$ is therefore the sum of, (i) the cross-ambiguity of the received spread pilot signal
$\sqrt{E_p} \, h_{\mbox{\scriptsize{eff}}}[k,l] \, *_{\sigma} \, x_{\mbox{\scriptsize{s,dd}}}[k,l] $ with $x_{\mbox{\scriptsize{s,dd}}}[k,l]$. Applying Theorem \ref{thm_crossambig2} (with $g[k,l] = h_{\mbox{\scriptsize{eff}}}[k,l]$, $a[k,l] = \sqrt{E_p} \, h_{\mbox{\scriptsize{eff}}}[k,l] \, *_{\sigma} \, x_{\mbox{\scriptsize{s,dd}}}[k,l]$, $b[k,l] = c[k,l] =  x_{\mbox{\scriptsize{s,dd}}}[k,l]$) gives the first term in the R.H.S. of (\ref{eqn352_ays}), (ii) the cross-ambiguity of the received data/information signal $\sqrt{E_d} \, h_{\mbox{\scriptsize{eff}}}[k,l] \, *_{\sigma} \, x_{\mbox{\scriptsize{d,dd}}}[k,l] $ with $x_{\mbox{\scriptsize{s,dd}}}[k,l]$. Applying Theorem \ref{thm_crossambig2} (with $g[k,l] = h_{\mbox{\scriptsize{eff}}}[k,l]$, $a[k,l] = \sqrt{E_d} \, h_{\mbox{\scriptsize{eff}}}[k,l] \, *_{\sigma} \, x_{\mbox{\scriptsize{d,dd}}}[k,l]$, $b[k,l] = x_{\mbox{\scriptsize{d,dd}}}[k,l]$, $c[k,l] = x_{\mbox{\scriptsize{s,dd}}}[k,l] $) gives the second term in the R.H.S. of (\ref{eqn352_ays}), and (iii) the cross-ambiguity between the noise signal $n_{\mbox{\scriptsize{dd}}}[k,l]$ and $x_{\mbox{\scriptsize{s,dd}}}[k,l]$, gives the third term in the R.H.S. of (\ref{eqn352_ays}).

\section{Derivation of the maximum likelihood (ML) estimate of $h_{\mbox{\scriptsize{eff}}}[k,l]$}
\label{app_prf_mlest}
The maximum likelihood (ML) estimate of $h_{\mbox{\scriptsize{eff}}}[k,l]$ is given by

{\vspace{-4mm}
\small
\begin{eqnarray}
\label{eqnmlest}
{\widehat h}_{\mbox{\scriptsize{eff}}}[k,l] & \hspace{-2mm} = & \hspace{-2mm} \arg \min_{h[k,l]}  \sum\limits_{k=0}^{M-1}  \sum\limits_{l=0}^{N-1} \left\vert y_{\mbox{\scriptsize{s,dd}}}[k,l] \, - \, h[k,l] \, *_{\sigma} \, x_{\mbox{\scriptsize{s,dd}}}[k,l] \right\vert^2 \nonumber \\
\end{eqnarray}\normalsize}where $y_{\mbox{\scriptsize{s,dd}}}[k,l]$ is the received pilot signal given by (\ref{eqn025448}).
Essentially, the ML estimate is such that it minimizes the energy of the estimation error signal $e_{\mbox{\scriptsize{dd}}}[k,l] =  (y_{\mbox{\scriptsize{s,dd}}}[k,l] \, - \, h[k,l] \, *_{\sigma} \, x_{\mbox{\scriptsize{s,dd}}}[k,l] )$. Since both $y_{\mbox{\scriptsize{s,dd}}}[k,l]$ and $ x_{\mbox{\scriptsize{s,dd}}}[k,l]$ are quasi-periodic and twisted convolution conserves quasi-periodicity, it follows that the error signal is also quasi-periodic. The expression in (\ref{eqnmlest}) then follows from the fact that the energy of any discrete quasi-periodic DD domain signal $e_{\mbox{\scriptsize{dd}}}[k,l]$ is given by $\sum\limits_{k=0}^{M-1} \sum\limits_{l=0}^{N-1} \left\vert e_{\mbox{\scriptsize{dd}}}[k,l] \right\vert^2$.

The expression for $h[k,l] \, *_{\sigma} \, x_{\mbox{\scriptsize{s,dd}}}[k,l]$ follows from the definition of twisted convolution and is given by (\ref{eqn916350}) (see top of next page).

\begin{figure*}
\vspace{-6mm}
\begin{eqnarray}
\label{eqn916350}
h[k,l] \, *_{\sigma} \, x_{\mbox{\scriptsize{s,dd}}}[k,l] & = & \sum\limits_{(k' , l') \in {\mathcal S}_{(0,0)} } 
h[k',l'] \, x_{\mbox{\scriptsize{s,dd}}}[k - k',l - l' ] \, e^{j 2 \pi \frac{l' (k - k')}{MN}} 
\end{eqnarray}
\vspace{-4mm}
 \begin{eqnarray*}
\hline
 \end{eqnarray*}
\end{figure*}
Further analysis of (\ref{eqnmlest}) gives (\ref{eqn88636}) (see top of next page).
\begin{figure*}
\vspace{-6mm}
\begin{eqnarray}
\label{eqn88636}
\arg \min_{h[k,l]}  \sum\limits_{k=0}^{M-1}  \sum\limits_{l=0}^{N-1} {\Big \vert} y_{\mbox{\scriptsize{s,dd}}}[k,l] \, - \, h[k,l] \, *_{\sigma} \, x_{\mbox{\scriptsize{s,dd}}}[k,l] {\Big \vert}^2 & & \nonumber \\
& & \hspace{-70mm} \mya \arg \min_{h[k,l]}  {\Bigg [} \sum\limits_{k=0}^{M-1}  \sum\limits_{l=0}^{N-1}  \left\vert  y_{\mbox{\scriptsize{s,dd}}}[k,l]  \right\vert^2 \, + \, \sum\limits_{k=0}^{M-1}  \sum\limits_{l=0}^{N-1}  \left\vert  h[k,l] \, *_{\sigma} \, x_{\mbox{\scriptsize{s,dd}}}[k,l]  \right\vert^2  \nonumber \\
& & \hspace{-70mm} \, - \,  2 \sum\limits_{k=0}^{M-1}  \sum\limits_{l=0}^{N-1} \Re\left( y_{\mbox{\scriptsize{s,dd}}}[k,l] \, \left( \sum\limits_{(k' , l') \in {\mathcal S}_{(0,0)} } 
h[k',l'] \, x_{\mbox{\scriptsize{s,dd}}}[k - k',l - l' ] \, e^{j 2 \pi \frac{l' (k - k')}{MN}} \right)^*  \, \right) \, {\Bigg ]}.
\end{eqnarray}
\vspace{-3mm}
 \begin{eqnarray*}
\hline
 \end{eqnarray*}
\end{figure*}
\begin{figure*}
\vspace{-6mm}
\small{
\begin{eqnarray}
\label{eqn9025440}
    \sum\limits_{k=0}^{M-1}  \sum\limits_{l=0}^{N-1}  \left\vert  h[k,l] \, *_{\sigma} \, x_{\mbox{\scriptsize{s,dd}}}[k,l]  \right\vert^2 & & \nonumber \\
       & & \hspace{-50mm}  = \sum\limits_{k=0}^{M-1}  \sum\limits_{l=0}^{N-1} \sum\limits_{(k_1' , l_1'), (k_2', l_2') \in {\mathcal S}_{(0,0)} } \hspace{-6mm} h[k_1', l_1'] \, h^*[k_2', l_2'] \, x_{\mbox{\scriptsize{s,dd}}}[k - k_1', l - l_1']  \, x_{\mbox{\scriptsize{s,dd}}}^*[k - k_2', l - l_2'] \, e^{j 2 \pi \frac{l_1' (k - k_1')}{MN}} \,  e^{-j 2 \pi \frac{l_2' (k - k_2')}{MN}} \nonumber \\
       & & \hspace{-50mm}  =  \hspace{-6mm} \sum\limits_{(k_1' , l_1'), (k_2', l_2') \in {\mathcal S}_{(0,0)} } \hspace{-6mm} h[k_1', l_1'] \, h^*[k_2', l_2'] \, \left[   \sum\limits_{k=0}^{M-1}  \sum\limits_{l=0}^{N-1} x_{\mbox{\scriptsize{s,dd}}}[k - k_1', l - l_1']  \, x_{\mbox{\scriptsize{s,dd}}}^*[k - k_2', l - l_2']  e^{j 2 \pi \frac{l_1' (k - k_1')}{MN}} \,  e^{-j 2 \pi \frac{l_2' (k - k_2')}{MN}} \right] \nonumber \\
       & & \hspace{-50mm}  =  \hspace{-6mm} \sum\limits_{(k_1' , l_1'), (k_2', l_2') \in {\mathcal S}_{(0,0)} } \hspace{-6mm} h[k_1', l_1'] \, h^*[k_2', l_2'] \, e^{j 2 \pi \frac{l_1'(k_2' - k_1')}{MN}} \underbrace{\left[   \sum\limits_{{\Tilde k}=-k_1'}^{M-1 - k_1'}  \sum\limits_{{\Tilde l}=-l_1'}^{N-1-l_1'} \hspace{-2mm} x_{\mbox{\scriptsize{s,dd}}}[{\Tilde k}, {\Tilde l}]  \, x_{\mbox{\scriptsize{s,dd}}}^*[{\Tilde k} - (k_2' - k_1'), {\Tilde l} - (l_2' - l_1')]  \, e^{-j 2 \pi \frac{(l_2' - l_1') ({\Tilde k} - (k_2' - k_1'))}{MN}}  \right]}_{= A_{x_s, x_s}[k_2' - k_1', l_2' - l_1'], \, \mbox{{ see Lemma}} \, \ref{lem_crossambig1} \,\, \mbox{{in Appendix}} \, \ref{app_prop_ambig} } \nonumber \\
       & & \hspace{-50mm}  =  \hspace{-6mm} \sum\limits_{(k_1' , l_1') \in {\mathcal S}_{(0,0)} } \hspace{-6mm}  \left\vert  h[k_1', l_1'] \right\vert^2 \,   A_{x_s, x_s}[0,0] \, + 
       \underbrace{\sum\limits_{\substack{(k_1' , l_1'), (k_2', l_2') \in {\mathcal S}_{(0,0)} \\ (k_2', l_2') \ne (k_1', l_1')}} \hspace{-6mm} h[k_1', l_1'] \, h^*[k_2', l_2'] \, e^{j 2 \pi \frac{l_1'(k_2' - k_1')}{MN}} \, A_{x_s, x_s}[k_2' - k_1', l_2' - l_1']}_{= 0,  \,\, \mbox{\small{see Lemma}} \, \ref{eqnlemma1} \,\, \mbox{\small{in Section}} \,  \ref{secpredictspreadpilot}} \, \,  \,  =   \sum\limits_{(k_1' , l_1') \in {\mathcal S}_{(0,0)} } \hspace{-6mm}  \left\vert  h[k_1', l_1'] \right\vert^2.
        \end{eqnarray}}\normalsize
        \vspace{-3mm}
 \begin{eqnarray*}
\hline
 \end{eqnarray*}
\end{figure*}
The first term $\sum\limits_{k=0}^{M-1}  \sum\limits_{l=0}^{N-1}  \left\vert  y_{\mbox{\scriptsize{s,dd}}}[k,l]  \right\vert^2 $ does not depend on $h[k,l]$ and is therefore irrelevant to the minimization. Next, in (\ref{eqn9025440}) it is shown that in the crystalline regime, the term $\sum\limits_{k=0}^{M-1}  \sum\limits_{l=0}^{N-1}  \left\vert  h[k,l] \, *_{\sigma} \, x_{\mbox{\scriptsize{s,dd}}}[k,l]  \right\vert^2$ is equal to the total energy of the taps of $h[k,l]$.
Using this in (\ref{eqn88636}) gives (\ref{eqn001337}).
\begin{figure*}
\vspace{-6mm}
    \begin{eqnarray}
        \label{eqn001337}
        {\widehat h}_{\mbox{\scriptsize{eff}}}[k,l] & \hspace{-2mm} = & \hspace{-2mm} \arg \min_{h[k,l]}  \sum\limits_{(k,l) \in {\mathcal S}_{(0,0)}} \hspace{-3mm} \left\vert h[k,l] \right\vert^2  \, \, - \,  2 \Re \left[ \sum\limits_{(k',l') \in {\mathcal S}_{(0,0)}}  \hspace{-4mm} h^*[k',l']  \underbrace{\left( \sum\limits_{k=0}^{M-1}  \sum\limits_{l=0}^{N-1} y_{\mbox{\scriptsize{s,dd}}}[k,l] \,  \, x_{\mbox{\scriptsize{s,dd}}}^*[k - k',l - l' ] \, e^{-j 2 \pi \frac{l' (k - k')}{MN}}  \right)}_{= A_{y_s, x_s}[k', l']}  \right] \nonumber \\
        & \hspace{-2mm} = & \hspace{-2mm} \arg \min_{h[k,l]}  \sum\limits_{(k,l) \in {\mathcal S}_{(0,0)}} \hspace{-3mm}  \left( \left\vert h[k,l] \right\vert^2  \, \, - \,  2 \Re {\Big [}  h^*[k,l] \, A_{y_s, x_s}[k, l] {\Big ]} \right) \nonumber \\
        & \hspace{-2mm} = & \hspace{-2mm}  \begin{cases}
            A_{y_s, x_s}[k, l]  &, (k,l) \in  {\mathcal S}_{(0,0)} \\
            0 &, \,\, \mbox{\small{otherwise}} \\
        \end{cases}.
    \end{eqnarray}
            \vspace{-3mm}
            \begin{eqnarray*}
            \hline
            \end{eqnarray*}
\end{figure*}

    \section{Proof of Theorem \ref{thm_askl1}}
    \label{prf_thm_askl1}
    We prove (\ref{eqnaskl4}) for a spread pilot obtained by filtering a point pilot located at $(k_p, l_p)$. For simplicity, we consider $(k_p,l_p) = (0,0)$, but the result is the same for any choice of point pilot. From (\ref{eqn_spreadpilotexpr}), the spread pilot corresponding to
$(k_p , l_p ) = (0,0)$ is given by
\begin{eqnarray}
\label{eqnxsddkl198}
    x_{\mbox{\scriptsize{s,dd}}}[k,l] & \hspace{-3mm} = &  \hspace{-3mm}  \sum\limits_{n=0}^{N-1} \sum\limits_{m=0}^{M-1} w[k - nM, l - mN] \, e^{j 2 \pi \frac{n l}{N}}. 
\end{eqnarray}
From (\ref{crossambig_1}),
the self-ambiguity function of the spread pilot
is given by

{\vspace{-4mm}
\small
\begin{eqnarray}
\label{eqn_autoambigss123}
    { A}_{x_s,x_s}[k,l]  & \hspace{-3mm} = &  \hspace{-4mm} \sum\limits_{k'=0}^{M-1} \sum\limits_{l'=0}^{N-1} \hspace{-1mm} x_{\mbox{\scriptsize{s,dd}}}[k',l'] \, x_{\mbox{\scriptsize{s,dd}}}^*[k' - k, l' - l] \, e^{-j 2 \pi \frac{l (k' - k)}{MN}}. \nonumber \\
    \end{eqnarray}\normalsize}By Lemma \ref{lem_crossambig1}, we can rewrite (\ref{eqn_autoambigss123}) by extending the summation over $k'$ to $k'=0, 1, \cdots, MN -1$ and over $l'$ to $l'=0, 1, \cdots, MN -1$, i.e.

    {\vspace{-4mm}
\small
\begin{eqnarray}
\label{eqn_autoambigss1234}
    { A}_{x_s,x_s}[k,l]  &  &  \nonumber \\
    & &  \hspace{-20mm} =  \frac{1}{MN} \hspace{-2mm} \sum\limits_{k'=0}^{MN-1} \sum\limits_{l'=0}^{MN-1} \hspace{-2mm} x_{\mbox{\scriptsize{s,dd}}}[k',l'] \, x_{\mbox{\scriptsize{s,dd}}}^*[k' - k, l' - l] \, e^{-j 2 \pi \frac{l (k' - k)}{MN}}. \nonumber \\
    \end{eqnarray}\normalsize}Substituting for $x_{\mbox{\scriptsize{s,dd}}}[k',l']$ using (\ref{eqnxsddkl198}) gives

    {\vspace{-4mm}
\small
\begin{eqnarray}
\label{paper3askleqn}
    { A}_{x_s,x_s}[k,l]  &  &  \nonumber \\
    &  &  \hspace{-20mm} =  \frac{1}{M N} 
 \hspace{-2mm} \sum\limits_{k'=0}^{MN-1} \sum\limits_{l'=0}^{MN-1} 
    \sum\limits_{n_1=0}^{N-1} 
    \sum\limits_{n_2=0}^{N-1}
    \sum\limits_{m_1=0}^{M-1}
    \sum\limits_{m_2=0}^{M-1}  {\Bigg [} w[k'- n_1 M, l'- m_1 N] \, \nonumber \\
     &  &  \hspace{-13mm}  w^*[k' - k - n_2 M, l'- l - m_2 N] \, e^{j 2 \pi \frac{n_1 l'}{N}} \, e^{-j 2 \pi \frac{n_2 (l' - l)}{N}} \,  \nonumber \\
     & & \hspace{-13mm} e^{-j 2 \pi \frac{l (k' - k)}{MN}} {\Bigg ]}.
\end{eqnarray}\normalsize}Substituting for the chirp filter using (\ref{eqnwpchirp})  gives (\ref{paper3eqn824y}) (see the top of the next page).
\begin{figure*}
\vspace{-6mm}
\begin{eqnarray}
\label{paper3eqn824y}
 { A}_{x_s,x_s}[k,l]  & \hspace{-3mm} = &  \frac{e^{j 2 \pi \frac{k l}{MN}}}{M^2 N^2} \sum\limits_{l'=0}^{MN-1} 
    \sum\limits_{n_1=0}^{N-1} 
    \sum\limits_{n_2=0}^{N-1}
    \sum\limits_{m_1=0}^{M-1}
    \sum\limits_{m_2=0}^{M-1} {\Bigg [} e^{j 2 \pi q \frac{(l'- m_1 N)^2 - (l'- l -m_2 N)^2}{MN}}  \,  e^{j 2 \pi \frac{n_1 l'}{N}} \, e^{-j 2 \pi \frac{n_2 (l' - l)}{N}} \nonumber \\
    & &  \left( \frac{1}{MN} \sum\limits_{k'= 0}^{MN -1}  e^{j 2 \pi q \frac{(k'- n_1 M)^2 - (k'- k -n_2 M)^2}{MN}}  \, e^{-j 2 \pi \frac{k' l}{MN}} \right) \, {\Bigg ]} \nonumber \\
    & \hspace{-3mm} = &  \frac{e^{j 2 \pi \frac{k l}{MN}}}{M^2 N^2} \sum\limits_{l'=0}^{MN-1} 
    \sum\limits_{n_1=0}^{N-1} 
    \sum\limits_{n_2=0}^{N-1}
    \sum\limits_{m_1=0}^{M-1}
    \sum\limits_{m_2=0}^{M-1} {\Bigg [} e^{j 2 \pi q \frac{(l + (m_2 - m_1)N) (2l' - l - m_1 N - m_2 N)}{MN}}  \,  e^{j 2 \pi \frac{n_1 l'}{N}} \, e^{-j 2 \pi \frac{n_2 (l' - l)}{N}} \nonumber \\
    & &  \left( \frac{1}{MN} \sum\limits_{k'= 0}^{MN -1}  e^{j 2 \pi q \frac{(k + (n_2 - n_1)M) (2k'- k - (n_1 + n_2)M)}{MN}}  \, e^{-j 2 \pi \frac{k' l}{MN}} \right) \, {\Bigg ]} \nonumber \\
    & \hspace{-3mm} = &  \frac{e^{j 2 \pi \frac{k l}{MN}}}{M N} 
    \sum\limits_{n_1=0}^{N-1} 
    \sum\limits_{n_2=0}^{N-1}
    \sum\limits_{m_1=0}^{M-1}
    \sum\limits_{m_2=0}^{M-1} {\Bigg [}  e^{-j 2 \pi q \frac{\left( l^2 + (m_2^2 - m_1^2)N^2 + 2 l m_2 N\right)}{MN}} \, e^{j 2 \pi \frac{n_2 l}{N}} \, e^{-j 2 \pi q \frac{\left(k^2 + (n_2^2 - n_1^2)M^2 + 2 k n_2 M\right)}{MN}}  \nonumber \\
    & & \underbrace{{\Bigg (} \frac{1}{MN} \sum\limits_{l'=0}^{MN-1}
     e^{j 2 \pi l' \frac{2 q\left( l + (m_2 - m_1)N\right) + (n_1 - n_2)M}{MN}} {\Bigg )}}_{= 1 \, \, \mbox{\small{when}} \,\, (\ref{eqnaskl4}), (\ref{newconditionm1m2eqn}) \,\, \mbox{\small{hold}} \,,\, \mbox{\small{zero otherwise}}} \,\,  \underbrace{\left( \frac{1}{MN} \sum\limits_{k'= 0}^{MN -1}  e^{j 2 \pi k' \frac{\left( \, 2q (k + (n_2 - n_1)M) - l \, \right)}{MN}} \right)}_{= 1 \,\,\, \mbox{\small{when}} \,\, (\ref{klconditioneqn1})  \,,\, (\ref{paper398247}) \,\, \mbox{\small{hold}}, \,\, \mbox{\small{zero otherwise}}} \, {\Bigg ]}.
\end{eqnarray}
\vspace{-3mm}
\begin{eqnarray*}
    \hline
\end{eqnarray*}
\end{figure*}
In (\ref{paper3eqn824y}),
the inner summation over $k'$ vanishes unless $(2qk - l) \equiv 0 $ (mod $M$) so that
\begin{eqnarray}
A_{x_s,x_s}[k,l] & = & 0 \,\,,\,\, \mbox{\small{if}} \, (2qk - l) \not \equiv 0 \, (\mbox{\small{mod $M$}}).
\end{eqnarray}
If $M$ divides $(2qk - l)$, then for every $n_1 = 0,1,\cdots, N-1$, there is a unique $n_2 \in \{0, 1, \cdots, N-1 \}$ such that
\begin{eqnarray}
\label{klconditioneqn1}
    (2qk - l) + 2q(n_2 - n_1)M  & \equiv & 0 \, (\mbox{\small{mod}} \, MN)
\end{eqnarray}that is given by
\begin{eqnarray}
\label{paper398247}
n_2 = \left[n_1  \, + \, (2q)^{-1} \frac{(l - 2qk)}{M} \right]_{N}.
\end{eqnarray}The inner summation vanishes unless $(k,l)$ satisfy
\begin{eqnarray}
\label{eqn865e4}
    2qk - l & \equiv & 0 \, (\mbox{\small{mod}} \, M)
\end{eqnarray}and $(n_1, n_2)$ satisfy (\ref{paper398247}).
When these conditions are satisfied, the inner summation equals $1$.

We need only evaluate the inner summation in (\ref{paper3eqn824y}) over $l'$, for $(k,l)$ satisfying (\ref{eqn865e4}) and $(n_1, n_2)$ satisfying (\ref{paper398247}).
This summation vanishes unless
\begin{eqnarray}
\label{jkh8263}
    2 q\left( l + (m_2 - m_1)N\right) + (n_1 - n_2)M & \equiv & 0 \, \, (\mbox{\small{mod}} \, MN). \nonumber \\
\end{eqnarray}
It follows from (\ref{paper398247}) that
\begin{eqnarray}
\label{eqn873552}
    M(n_2 - n_1) & \equiv & (2q)^{-1} l -  k \,\,\, (\mbox{\small{mod}} \, MN). \nonumber \\
\end{eqnarray}
Setting $\theta = \left[ (2q)^{-1} - 2q \right]_{MN}$, it follows from (\ref{jkh8263}) and (\ref{eqn873552}) that
\begin{eqnarray}
\label{eqn824338}
2q (m_2 - m_1)N & \equiv &  \theta l - k \,\,\,\, (\mbox{\small{modulo}} \, MN).
\end{eqnarray}so that
\begin{eqnarray}
\label{eqny6443}
    k - \theta l & \equiv & 0 \,\,\, (\mbox{\small{modulo}} \, N).
\end{eqnarray}
Note that (\ref{eqn865e4}) and (\ref{eqny6443}) are the two arithmetic conditions appearing in Theorem \ref{thm_askl1} as (\ref{eqnaskl4}).
Thus the inner summation over $l'$ vanishes unless $(k,l)$ satisfy (\ref{eqnaskl4}).
In addition, it follows from
(\ref{eqn873552}) that
\begin{eqnarray}
    n_2 - n_1 & \equiv & (2q)^{-1} \, \frac{l - 2qk}{M} \,\,\, (\mbox{\small{mod}} \, N),
\end{eqnarray}and it follows
from (\ref{eqn824338}) that

{\vspace{-4mm}
\small
\begin{eqnarray}
\label{newconditionm1m2eqn}
2q (m_2 - m_1) & \hspace{-2mm} \equiv &  \hspace{-2mm} \frac{ \theta l -  k  }{N} \,\, \mbox{\small{modulo}} \,\, M. \nonumber \\
\end{eqnarray}\normalsize}These conditions appear in (\ref{paper3eqn824y}).
We assume $(k,l)$ satisfy (\ref{eqnaskl4}), and rewrite (\ref{paper3eqn824y}) as the product of two terms in (\ref{eqnnext92736}). We simplify the sum over $n_1$ by substituting for $n_2 M$ using (\ref{eqn873552}) to obtain (\ref{uywrnneqn}) (see next page). We simplify the sum over $m_1$ by substituting for $2q m_2 N$ using (\ref{eqn824338})
to obtain (\ref{uywrnneqn2}) (see next page). Combining (\ref{uywrnneqn}) and (\ref{uywrnneqn2}) in (\ref{eqnnext92736}) we obtain
Theorem \ref{thm_askl1}.


\begin{figure*}
\vspace{-6mm}
    \begin{eqnarray}
     \label{eqnnext92736}
     A_{x_s,x_s}[k,l] & = & e^{j 2 \pi \frac{k l}{MN}} 
    {\Bigg (} \frac{1}{N} \hspace{-5mm} \sum\limits_{\substack{n_1=0, \\ n_2 \, \mbox{\small{satisfies}} (\ref{paper398247})} }^{N-1} 
    \hspace{-8mm} e^{j 2 \pi \frac{n_2 l}{N}} \, e^{-j 2 \pi q \frac{\left(k^2 + (n_2^2 - n_1^2)M^2 + 2 k n_2 M\right)}{MN}} {\Bigg )}    {\Bigg (} 
    \frac{1}{M} \hspace{-5mm} 
    \sum\limits_{\substack{m_1=0, \\ m_2 \,  \mbox{\small{satisfies}} (\ref{newconditionm1m2eqn}) }}^{M-1}  \hspace{-8mm} 
      e^{-j 2 \pi q \frac{\left( l^2 + (m_2^2 - m_1^2)N^2 + 2 l m_2 N\right)}{MN}}  {\Bigg )}.
    \end{eqnarray}
    \vspace{-3mm}
     \begin{eqnarray*}
\hline
 \end{eqnarray*}
\end{figure*}



\begin{figure*}
\vspace{-6mm}
\begin{eqnarray}
\label{uywrnneqn}
 \frac{1}{N} \hspace{-5mm} \sum\limits_{\substack{n_1=0, \\ n_2 \, \mbox{\small{satisfies}} (\ref{paper398247})} }^{N-1} 
    \hspace{-8mm}  
    e^{j 2 \pi \frac{\left( n_2  M l  - q\left( k^2 + (n_2^2 - n_1^2)M^2 + 2 k n_2 M \right) \, \right)}{MN}} &  & \nonumber \\
    & & \hspace{-58mm} = 
    e^{-j 2 \pi \frac{q k^2}{MN} } \frac{1}{N} \sum\limits_{\substack{n_1=0}}^{N-1} e^{j 2 \pi \frac{(l - 2k q) (n_1 M + (2q)^{-1} l - k)}{MN} } \, e^{- j 2 \pi \frac{q}{MN} \left( ((2q)^{-1}l - k)^2  + 2 n_1 M ((2q)^{-1}l - k) \right) }
    \nonumber \\
    & & \hspace{-58mm} = e^{j 2 \pi \frac{q }{MN} \left(  ((2q)^{-1}l - k)^2 - k^2 \right)} \underbrace{\frac{1}{N} \sum\limits_{\substack{n_1=0}}^{N-1} e^{j 2 \pi \frac{n_1}{N} \left[  (l - 2qk) - 2q ((2q)^{-1}l - k)\right]}}_{= 1, \, \mbox{\small{since}} \,\, 2q((2q)^{-1} l - k) \equiv (l - 2q k) \, (\mbox{\small{modulo}} \, MN)} \nonumber \\
    & & \hspace{-58mm} = e^{j 2 \pi \frac{q }{MN} \left(  ((2q)^{-1}l - k)^2 - k^2 \right)}. 
\end{eqnarray}
\vspace{-3mm}
 \begin{eqnarray*}
\hline
 \end{eqnarray*}
\end{figure*}

\begin{figure*}
\vspace{-6mm}
\begin{eqnarray}
\label{uywrnneqn2}
 \frac{1}{M} \hspace{-5mm} \sum\limits_{\substack{m_1=0, \\ m_2 \, \mbox{\small{satisfies}} (\ref{newconditionm1m2eqn})} }^{M-1} 
 e^{-j 2 \pi q \frac{\left( l^2 + (m_2^2 - m_1^2)N^2 + 2 l m_2 N\right)}{MN}}  &  & \nonumber \\
 & & \hspace{-70mm} = \frac{1}{M} \sum\limits_{m_1 = 0}^{M-1}  \hspace{-1.5mm} e^{-j 2 \pi q \frac{l^2}{MN}} \, e^{-j 2 \pi l \frac{\left(\, 2q m_1 N + \theta l - k \, \right) }{MN}} \, e^{-j 2 \pi (4q)^{-1} \frac{\left(  \theta l - k \right)  \, \left( 4 q m_1 N +   \theta l - k \right) }{ MN}} \nonumber \\
 & & \hspace{-70mm} =  e^{-j 2 \pi q \frac{l^2}{MN}} \, e^{-j 2 \pi l \frac{\left( \theta l - k  \, \right) }{MN}} \, e^{-j 2 \pi (4q)^{-1} \frac{\left(   \theta l - k \right)^2}{ MN}} \, \left[ \frac{1}{M} \sum\limits_{m_1=0}^{M-1} e^{-j 2 \pi m_1 \frac{ \left( 2ql   +  \theta l - k \right) }{M}} \right] \nonumber \\
 & & \hspace{-70mm} =   e^{-j 2 \pi q \frac{l^2}{MN}} \, e^{-j 2 \pi l \frac{\left( \theta l - k  \, \right) }{MN}} \, e^{-j 2 \pi (4q)^{-1} \frac{\left(   \theta l - k \right)^2}{ MN}}  \, \underbrace{\left[ \frac{1}{M} \sum\limits_{m_1=0}^{M-1} e^{-j 2 \pi m_1 (2q)^{-1} \frac{ \left( l  -  2 q k\right) }{M}} \right]}_{= 1 \, \mbox{\small{since}} \, (l - 2qk) \equiv 0 \, (\mbox{\small{modulo}} M) \,,\, \mbox{\small{see}} (\ref{eqnaskl4})} \,\,  =  \,\, e^{-j 2 \pi \frac{(4q)^{-1}}{MN} \left( 2ql + \theta l - k \right)^2 }
 \end{eqnarray}
 \vspace{-3mm}
 \begin{eqnarray*}
\hline
 \end{eqnarray*}
 \end{figure*}

\section{Average energy of DD noise samples}
\label{prfnddklenergy}
We follow the noise through the
Zak-OTFS signal processing architecture described in Section \ref{seczakotfsreceiver}. At the receiver, the Zak transform ${\mathcal Z}_t$ converts AWGN $n_{\mbox{\scriptsize{td}}}(t)$
with PSD $N_0$ to its DD representation $n_{\mbox{\scriptsize{dd}}}(\tau, \nu) = {\mathcal Z}_t\left(n_{\mbox{\scriptsize{td}}}(t)\right)$. Pulse shaping with $w_{rx}(\tau, \nu)$ results in $ n_{\mbox{\scriptsize{dd}}}^{w_{rx}}(\tau, \nu) = w_{rx}(\tau, \nu) *_{\sigma} n_{\mbox{\scriptsize{dd}}}(\tau, \nu)$ which is then sampled on the information grid $\Lambda_{\mbox{\scriptsize{dd}}}$ to give the discrete DD domain quasi-periodic noise signal $n_{\mbox{\scriptsize{dd}}}[k,l]$.

In this Appendix, we show that the expected energy of each noise sample $n_{\mbox{\scriptsize{dd}}}[k,l]$ 
satisfies ${\mathbb E}\left[ \left\vert n_{\mbox{\scriptsize{dd}}}[k,l] \right\vert^2 \right] = N_0$.
We consider receive pulse shaping filters $w_{rx}(\tau, \nu)$ that factor as $w_{rx}(\tau, \nu) = w_1(\tau) \, w_2(\nu)$ where $w_1(\tau)$, $w_2({\nu})$ are unit energy pulses along the delay and Doppler axes respectively.
We assume $w_1(\tau)$, $w_2(\nu)$ satisfy the Nyquist no-ISI criterion for information spacing $1/B$ and $1/T$ along the delay and Doppler domain respectively ($\int \left\vert w_1(\tau) \right\vert^2 d\tau = \int \left\vert w_2(\nu) \right\vert^2 d\nu = 1$).

The DD representation of AWGN $n_{\mbox{\scriptsize{td}}}(t)$
is given by
\begin{eqnarray}
\label{eqn8824502}
    n_{\mbox{\scriptsize{dd}}}(\tau, \nu) & = & \sqrt{\tau_p} \sum\limits_{n \in {\mathbb Z}} n_{\mbox{\scriptsize{td}}}(\tau + n \tau_p) \, e^{-j 2 \pi n \nu \tau_p}.
\end{eqnarray}The pulse shape filtered noise is given by

{\vspace{-4mm}
\small
\begin{eqnarray}
\label{eqn8824503}
    n_{\mbox{\scriptsize{dd}}}^{w_{rx}}(\tau, \nu) & = & w_{rx}(\tau, \nu) *_{\sigma} n_{\mbox{\scriptsize{dd}}}(\tau, \nu) \nonumber \\
    &  &  \hspace{-23mm} =  \iint w_{rx}(\tau', \nu') n_{\mbox{\scriptsize{dd}}}(\tau - \tau', \nu - \nu') \, e^{j 2 \pi \nu' (\tau - \tau') } \, d\tau' \, d\nu' 
\end{eqnarray}\normalsize}
The noise signal $n_{\mbox{\scriptsize{dd}}}[k,l]$ is therefore given by

{\vspace{-4mm}
\small
\begin{eqnarray}
\label{eqn921664}
n_{\mbox{\scriptsize{dd}}}[k,l] & = & n_{\mbox{\scriptsize{dd}}}^{w_{rx}}\left(\frac{k}{B}, \frac{l}{T} \right)  \nonumber \\
& & \hspace{-18mm} \mya \sqrt{\tau_p} \sum\limits_{n \in {\mathbb Z}} \int w_1(\tau')  e^{-j 2 \pi n \frac{l \tau_p }{T}} n_{\mbox{\scriptsize{td}}}\left(\frac{k}{B} - \tau' + n \tau_p \right) \nonumber \\
& & \left[ \int w_2(\nu') e^{j 2 \pi \nu' \left(\frac{k}{B}  - \tau' + n \tau_p\right)} \, d\nu' \right] \, d\tau' \nonumber \\
& & \hspace{-18mm} \myb \sqrt{\tau_p} \sum\limits_{n \in {\mathbb Z}} e^{-j 2 \pi \frac{n l}{N}} \int w_1(\tau) \, n_{\mbox{\scriptsize{td}}}\left(\frac{k}{B} - \tau + n \tau_p \right) \nonumber \\
& & \hspace{10mm} W_2\left( \frac{k}{B} - \tau + n \tau_p \right) \, d\tau
\end{eqnarray}\normalsize}where
\begin{eqnarray}
    W_2(t) & = & \int w_2(\nu) \, e^{j 2 \pi \nu t} \, d\nu
\end{eqnarray}is the inverse Fourier transform of $w_2(\cdot)$. Step (a) follows from (\ref{eqn8824503}) after using (\ref{eqn8824502}) to substitute for $n_{\mbox{\scriptsize{dd}}}(\tau, \nu)$. Step (b) follows from the identity $T = N \tau_p$ after changing the integration variable from $\tau'$ to $\tau$.

It follows from (\ref{eqn921664}) that
$n_{\mbox{\scriptsize{dd}}}[k,l]$ is zero mean and that ${\mathbb E}\left[ \left\vert n_{\mbox{\scriptsize{dd}}}[k,l] \right\vert^2 \right]$ is given by (\ref{eqn264016}) (see top of next page).
Step (a) in (\ref{eqn264016}) follows from the fact that the autocorrelation ${\mathbb E}\left[ n_{\mbox{\scriptsize{td}}}(t) \, n_{\mbox{\scriptsize{td}}}^*(t - \tau) \right] = N_0 \delta(\tau)$.
Next, we observe that for large $M$, the spread of $w_1(\tau)$
along the delay axis is significantly less than $\tau_p$ (since $M = B \tau_p$, the spread $1/B = \tau_p/M \ll \tau_p$). Hence for $n_1 \ne n_2$, the delay domain supports of $w_1(\tau_1)$ and $w_1^*(\tau_1 + (n_2 - n_1) \tau_p)$ do not overlap. Step (b) in (\ref{eqn264016}) now follows from the fact that we can ignore all contributions to the summation for which $n_1 \ne n_2$.

The spread of the pulse $w_2(\nu)$ along the Doppler domain is roughly $1/T$. Therefore the spread of its inverse Fourier transform $W_2(t)$ is roughly $T = N \tau_p$. 
\begin{figure*}
\vspace{-8mm}
{\small
    \begin{eqnarray}
    \label{eqn264016}
        {\mathbb E}\left[ \left\vert n_{\mbox{\scriptsize{dd}}}[k,l] \right\vert^2 \right] & = & \tau_p  \sum\limits_{n_1 \in {\mathbb Z}}  \sum\limits_{n_2 \in {\mathbb Z}} e^{j 2 \pi \frac{(n_2 - n_1)l}{N}} \iint {\Bigg (} w_1(\tau_1) \, w_1^*(\tau_2) \, {\mathbb E}\left[ n_{\mbox{\scriptsize{td}}}\left(\frac{k}{B} - \tau_1 + n_1 \tau_p \right) \, n_{\mbox{\scriptsize{td}}}^*\left(\frac{k}{B} - \tau_2 + n_2 \tau_p \right)  \right] \nonumber \\
        & & \hspace{50mm} W_2\left( \frac{k}{B} - \tau_1 + n_1 \tau_p \right) \, W_2^*\left( \frac{k}{B} - \tau_2 + n_2 \tau_p \right) {\Bigg )} \, d\tau_1 \, d\tau_2 \nonumber \\
        & & \hspace{-25mm} \mya N_0 \tau_p  \sum\limits_{n_1 \in {\mathbb Z}}  \sum\limits_{n_2 \in {\mathbb Z}} e^{j 2 \pi \frac{(n_2 - n_1)l}{N}} \iint {\Bigg (} w_1(\tau_1) \, w_1^*(\tau_2) \, \delta((n_1 - n_2) \tau_p  + \tau_2 - \tau_1)   W_2\left( \frac{k}{B} - \tau_1 + n_1 \tau_p \right) \, W_2^*\left( \frac{k}{B} - \tau_2 + n_2 \tau_p \right) {\Bigg )} \, d\tau_1 \, d\tau_2 \nonumber \\
        & & \hspace{-25mm}  = N_0 \tau_p  \sum\limits_{n_1 \in {\mathbb Z}}  \sum\limits_{n_2 \in {\mathbb Z}} e^{j 2 \pi \frac{(n_2 - n_1)l}{N}}  \int  w_1(\tau_1) \, w_1^*(\tau_1 + (n_2 - n_1) \tau_p) \, \left\vert  W_2\left( \frac{k}{B} - \tau_1 + n_1 \tau_p \right) \right\vert^2 \, d\tau_1 \nonumber \\
        & & \hspace{-25mm}  \myb  N_0 \tau_p  \sum\limits_{n_1 \in {\mathbb Z}}   \int  \left\vert w_1(\tau_1) \right\vert^2  \, \left\vert  W_2\left( \frac{k}{B} - \tau_1 + n_1 \tau_p \right) \right\vert^2 \, d\tau_1
    \end{eqnarray}\normalsize}
    \vspace{-3mm}
     \begin{eqnarray*}
\hline
 \end{eqnarray*}
\end{figure*} Since $w_2(\nu)$ satisfies the Nyquist no-ISI criterion along the Doppler domain with symbol spacing $1/T$, it follows that its inverse Fourier transform (i.e., $W_2(t)$) must satisfy
\begin{eqnarray}
\label{eqn815339}
\Tilde{W_2}(t) \, \Define  \, \sum\limits_{n \in {\mathbb Z}} \left\vert W_2\left(t + nT \right) \right\vert^2 & = & \frac{1}{T}
\end{eqnarray}for all $t$. The sinc and RRC pulses used in this paper are examples of Nyquist pulses. Using (\ref{eqn815339}), we show that for all $t$

{\vspace{-4mm}
\small
\begin{eqnarray}
\label{eqn723400}
    \sum\limits_{n \in {\mathbb Z}} \left\vert W_2\left(t + n \tau_p \right) \right\vert^2 & \mya & \sum\limits_{n \in {\mathbb Z}} \left\vert W_2\left(t + n \frac{T}{N} \right) \right\vert^2  \nonumber \\
& & \hspace{-20mm}  \myb \sum\limits_{q=0}^{N-1}  \sum\limits_{m \in {\mathbb Z}} \left\vert W_2\left(t + \frac{q T}{N} + m T \right) \right\vert^2   \nonumber \\
    & & \hspace{-20mm} =  \sum\limits_{q=0}^{N-1}  \underbrace{\Tilde{W_2}\left(t + \frac{q T}{N} \right)}_{= \frac{1}{T}, \, \mbox{\small{see}} \, (\ref{eqn815339})} \, = \, \frac{N}{T}
\end{eqnarray}\normalsize}Step (a) follows from the identity $T = N \tau_p$. Step (b) results from replacing the summation index $n$ by $(mN + q)$ where $q  \equiv n$ (mod $N$). Combining (\ref{eqn264016}) and (\ref{eqn723400}) we obtain

{\vspace{-4mm}
\small
\begin{eqnarray}
\label{eqn12845529}
    {\mathbb E}\left[ \left\vert n_{\mbox{\scriptsize{dd}}}[k,l] \right\vert^2 \right] & \hspace{-3mm} = &  \hspace{-3mm} N_0 \tau_p \hspace{-1mm} \sum\limits_{n_1 \in {\mathbb Z}}   \int  \left\vert w_1(\tau_1) \right\vert^2  \, \left\vert  W_2\left( \frac{k}{B} - \tau_1 + n_1 \tau_p \right) \right\vert^2  d\tau_1 \nonumber \\
    & & \hspace{-18mm} = N_0 \tau_p \hspace{-1mm} \int \left\vert w_1(\tau_1) \right\vert^2  \, \underbrace{{\Bigg [}  \sum\limits_{n_1 \in {\mathbb Z}} \left\vert  W_2\left( \frac{k}{B} - \tau_1 + n_1 \tau_p \right) \right\vert^2  {\Bigg ]}}_{= \frac{N}{T}, \mbox{\small{see}} \, (\ref{eqn723400})} d\tau_1 \nonumber \\
     & & \hspace{-18mm} \mya N_0 \tau_p \frac{N}{T}  \, \myb \, N_0,
\end{eqnarray}\normalsize}Step (a) follows from the fact that $w_1(\tau)$ has unit energy, and step (b) follows the identity $T = N \tau_p$.

 \section{Proof of Theorem \ref{thm_927482}}
\label{prf_app_thm_927482}
In this Appendix we calculate the expected interference from data transmission to sensing. We start by combining (\ref{eqnxddklinf}) and (\ref{expradskl}) to give
(\ref{exprsimpleadskl}).
The information symbols $x[k,l]$ have unit energy and are statistically independent. Hence the mean squared interference energy is given by
(\ref{eqnmeansqrdval}) on next page. Through (\ref{ugfeu7772}), we simplify the inner summation in (\ref{eqnmeansqrdval}) using
 Lemma \ref{lem_crossambig1} in Appendix \ref{app_prop_ambig}
 (the self-ambiguity function $A_{x_s, x_s}[k,l]$ does not depend on the period over which the summation is carried out). We now obtain (\ref{eqn98635}) by substituting (\ref{ugfeu7772}) in (\ref{eqnmeansqrdval}).
 Step (a) in (\ref{eqn98635}) follows from Lemma \ref{eqnlemma1} when the (weak) crystallization condition is satisfied. 


\begin{figure*}
\vspace{-8mm}
\begin{eqnarray}
\label{exprsimpleadskl}
\sqrt{E_d}   \,  h_{\mbox{\scriptsize{eff}}}[k,l] *_{\sigma} A_{x_d,x_s}[k,l] &  & \nonumber \\
& & \hspace{-27mm} = \sqrt{\frac{E_d}{MN}} \hspace{-2mm} \sum\limits_{(k',l') \in {\mathcal S}_{(0,0)}} \hspace{-5mm} h_{\mbox{\scriptsize{eff}}}[k',l'] \, \sum\limits_{{\Tilde k} = 0}^{M-1} \sum\limits_{{\Tilde l} = 0}^{N-1}   x[{\Tilde k}, {\Tilde l}] \, x_{\mbox{\scriptsize{s,dd}}}^*[{\Tilde k} - k + k', {\Tilde l} - l + l'] \, e^{-j 2 \pi \frac{(l - l') ({\Tilde k} -k + k')}{MN} } \, e^{j 2 \pi \frac{l'(k - k') }{MN}}  \nonumber \\
& & \hspace{-32mm} \mya \sqrt{\frac{E_d}{MN}} \sum\limits_{{\Tilde k} = 0}^{M-1} \sum\limits_{{\Tilde l} = 0}^{N-1}   x[{\Tilde k}, {\Tilde l}] \, {\Bigg [} \sum\limits_{(k',l') \in {\mathcal S}_{(0,0)}}  \hspace{-5mm} h_{\mbox{\scriptsize{eff}}}[k',l']  x_{\mbox{\scriptsize{s,dd}}}^*[{\Tilde k} - k + k', {\Tilde l} - l + l'] \, e^{-j 2 \pi \frac{(l - l') ({\Tilde k} -k + k')}{MN} } \, e^{j 2 \pi \frac{l'(k - k') }{MN}}  {\Bigg ]}
\end{eqnarray}
\vspace{-3mm}
\begin{eqnarray*}
    \hline
\end{eqnarray*}
\end{figure*}

\begin{figure*}
\vspace{-6mm}
    \begin{eqnarray}
    \label{eqnmeansqrdval}
       {\mathbb E}\left[ \left\vert  \sqrt{E_d}   \,  h_{\mbox{\scriptsize{eff}}}[k,l] *_{\sigma} A_{x_d,x_s}[k,l]
  \right\vert^2 \right] &  = & \frac{E_d}{MN} \sum\limits_{{\Tilde k} = 0}^{M-1} \sum\limits_{{\Tilde l} = 0}^{N-1}  \sum\limits_{(k_1',l_1') \in {\mathcal S}_{(0,0)}}  \hspace{-1mm} \sum\limits_{(k_2',l_2') \in {\mathcal S}_{(0,0)}}  \hspace{-1mm}  {\Bigg [} h_{\mbox{\scriptsize{eff}}}[k_1',l_1'] h_{\mbox{\scriptsize{eff}}}^*[k_2',l_2'] \nonumber \\
  & & x_{\mbox{\scriptsize{s,dd}}}^*[{\Tilde k} -k + k_1' , {\Tilde l} -l + l_1'] \, x_{\mbox{\scriptsize{s,dd}}}[{\Tilde k} -k + k_2'] , {\Tilde l} -l + l_2'] \,  e^{j 2 \pi l_1' \frac{(k - k_1')}{MN}} \,  e^{-j 2 \pi l_2' \frac{(k - k_2')}{MN}}  \nonumber \\
  & & e^{-j 2 \pi \frac{(l - l_1') ({\Tilde k} - k + k_1')}{MN} } \, e^{j 2 \pi \frac{(l - l_2') ({\Tilde k} - k + k_2')}{MN} }   {\Bigg ]} \nonumber \\
  & & \hspace{-60mm}  =  \frac{E_d}{MN}  \sum\limits_{(k_1',l_1') \in {\mathcal S}_{(0,0)}}  \hspace{-1mm} \sum\limits_{(k_2',l_2') \in {\mathcal S}_{(0,0)}}  \hspace{-5mm} h_{\mbox{\scriptsize{eff}}}[k_1',l_1'] h_{\mbox{\scriptsize{eff}}}^*[k_2',l_2'] {\Bigg [ }  \sum\limits_{{\Tilde k} = 0}^{M-1} \sum\limits_{{\Tilde l} = 0}^{N-1}  x_{\mbox{\scriptsize{s,dd}}}^*[{\Tilde k} -k + k_1' , {\Tilde l} -l + l_1'] \, x_{\mbox{\scriptsize{s,dd}}}[{\Tilde k} -k + k_2' , {\Tilde l} -l + l_2'] \,  e^{j 2 \pi l_1' \frac{(k - k_1')}{MN}} \,  \nonumber \\
  & &  \hspace{20mm}  e^{-j 2 \pi l_2' \frac{(k - k_2')}{MN}} \,  e^{-j 2 \pi \frac{(l - l_1') ({\Tilde k} - k + k_1')}{MN} } \, e^{j 2 \pi \frac{(l - l_2') ({\Tilde k} - k + k_2')}{MN} }  {\Bigg ]}
    \end{eqnarray}
    \vspace{-3mm}
    \begin{eqnarray*}
    \hline
\end{eqnarray*}
\end{figure*}

\begin{figure*}
\vspace{-6mm}
    \begin{eqnarray}
    \label{ugfeu7772}
 \sum\limits_{{\Tilde k} = 0}^{M-1} \sum\limits_{{\Tilde l} = 0}^{N-1}  x_{\mbox{\scriptsize{s,dd}}}^*[{\Tilde k} -k + k_1' , {\Tilde l} -l + l_1'] \, x_{\mbox{\scriptsize{s,dd}}}[{\Tilde k} -k + k_2' , {\Tilde l} -l + l_2'] \,  e^{j 2 \pi l_1' \frac{(k - k_1')}{MN}} \, e^{-j 2 \pi l_2' \frac{(k - k_2')}{MN}} \,  e^{-j 2 \pi \frac{(l - l_1') ({\Tilde k} - k + k_1')}{MN} } \, e^{j 2 \pi \frac{(l - l_2') ({\Tilde k} - k + k_2')}{MN} } & & \nonumber \\
 & & \hspace{-180mm} = e^{j 2 \pi \frac{(l_2'- l_1') (k_1'- k)}{MN}} \, e^{j 2 \pi \frac{l (k_2'- k_1')}{MN}} \underbrace{\sum\limits_{k_0 = k_2'- k}^{M-1 +k_2' - k} \sum\limits_{l_0 = l_2'- l}^{N-1 + l_2'- l} x_{\mbox{\scriptsize{s,dd}}}[k_0, l_0] \, x_{\mbox{\scriptsize{s,dd}}}^*[k_0 + k_1'- k_2', l_0+l_1'- l_2'] \, e^{-j 2 \pi \frac{(l_2'- l_1') (k_0 + k_1'- k_2')}{MN}}}_{= A_{x_s,x_s}[k_2' - k_1' , l_2' - l_1']} \nonumber \\
  & & \hspace{-180mm} = e^{j 2 \pi \frac{(l_2'- l_1') (k_1'- k)}{MN}} \, e^{j 2 \pi \frac{l (k_2'- k_1')}{MN}} \, A_{x_s,x_s}[k_2' - k_1' , l_2' - l_1']
    \end{eqnarray}
    \vspace{-3mm}
        \begin{eqnarray*}
    \hline
\end{eqnarray*}
\end{figure*}

\begin{figure*}
\vspace{-7mm}
    \begin{eqnarray}
    \label{eqn98635}
    {\mathbb E}\left[ \left\vert  \sqrt{E_d}   \,  h_{\mbox{\scriptsize{eff}}}[k,l] *_{\sigma} A_{x_d,x_s}[k,l]
  \right\vert^2 \right] & &  \nonumber \\
  & & \hspace{-40mm} = \frac{E_d}{MN} \sum\limits_{(k_1',l_1') \in {\mathcal S}_{(0,0)}}  \hspace{-1mm} \sum\limits_{(k_2',l_2') \in {\mathcal S}_{(0,0)}}  \hspace{-5mm} h_{\mbox{\scriptsize{eff}}}[k_1',l_1'] h_{\mbox{\scriptsize{eff}}}^*[k_2',l_2'] \,  e^{j 2 \pi \frac{(l_2'- l_1') (k_1'- k)}{MN}} \, e^{j 2 \pi \frac{l (k_2'- k_1')}{MN}} \, A_{x_s,x_s}[k_2' - k_1' , l_2' - l_1'] \nonumber \\
  & & \hspace{-40mm} = \frac{E_d}{MN}  \sum\limits_{(k_1',l_1') \in {\mathcal S}_{(0,0)}} \hspace{-5mm} \left\vert  h_{\mbox{\scriptsize{eff}}}[k_1',l_1'] \right\vert^2  \nonumber \\
  & & \hspace{-40mm} +  \underbrace{\sum\limits_{(k_1',l_1') \in {\mathcal S}_{(0,0)}}  \hspace{-1mm} \sum\limits_{\substack{(k_2',l_2') \in {\mathcal S}_{(0,0)} \\  (k_2',l_2') \neq (k_1',l_1') }}  \hspace{-5mm} h_{\mbox{\scriptsize{eff}}}[k_1',l_1'] h_{\mbox{\scriptsize{eff}}}^*[k_2',l_2'] \,  e^{j 2 \pi \frac{(l_2'- l_1') (k_1'- k)}{MN}} \, e^{j 2 \pi \frac{l (k_2'- k_1')}{MN}} \, A_{x_s,x_s}[k_2' - k_1' , l_2' - l_1']}_{= 0 , \,\, \mbox{\small{see Lemma}} \, \ref{eqnlemma1} \,\, \mbox{\small{in Section}} \,  \ref{secpredictspreadpilot}} \nonumber \\
  & & \hspace{-40mm} \mya  \frac{E_d}{MN}  \sum\limits_{(k_1',l_1') \in {\mathcal S}_{(0,0)}} \hspace{-5mm} \left\vert  h_{\mbox{\scriptsize{eff}}}[k_1',l_1'] \right\vert^2.
\end{eqnarray}
    \vspace{-3mm}
     \begin{eqnarray*}
\hline
 \end{eqnarray*}
\end{figure*}


\begin{thebibliography}{1}
\bibitem{Nee2000}
R. V. Nee, and R. Prasad, “OFDM for Wireless Multimedia Communications,” Artech House Inc., 2000.

\bibitem{Wang2006}
T. Wang, J. G. Proakis, E. Masry and J. R. Zeidler, “Performance degradation of OFDM systems due to Doppler spreading,” IEEE Trans. on Wireless Commun., vol. 5, no. 6, June 2006.

\bibitem{otfsbook}
{S. K. Mohammed, R. Hadani and A. Chockalingam, ``OTFS Modulation: Theory and Applications," IEEE Press and Wiley, Nov. 2024.}

\bibitem{zakotfs1}
S. K. Mohammed, R. Hadani, A. Chockalingam and R. Calderbank, “OTFS - A mathematical foundation for communication and radar sensing in the delay-Doppler domain," IEEE BITS the Information Theory Magazine, vol. 2, no. 2, pp. 36–55, 1 Nov. 2022.

\bibitem{zakotfs2}
S.K. Mohammed, R. Hadani, A. Chockalingam, and R. Calderbank, OTFS – Predictability in the delay-Doppler domain and its value to communications and radar sensing, IEEE BITS the Information Theory Magazine, Early Access, September 2023.

\bibitem{zakotfsbook}
S.K. Mohammed, R. Hadani, A. Chockalingam, ``OTFS Modulation: Theory and Applications," Wiley publications and IEEE press, August 2024.

\bibitem{YR2013}
Y. Rahmatallah and S. Mohan, “Peak-to-average power ratio reduction in OFDM systems: A survey and taxonomy,” IEEE Commun. Surveys Tuts., vol. 15, no. 4, pp. 1567–1592, 4th Quart., 2013.

\bibitem{Thaj2022}
T. Thaj, E. Viterbo and Y. Hong, “General I/O relations and low-complexity universal MRC detection for all OTFS variants,” IEEE Access, vol. 10, pp. 96026-96037, 2022.

\bibitem{Hadani2017}
R. Hadani, S. Rakib, M. Tsatsanis, A. Monk, A. J. Goldsmith, A. F. Molisch, and R. Calderbank, “Orthogonal time frequency space modulation,” Proc. IEEE WCNC’2017, pp. 1-6, Mar. 2017.

\bibitem{Hadani2018}
R. Hadani and A. Monk, “OTFS: a new generation of modulation addressing the challenges of 5G,” arXiv:1802.02623[cs.IT], Feb. 2018.

\bibitem{Bestreadings2022}
“Best Readings in Orthogonal Time Frequency Space (OTFS) and Delay Doppler Signal Processing,” June 2022. https://www.comsoc.org/publications/best-readings/orthogonal-timefrequency-space-otfs-and-delay-doppler-signal-processing

\bibitem{Surabhi2019}
G. D. Surabhi, R. M. Augustine, and A. Chockalingam, “Peak-to-average power ratio of OTFS modulation,” IEEE Commun. Lett., vol. 23, no. 6, pp. 999-1002, Jun. 2019.

\bibitem{Wei2022}
P. Wei, Y. Xiao, W. Feng, N. Ge, and M. Xiao, “Charactering the peak-to-average power ratio of OTFS signals: A large system analysis,” IEEE Trans. Wireless Commun., vol. 21, no. 6, pp. 3705-3720, Jun. 2022.

\bibitem{Hossain2020}
M. N. Hossain, Y. Sugiura, T. Shimamura, and H.-G. Ryu, “DFT-spread OTFS communication system with the reductions of PAPR and nonlinear degradation,” Wireless Pers. Commun., vol. 115, no. 3, pp. 2211-2228, Aug. 2020.

\bibitem{Palmer2013}
J.E. Palmer, H.A. Harms, S.J. Searle and L.M. Davis, DVB-T Passive Radar Signal Processing, IEEE Transactions on Signal Processing, vol. 61, no. 8, pp. 2116-2226, April 2013.

\bibitem{Howard2004}
S. D. Howard, S. Suvorova and W. Moran, ``Waveform libraries for radar tracking applications," 2004 International Waveform Diversity \& Design Conference, Edinburgh, UK, 2004.


		\bibitem{SKM4}
		S.~K.~Mohammed, ``Derivation of OTFS Modulation from First Principles," \emph{IEEE Trans. on Vehicular Tech.}, vol. 70, no. 8, August 2021. 

 
\bibitem{EVAITU} 
ITU-R M.1225, ``Guidelines for evaluation of radio transmission technologies for IMT-2000,'' {\it International Telecommunication Union Radio communication}, 1997. 
		
  \bibitem{channel}
		P.~Raviteja, K.~T.~.Phan, Y.~Hong and E.~Viterbo, ``Interference Cancellation and Iterative Detection for
		Orthogonal Time Frequency Space Modulation,'' {\em IEEE Trans. on Wireless Comm.}, vol. 17, no. 10, Oct. 2018.   
		
  \bibitem{OTFSMP}
		L.~Gaudio, M.~Kobayashi, G.~Caire and G.~Colavolpe, ``On the Effectiveness of OTFS for Joint Radar Parameter Estimation and
		Communication,'' {\em IEEE Trans. on Wireless Comm.}, vol. 19, no. 9, Sept. 2020.  
	
        \bibitem{Bello}
        P.~A.~Bello, ``Characterization of Randomly Time-Variant Linear Channels," {\em IEEE Trans. Comm. Syst.}, vol. 11, pp. 360-393, 1963. 

\bibitem{Lampel22}
F. Lampel, A. Avarado and F. M. J. Willems, ``On OTFS using the Discrete Zak Transform," 2022 IEEE International Conference on Communications Workshops (ICC Workshops), Seoul, Korea, Republic of, 2022.


\bibitem{Hanly23}
S. Gopalam, I. B. Collings, S. V. Hanly, H. Inaltekin, S. R. B. Pillai and P. Whiting, ``Zak-OTFS Implementation via Time and Frequency Windowing," \emph{IEEE Transactions on Communications}, Early Access, Feb. 2024.



	\end{thebibliography}
\end{document}